**Evaluating the sustainability of a *de facto* harvest strategy for British Columbia's Spot Prawn (*Pandalus platyceros*) fishery in the presence of environmental drivers of recruitment and hyperstable catch rates**

Steven P. Rossi *[a] (sprossi@landmarkfisheries.com)
Sean P. Cox [a,b] (spcox@sfu.ca)
Samuel D.N. Johnson [a] (sdnjohnson@landmarkfisheries.com)
Ashleen J. Benson [a] (abenson@landmarkfisheries.com)

*Corresponding author
[a] Landmark Fisheries Research, 211-2414 St. Johns Street, Port Moody, BC, V3H 2B1, Canada
[b] School of Resource and Environmental Management, Simon Fraser University, 8888 University Drive, Burnaby, BC, V5A 1S6, Canada

Abstract

The Spot Prawn trap fishery off the west coast of British Columbia (BC) is managed using a fixed escapement strategy that aims to prevent recruitment overfishing while maximizing expected long-term yield by closing the fishery when the catch rate of spawners, projected to the following spring, drops below 1.7 spawners per trap (the *de jure* rule). We develop a management strategy evaluation framework for BC's Spot Prawn fishery that examines the expected performance of the management procedure implemented in practice (the *de facto* rule), which was significantly more conservative than the *de jure* rule, usually closing the fishery when spawner catch rates were at least twice as high as specified by the *de jure* rule. Simulations indicate that the *de facto* spawner index rule using average empirical March 31$^{st}$ targets from 2000-2019 maintains most stocks near or above 0.8 $B_{MSY}$ with or without accounting for environmental effects and/or increasing future SST on recruitment. Abundance indices were found to be strongly hyperstable, with fishing efficiency 1.5 - 3.0 times higher under low biomass than high biomass.

# 1 Introduction

Fixed escapement harvest strategies are commonly used to maximize long-term yield for fisheries that target many independent populations (Punt, 2010). For example, many commercial salmon fisheries along the west coast of North America implement fixed escapement strategies via data-intensive in-season management systems that adjust catches almost in real-time as fish migrate through gauntlet fisheries (Eggers, 1992; Michielsens and Cave, 2019). In some cases, like sockeye salmon (*Onchorhynchus nerka*) fisheries in the Fraser River, British Columbia and Bristol Bay, Alaska, fisheries occur near enough to spawning grounds that fish may be present in both escapement counts and fisheries, allowing rapid adjustment of catch and escapement in response to highly variable salmon returns.

The commercial Spot Prawn (*Pandalus platyceros*) fishery in British Columbia (BC), Canada is also managed using a fixed escapement strategy that aims to maximize long-term average yield by allowing an optimal number of female prawns to escape fisheries. However, unlike salmon fisheries, (i) spawning prawns are not actively migrating and (ii) fishing in May-June and spawning activity in March are too separated in time for in-season feedback between observed escapement and effort controls. Therefore, rather than relying on real-time feedback, fishery managers forecast the expected number of female prawns escaping to spawn the following March 31$^{st}$ based on a commercial fishery catch rate index of female prawns in May-June (hereafter the "Spawner Index") and an assumed natural mortality rate. In theory, this Spawner Index forecasting approach could achieve a target fixed escapement goal; however, uncertainty in fishery catchability, natural mortality, and recruitment create questions as to whether the approach is robust in practice. In fact, commercial Spot Prawn landings have declined 40-60% over the past decade, leaving commercial, recreational, and First Nations stakeholders concerned that the current Spawner Index targets may not be achieving the intended conservation and economic objectives for the fishery.

This paper develops a management strategy evaluation (MSE) framework for BC's Spot Prawn fishery that examines the expected performance of a *de facto* spawner index approach given our current understanding of Spot Prawn population dynamics, recruitment processes, fishery dynamics, and the role of environmental processes. Specifically, we first develop and fit a stage-structured Bayesian operating model of Spot Prawn dynamics to observed catch, catch rates, and compositional data from fisheries and surveys to estimate key population and management parameters, biological reference points, and current population status with respect to those reference points. Then, we use the operating model to simulate performance of the spawner index management approach under a range of uncertainties about population processes, observation model assumptions, and future climate change.

## 2  Methods

Spot Prawns are landed in Pacific Fishery Management Areas (hereafter "PFMAs") along the BC coast by commercial (typically May-June) and recreational (year-round) fisheries. Modelling the PFMA-specific dynamics of BC spot prawns was not feasible given the large number of PFMAs and sub-areas within each as there is only sporadic survey and spawner index sampling coverage at these small spatial scales. Therefore, we aggregated PFMAs into larger "regions" that could be modelled more tractably and that had somewhat comprehensive data coverage (Table 1). These regions – defined as Strait of Georgia (SOG), North Vancouver Island (NVI), West Coast Vancouver Island (WCVI), and Northern Shelf (NS) – generally correspond to the major biogeographic planning regions in the Canada-British Columbia MPA strategy where the prawn fishery operates (Figure 1; Doherty & Benson, 2018).

### 2.1  Data inputs

Spot Prawns are targeted by trap fisheries in rocky areas along the BC coast at depths of 40 m to 100 m. Coastwide total landings increased from 3.9 million lbs in 2000 to over 7.5 millions lbs in 2010, but have since declined to around 3.5 million lbs in recent years (Figure 2a). Most catches from 2000 to 2019 were taken in the SOG (52%), followed by the NS (18%), NVI (17%), and WCVI (12%).

Biological monitoring data available for the 2000-2019 period (except where noted below) for modelling Spot Prawn population dynamics and fisheries include (i) weekly total trap fishing catch and effort by region; (ii) weekly sub-samples of fishery catches assigned to six stages: 0) Juvenile, 1) Male, 2) Transitional, 3) Female, 4) Egged Female, and 5) Spent Female; and (iii) estimated month, year, and region-specific recreational catch (2012-2019). Spawner (transitional and female stages) frequencies in commercial fishery samples were adjusted based on the relative efficiency of each trap type used in the fishery. The adjusted frequencies were then used to calculate region-specific weekly spawner indices as the number of adjusted spawners divided by the total number of traps sampled. Weekly stage-composition observations for the commercial fishery were converted to proportion-at-stage by dividing each stage's total weekly sample size by the total sample size over all stages for that week. Few Egged or Spent Females were landed in the commercial fishery, so we combined stages 3-5 into a single female stage for the commercial stage-composition. We assumed that recreational landings for 2000-2011 were equal to their 2012-2019 mean.

**2.2     Age-/stage-structured population dynamics model**

We developed a spatial, age-/stage-structured model for the BC Spot Prawn fishery to represent both the observational data above, as well as a protandrous hermaphroditic life history, meaning they begin life as Males and transition later in life to Females, at which stage they spawn and die shortly after (Figure 3). We use age-/stage- terminology because Juvenile and Male stages correspond to ages 1 and 2, respectively, while the Transitional and Female stages co-occur during age-3 and some residual abundance occurs at egged/spent Female stages in their fourth year (although these are assumed to die soon after observations are taken). Although the Transitional and Egged/Spent stages are not particularly important for population dynamics, these stages may comprise a substantial proportion of the in-season stage-composition samples in some years and, therefore, are important in the model fitting process described below.

We modelled the dynamics of prawns from the Juvenile stage one year after hatching (i.e., Recruitment; Figure 3) through their first appearance as Males at age-2 on adult grounds, transition to Females between ages 2 and 3 or during their third year (Transitional) until their death or disappearance from adult grounds three years later. Model parameters are all presented on an annual time step but are scaled when necessary to represent shorter term dynamics. In particular, the model computations within a year are partitioned into an in-season (May-August) discrete weekly phase and post-season (September-March) continuous phase. The first day of each model year is set to the first day with observed commercial landings in any region, which was between Apr 30 and May 12 (Figure 2b). Recreational fishing mortality in each region is assumed to be constant throughout the year and is applied in both the in-season and postseason model phases.

Model notation and equations are given in Tables 2 and 3, respectively. We use a multilevel modelling approach in which parameters for fishing mortality, recruitment, stage-transitions, selectivity, initial abundance, and observation error variance are estimated at the region level, while parameters for natural mortality are modelled at the coastwide level. Note that the following description leaves out some region, year, and week indices to simplify the notation where possible.

The Spot Prawn population model is driven by a Ricker stock-recruitment function, parameterized via region-specific stock recruitment steepness ($h_r$) and unfished recruitment ($R_r^{(0)}$) parameters. These, along with unfished spawner biomass-per-recruit define the Ricker function ($\alpha, \beta$) parameters in the absence of environmental effects (OM.1-OM.3). Region-specific unfished equilibrium numbers-at-age are then derived via $R_r^{(0)}$ and survivorship-at-age (OM.4-OM.7). Note that the unfished recruitment is used to initialize Juvenile numbers in both of the first two model years (OM.4 and OM.7). Subsequent Juvenile abundances are determined via the Ricker stock-recruitment function adjusted for environmental effects and lognormal process noise (OM.8).

Changes in abundance-at-age between years in OM.9-OM.11 generally depend on post-season season total mortality rates (OM.20), which are the sum of natural mortality ($M$), recreational fishing mortality ($g$), and the length of the post-season phase ($n^{(w)}$). After abundances are updated, we assume that some region- and year-specific proportion ($p_{r,y}$) of Male prawns transition to Female between ages 2 and 3 by the commercial fishery opening and then, at the end of each fishing week, a constant proportion of the remaining Transitionals become Female ($t_r$).

In-season week-to-week dynamics are driven by total mortality (OM.17), which is the sum of weekly natural mortality ($M/52$), commercial fishing mortality ($F$; OM.14), and recreational fishing mortality ($G$; OM.16). Weekly commercial fishing mortality is modelled as the product of regional weekly total fishing effort (*E*; number of traps), fishing efficiency (*f*), and stage-based fishery selectivity ($\kappa_{s,r}^{(F)}$). Preliminary analyses estimating random effects for fishing efficiency showed a strong negative effect of vulnerable density on trap fishery efficiency at the region level. Therefore, we modelled fishing efficiency via region-specific depensatory density-dependence functions combined with region-/year-specific random effects (OM.13). Age-3 (i.e., Transitional and Female) prawns were assumed to be fully selected by the fishery, while partial selectivity for age-2 Male prawns were estimated using region-specific parameters. For recreational fisheries, we assumed that weekly fishing mortality was constant over the in-season phase via parameter *g*. We fixed recreational selectivity $\kappa_s^{(G)} = 1$ (i.e., fully-selected) for age-2 and age-3 prawns and $\kappa_s^{(G)} = 0$ for age-1 because stage-composition data are not collected for this fishery. Spawner abundance at the beginning of each week (OM.21), which is a critical state variable because it used later in the likelihood for observed spawner abundance indices, was the sum of Transitional and Female abundance.

Total mortality over the post-season continuous phase (OM.20) is the sum of recreational fishing mortality occurring outside the commercial season and a fraction of annual natural mortality *M* adjusted for the year-specific length of the fishing season.

### 2.2.1 Stock-recruitment submodels

The Ricker stock-recruitment function described in OM.8 is extended to account for environmental covariates to characterize the relationship between spawning stock biomass, age-1 recruitment, and potential impacts of climate change on recruitment success. We considered SST, the Pacific Decadal Oscillation (PDO), and North Pacific Gyre Oscillation (NPGO; Di Lorenzo et al., 2008) as environmental covariates, although we dropped the PDO because it was highly correlated with SST. In OM.8, $\alpha$ and $\beta$ are the standard Ricker parameters representing region-specific fecundity (i.e., maximum recruits-per-spawner) and density-dependence, respectively, $X_{i,r,t}$ is environmental index *i* in region *r* and year *t*, and $x_{i,r}$ is a coefficient scaling the effect of covariate *i* on log-recruits-per-spawner. The structure of the process error component ($\varepsilon_{r,y}^{(R)}$) is described below in the model fitting section. SST was obtained from the NOAA Extended Reconstructed Sea Surface Temperature (ERSST) dataset (Huang et al., 2017) and NPGO indices were obtained from http://www.o3d.org/npgo/npgo.php. Each environmental index time-series was standardized to have mean 0 and variance 1.

To test the effects of environmental covariates on prawn recruitment, we defined two OMs based on inclusion of the effects of climate-change on the stock-recruitment function:

- **Base** – No environmental covariates
- **Full** – SST and NPGO included as environmental covariates

## 2.3    Fitting the prawn model to observational data

Model-predicted spawner indices (*I*) were calculated as spawners-per-trap, multiplied by a region-level catchability effect (*q*) (OM.22), while predicted stage-composition was proportional to the selected abundance at each stage (OM.24-OM.25).

Likelihood and prior distributions are listed in Table 4. Weekly spawner indices sampled from commercial fisheries, weekly landings from commercial fisheries, and annual landings from recreational fisheries were assumed to be lognormally distributed with estimated region-specific standard deviations (L.1-L.3). The weekly stage-composition of prawns sampled from commercial fisheries was assumed to follow logistic-normal distributions (L.4).

Recreational fishing mortality (*G*) was assumed to vary as a random walk over log space (OM.27), i.e., for *y* > 1, recruitment (*R*) was calculated as

$$\log G_{r,y} = \log G_{r,y-1} + \varepsilon_{r,y}^{(G)}$$

where process errors $\varepsilon_{r,y}^{(G)}$ are assumed to arise from a mean-zero normal distributions with standard deviation $\sigma_G$. We fixed $\sigma_G$ at a small value (0.025) to prevent the random walk in recreational fishing mortality from absorbing noise due to sparsity of the recreational catch data. The proportion of age-3 prawns that were female at the fishery opening (*p*) were estimated using a similar process, except *p* followed a random walk over logit space (OM.27) and the standard deviation $\sigma_p$ on the process error likelihood was estimated by region. Recruitment and fishing efficiency deviations were assumed *iid* from zero-mean normal distributions with estimated variance by region (P.1-P.2). Weakly informative priors were specified for all other model parameters to promote convergence by ruling out implausible model values.

Natural mortality *M* was fixed at the mean of previously estimated levels from Howe Sound for 1989-1995 (mean: 0.9 yr$^{-1}$; SD: 0.4 yr$^{-1}$; range: 0.4-1.5 yr$^{-1}$). We tested OM sensitivity to values of *M* plus and minus one standard deviation from the mean, because initial trials failed to estimate *M* using an informative normal prior constructed from the above estimates.

## 2.4 Optimization

Operating model parameters were estimated by fitting the model to observed spawner indices, stage-at-age data, and landings for 2000-2019. The model was implemented using the Template Model Builder (`tmb`) package (Kristensen et al. 2016) within R version 4.1.2 (R Core Team, 2021). Bayes posterior distributions for parameters and predictive distributions for latent quantities of interest were generated via a Hamiltonian Monte Carlo (HMC; Duane et al. 1987, Neal 2011) algorithm known as the no-U-turn sampler (Hoffman and Gelman 2014) in the `tmbstan` R package (Kristensen 2018).

## 2.5 Harvest strategy simulations

The fitted operating model was used to evaluate the sustainability of the existing prawn harvest strategy as implemented via the spawner index control rule. Projections of the Spot Prawn fishery from 2020 to 2049 were based on posterior samples, the spawner index control rule (next section), and simulated environmental indices. Annual weight-at-age was drawn randomly from weight-at-age data observed in the SOG Fall and Spring fishery independent surveys from 2015 - 2019. Annual recreational fishing mortality and the proportion of age-3 prawns that were Female at the fishery opening were assumed constant and set equal to estimated 2019 levels. We simulated process error in recruitment, accounting for the average bias in historical recruitment process errors. Process errors for *p*, and *g* were set to 0. Fishing efficiency (*f*) was calculated using the hyperstability relationship estimated from the historical data with all process errors set to zero. In addition, *f* was capped at the highest estimate from 2000-2019 to prevent deviations outside the historical range.

Standardized SST time-series were almost perfectly correlated among regions (Figure A.1), so we used a single series of annual deviations from region-specific mean SST in the projections. We projected SST deviations forward in time using a zero-mean white-noise autoregressive integrated moving average (ARIMA) model, which was identified as the optimal model for the historical data via model selection. We tested the sensitivity to climate-warming under an alternative projection scenario using the Full OM, in which mean SST rose linearly by

1º C from 2020 to 2049. Similarly, we projected the NPGO index using an ARIMA(2,0,0) with lag-1 coefficient 0.76 (SE: 0.12) and lag-2 coefficient -0.19 (SE: 0.12) based on fits to historical NPGO indices (1950-2019).

**Minimum monthly index harvest control rule for commercial fisheries**

As noted above, the Spot Prawn fishery is managed using a fixed escapement strategy that aims to prevent recruitment overfishing, while maximizing expected long-term yield (Boutillier & Bond, 2000). This policy is implemented by attempting to maintain monthly spawning abundance above a threshold that decreases exponentially in proportion to the natural mortality rate of female spawners (Figure 4). Spawner indices within the harvest strategy are measured via observer catch sampling and are assumed to be directly proportional to spawning abundance; therefore, if the spawner indices within a region drop below a minimum monthly index (MMI) threshold ($I^*$) at any time during the in-season commercial fishery phase, the commercial fishery in that region closes. Note that this definition is cast in the context of this operating model spatio-temporal scale (regional), but actually operates at a finer sub-area scale within the regions. Further, analysis of the actual closure decisions suggest that some additional rules also influence closure choices, but we do not attempt to model those decisions here.

The MMI specified in the formal fishery management plan (the *de jure* rule; DFO 2022) aims to maintain at least 1.7 spawners per trap by March 31 the following year. The MMI during the in-season phase, when the commercial fishery takes place, is back-calculated from this target, accounting for continuous natural mortality between the closure of the fishery and spawning (Figure 4). As a precautionary buffer, fisheries close when the weekly spawner index is less than 1.1 times the MMI threshold, though in some areas this buffer multiplier ($b$) is as high as 1.25 (Howe Sound, Indian Arm, Powell River, Malaspina Strait/Lower Jervis and Nanaimo) or 1.5 (Saanich Inlet, Stuart Channel, Alberni Inlet).

The spawner index control rule used in the projections aimed to mimic the MMI rule as implemented in practice historically, except operating on regions rather than PFMAs or sub-areas. At a region level, effort was fairly stable for most of the season, before declining sharply in the last two weeks (e.g., Figure A.2). For each region and year, we considered the fishery "closed" when daily landings declined by 80% or more (Appendix B). The commercial fishery has closed at progressively earlier dates and higher spawner index levels since 2000, with all closures since 2010 occurring in June (Figure 2b). Since 2010, the spawner index at closure has been approximately twice the MMI threshold (Figure 2c). In the projections, we used the realized mean spawner index at closing for 2000-2019 to determine when fisheries closed, rather than the documented rule of applying a $b$=1.1-1.5 multiplier. The mean spawner indices at closing implied multipliers of 2.56 for SOG, 2.46 for NVI, 2.90 for WCVI, and 2.26 for NS, essentially reflecting a far more conservative approach than stated in the fishery management plan. We refer to this historical average MMI rule as the *de facto* spawner index rule.

Given the relative stability of effort at the region level over the fishing season, we assumed that effort in the projections was region and year-specific, but constant over weeks of the open season. We determined the annual weekly effort by first fitting region-specific linear regressions between observed landings in a fishing season and the mean effort across the first four weeks of the following fishing season (Figure A.3). The effect ($\beta$) of catch on effort in the following season was significant for SOG ($\beta$=2.81e-5, $p$<0.05) and WCVI ($\beta$=5.11e-5, $p$<0.05), so we projected effort in those regions via the fitted linear model. Linear regressions were not significant for NVI ($p$=0.254) and NS ($p$=0.481), so we projected effort in those regions based on mean historical effort (NVI: 78,600 traps; NS: 76,100 traps).

**Reference points**

To estimate reference points, we projected spot prawn biomass and fishery yield using the simulation process described above (e.g., constant effort within a region/year, with fisheries closing when $I < bI^*$) over a range of $b$ values, from $b$=0.1 to $b$=4.0. Smaller values of $b$ imply a later fishery closing time and thus higher cumulative effort, so this formulation is analogous to running simulations over a range of effort. For each $b$, the simulations were repeated over a sample of 1000 draws from both the Base and Full model posteriors, each averaged over three natural mortality scenarios (explained below). From the resulting yield curves, we estimated maximum sustainable yield (MSY) and the corresponding biomass ($B_{MSY}$) for each region as well as the coast-wide aggregate.

**Performance metrics**

The *de facto* spawner index rule was evaluated based on a suite of conservation and catch performance metrics (Table 5), including:

- Probability that biomass is below the LRP (pLRP);
- Probability that biomass is greater than the USR (pHealthy);
- Mean catch for 2020-2029 (avgC10)
- Mean catch for 2020-2049 (avgC30)
- Average annual variation in catch for 2020-2049 (AAV)
- Mean number of weeks the fishery is open (nOpen)
- Mean cumulative annual effort for 2020-2049 (eff)

At the present time, there are no formally adopted biological reference points as specified by Canadian fisheries policy. Therefore, we chose the limit reference point (LRP) of $0.3B_{MSY}$ based on the default North Atlantic Fishery Organization LRP (Hvingel and Orr, 2011) and an upper stock reference (USR) of $0.8B_{MSY}$. The LRP defines the upper bound of the Critical zone, while the USR defines the lower bound of the Healthy stock status zone (DFO 2006).

## 2.7 Model averaging

The natural mortality of Spot Prawns is difficult to estimate given their short life span and duration in fishery samples. We included uncertainty in M by combining posterior samples from model runs under each of the three assumed *M* values (0.5, 0.9, 1.3) to create ensemble estimates for each stock-recruitment scenario. For a given model variable or performance metric $\chi$, we calculate an ensemble or average $\chi$ as

$$\bar{\chi} = \left(\lambda_1 \chi^{(M=0.5)} + \lambda_2 \chi^{(M=0.9)} + \lambda_3 \chi^{(M=1.3)}\right)/3$$

where $\chi^{(M=m)}$ represents the variable or performance metric associated with the *M=m* run and $\lambda$ is the weight associated with each *M*. Weights for *M* were set to their respective relative probabilities under a Normal distribution with mean $M = 0.9$ yr$^{-1}$ and standard deviation 0.4 yr$^{-1}$, implying relative weights $\lambda_1 = \lambda_3 = 0.274$ and $\lambda_2 = 0.452$ (Boutillier and Bond 2000).

## 3 Results

## 3.1 Model fits

Spawner index residual standard errors ranged from 0.15 in SOG to 0.32 in NVI for both the Base and Full OMs (Table 6). Observation error for spawner indices, commercial landings, and stock-composition was smallest in SOG (Table 6) mainly because of higher fishing effort and sampling intensity in this region. In some relatively rare region-year combinations where there is no clear in-season decline of observed spawner indices, the model captures only the mean of the data, but not the in-season trend (e.g., Figure 5, NS-2009).

Each OM fit the observed landings closely in all regions (Table 6, Figure A.2). Model-estimated recreational landings matched mean observed catches during the period 2012-2019 in which landings data was recorded (Figure A.4), though the estimated landings are less variable than observed, suggesting closer model fits could be achieved with a less restrictive random-walk in recreational fishing mortality. Recreational catches prior to 2012 deviated from the assumed constant mean, however this deviation was considered acceptable given the lack of data from this time period.

The mean stage-composition in the model matched the data (Figure A.5). Each OM captured the general in-season trend of the stage-composition, namely, decreasing relative abundance of Transitionals and increasing relative abundance of Males and Females over the course of the fishing season (Figure A.6-A.9). Models fits to the stage-composition data improved under the Full model including environmental effects in the stock-recruitment function as indicated by lower residual standard errors (Table 6).

## 3.2 Parameter estimates

Including environmental covariates in the Full model had little impact on most fishing efficiency, proportion female, stage-transition, and catchability parameters (Table 7); therefore, we only present these estimates for the Base model. Stock-recruitment parameters are presented below for both Base and Full models.

Relatively strong hyperstability in fishing efficiency (i.e., fishing efficiency increases as abundance decreases) was present in all regions (Table 7; Figure 6) resulting in relatively stable proportional harvest rates of 60% in SOG and NVI, 20-30% in NS, and 20-40% in WCVI (Figure 7; bottom row).

The estimated proportion of age-3 prawns that were Female on June 1 in SOG, NVI, and WCVI declined by 27-35% between 2000 and 2019 (Figure 7, top row), suggesting a trend toward later maturity and increasing prevalence of Transitional prawns in the fishery catches. The proportion of age-3 prawns in NS that were Female on June 1 declined by 30% between 2003 and 2009 but has since increased by 13% (Figure 7, top row). Prawns in NS matured later than prawns in other regions across all model years, except for two years (2014 and 2015) of exceptionally late maturity in NVI (Figure 7, top row). Transition probabilities 0.12-0.15 wk$^{-1}$ implied approximately 19-24 weeks to complete >95% of transitions for SOG, NVI, and WCVI, while 0.06 wk$^{-1}$ for NS implied approximately 50 weeks to complete >95% of transitions, which suggests a slower maturation process overall in northern BC.

In general, stock-recruitment steepness ($h$) was lower (i.e., less productive) and more variable among regions for the Full model including environmental covariates compared to the Base model, while unfished recruitment ($R^{(0)}$) showed no consistent directional differences between models (Table 7).

Estimated spawning biomass grew in each region in between 2002 and 2009, and began to decline thereafter (Figure 7, middle row). Recent spawning biomass is estimated to be relatively low, though SOG biomass has increased by more than 80% since 2014 (Figure 7, middle row).

Estimated recruitment patterns varied across regions, but in general have trended lower over the long term, except for recent years in SOG and NVI where recruitment has been increasing (Figure A.10, left column). Recruitment patterns appear to be driven by changes in productivity as indicated by estimates of recruits-per-spawner (Figure A.10, right column). SOG and NVI productivity rebounded strongly in the last decade, whereas weak productivity has persisted in WCVI and NS.

Productivity at low spawner biomass (i.e., stock-recruitment steepness) was similar across regions in the Base model, but varied widely in the Full model, with relatively high productivity in WCVI and relatively low productivity in NS (Figure 8; Table 7). Including environmental effects on productivity resulted in lower steepness estimates than models with no environmental effects. SST was associated with slightly higher productivity, although the effect was somewhat marginal as posterior probability distributions included 0 (Figure 9). In contrast, a negative relationship between NPGO and productivity was estimated for all regions with >99% of probability mass less than zero (Figure 9). When both SST and NPGO covariates were included, the negative NPGO covariate tended to dominate the net effect. Overcompensation (i.e., reduced recruitment at high spawner biomass) was evident for WCVI and NS in the Base model and WCVI in the Full model (Figure 8).

### 3.3 Reference points

Yield in SOG and NVI showed the strongest influence of the March 31$^{st}$ target, because high average annual fishing effort in those regions is capable of depleting those stocks to their specified targets regardless of how low the target index might be (Figure 10). In contrast, the annual average fishing effort is not high enough in more remote regions, such as WCVI and NS, to deplete the stock to low March 31$^{st}$ targets. Therefore, yield in these regions does not decline very much for low March 31$^{st}$ targets.

Natural mortality scenarios had relatively small impacts on the region-specific relationships between the March 31$^{st}$ target spawner index and available yield, as well as the implied optimal March 31$^{st}$ target spawner index. In general, the lower natural mortality rate (0.5 yr$^{-1}$) generated slightly higher peak yields across regions, but also required the highest target spawner indices, while the highest mortality scenario (1.3 yr$^{-1}$) generated the lowest peak yield at lower March 31$^{st}$ targets.

The Full Model showed larger and more variable effects of climate indices compared to the natural mortality scenarios. Including climatic influences on the stock-recruitment relationship shifted both peak yield and the optimal March 31$^{st}$ target higher for SOG and NVI, and lower for WCVI and NS (Figure 10). The Full Model also showed greater differences among natural mortality scenarios for SOG and NVI compared to very little difference for WCVI and NS.

The coastwide yield, which is just the sum of region-specific yields, is mainly dominated by SOG and NVI, so it does show a depletion effect for low March 31$^{st}$ targets; however, the lack of depletion effect for WCVI and NS shifts the optimal March 31$^{st}$ target considerably lower than the implied values for SOG and NVI.

The coastwide MSY was 4.0 million lbs in the Base model and 3.7 million lbs in the Full model, with SOG contributing 49-55% of those totals (Table 8). Estimated spawning biomass for SOG, NVI, and NS were consistently in the Healthy zone, while harvest rates were generally near or below U$_{MSY}$ (Figure 11). In contrast, WCVI biomass was occasionally near or below the LRP, despite occasional pulses of high biomass (Figure 11).

### 3.5 Harvest strategy simulations

Under the Base operating model scenario, the *de facto* spawner index rule maintained Spot Prawn biomass in the Healthy zone (>0.8$B_{MSY}$) at least 55% of the time across all regions and more than 88% of the time at the coastwide aggregate level (Table 9). The probability of Spot Prawn biomass below the LRP was 2-15 times higher for the WCVI region (11%) compared to other regions under the *de facto* spawner index rule.

Projections of the *de facto* spawner index rule under the Full operating model with constant mean SST were less optimistic, with only 39-66% of simulated biomasses in the Healthy zone and 2-30% below the LRP (Table 9, SST-Const). Results under a 1º C increase in mean SST over the projection period were slightly more optimistic in SOG than the Full operating model with constant mean SST, but less optimistic for NVI, WCVI, and NS, with simulated biomasses in the Healthy zone between 31-50% in the regions (Table 9, SST-Inc).

The lower limits of projected biomass under *de facto* management were similar under the Base and Full model, though the Full model produced higher biomass targets. Therefore, while biomass only tended to drop below the LRP in WCVI and NS in the Base model projections, biomass in all regions was prone to in Full *de facto* projections.

The Full model estimated a negative effect of NPGO on recruitment and the NPGO index has strongly declined since 2012, with 2018 and 2019 values the lowest on record. This produces a strongly positive effect of recruitment across all regions in 2019 and 2020, which leads to high biomass in the opening years of the projection (Figure 12). This strong recruitment effect is partly caused by a lack of data for these year classes, as these prawns have yet to be landed to be landed as 2- or 3-year-olds, therefore there is relatively information about the size of these cohorts. These recruitment estimates may be revised significantly downward as new data are added.

## 4     Discussion

We developed an MSE framework to evaluate the *de facto* spawner index management procedure for BC's Spot Prawn fishery. Simulations indicate that the *de facto* spawner index rule using average empirical March 31$^{st}$ targets from 2000-2019 maintains stocks near or above 0.8 $B_{MSY}$ with or without accounting for environmental effects and/or increasing future SST on recruitment. The only exception occurred in the WCVI region, where the stock was Healthy less than 50% of the time and below the LRP 32% and 39% of time under environmental drivers and increasing SST, respectively. Stock status for WCVI and NS appears to be partially impaired by productivity, which declined to low levels across all regions in the early 2010s and has largely failed to recover for WCVI and NS.

The *de jure* March MMI target of 1.7 female spawners per trap was originally derived by comparing fishery-independent survey catches in Knight Inlet and Kingcome Inlet with historical recruitment indices from commercial observations (Cadrin et al., 2004; Smith 2013). This MMI target was later increased by 10-50% to provide a buffer for recreational landings and because spawner-recruitment analysis of fishery independent survey data from Howe Sound suggested a higher March MMI target of 3.9 spawners per trap (Bond and Boutillier, 2000). Our study demonstrates that the *de facto* rule implied MMI targets of 3.8 to 4.9 spawners per trap. The more conservative *de facto* rule produced yields that were 30% than the *de jure* rule in SOG and 16% higher coastwide. The reasons for the discrepancy between the *de jure* and *de facto* rules was not investigated and would need to be understood before formally changing harvest control rules.

A major initial goal of this research was to incorporate climate change into a management strategy evaluation framework for BC Spot Prawn. Environmental factors may primarily affect fish populations by altering early life (i.e., pre-recruit) survival, so we explored possible environment-recruitment correlations, though we propose no specific mechanism by which environment may alter pre-recruit survival. The relationship between recruitment and temperature is expected to be positive at the colder limits of a species range and negative at the warmer limit of the range (Myers 1998). Spot Prawn range from Unalaska (Butler, 1980) to Baja California (Ortega-Lizárraga et al., 2021), with productive populations concentrated between the Canada/Alaska border and San Diego (Lowry 2007). BC Spot Prawn therefore inhabit the upper half of the Spot Prawn range, with some NS subpopulations existing near the upper limits of the range. We estimated a positive SST-recruitment relationship in all regions when NPGO effects were fixed at 0; however, once we accounted for NPGO, we estimated the SST-recruitment relationship to be neutral in SOG and NVI and mostly negative in NS.

While many studies have correlated recruitment with a variety of environmental indices, including temperature (e.g., Flowers and Saila, 1972), salinity (e.g., Vance et al., 1985), and wind (e.g., Rae 1957), use of environmental indices to predict recruitment in practice is rare (Myers 1998). Establishing environment-recruitment relationships is difficult because correlations could be masked by other factors (e.g., nonlinearity, multi-dimensionality, indirect effects) or because detected correlations are spurious (e.g., failure to account for spatial heterogeneity or time lags) (Keyl and Wolff, 2008), leading to correlations often breaking down upon being retested with additional data (Myers 1998). Our study exemplifies the potential pitfalls of incorporating environmental effects into recruitment predictions without sufficient scrutiny. The strong estimated effect of NPGO in our model, coupled with record low estimates of NPGO in the terminal model years, leads to overly optimistic projections of spawning biomass. Given the relatively short timeframe that these NPGO-recruitment relationships were estimated over, it is likely that these relationships could change as new data is observed. In particularly it may be expected that the model will estimate a more moderate impact of NPGO on recruitment as new data is added, unless very high recruitment is observed after 2020.

The *de jure* MMI rule assumes that Spawner Index data collected from the fishery is directly proportional to spawner abundance; however, our analysis found that a negative relationship between abundance and fishing efficiency caused catch rates to remain high amid declining abundance. This hyperstability in the Spawner Index creates conservation risks by causing management decisions to be based on overly optimistic impressions of stock status. The *de facto* implementation of the MMI rule has possibly offset the risks of incorrectly assuming CPUE varies linearly with abundance by using much more conservative Spawner Index targets (Figure 10).

Environmental processes may influence vulnerability to fishing or catchability by altering species distributions (e.g., Engelhard et al., 2014) or foraging behaviours. For instance, ambient temperature regulation of metabolic and digestive rates (Gillooly et al., 2001) increases feeding frequency and may increase catchability or vulnerability to fisheries. Trap catchability was positively correlated with mobility in Antillean reef fishes (Robichaud et al., 2000) and with temperature in several reef fishes along the southeast US Atlantic coast (Bacheler and Shertzer, 2020). Experiments demonstrating a positive relationship between temperature and activity in Moreton Bay Brown Tiger Prawn (*Penaeus esculentus*; Hill, 1985) have been incorporated into catchability estimates for that stock (Kienzle et al., 2016). In contrast, temperature had little effect on trawl catchability of three species of juvenile penaeid prawns in Australia (Vance and Staples, 1992) and on trawl catchability of Smooth Pink Shrimp (*Pandalus jordani*) off Vancouver Island (Perry et al., 2000). We did not explicitly account for effects of temperature on catchability or fishing efficiency, though deviations of fishing efficiency from its underlying relationship with abundance may implicitly account for environmental effects, including from temperature. There was no clear relationship between estimated fishing efficiency deviations and SST; deviations and SST had a weakly negative correlation in SOG and WCVI and a weakly positive correlation in NVI and NS (Figure A.11). In all cases the relationships between fishing efficiency deviations and temperature were noisy and linear models fitted to these relationships explained little of the variance (Figure A.11). Environmental processes can also influence the mortality of recruited prawns. For instance, temperature-induced changes in foraging patterns may affect availability not just to fisheries, but also to predators.

We estimated a trend of age-3 prawns maturing later in the summer. It is not clear if this trend reflects a response to changes in reproductive capacity or reflects a change in growth rates (e.g., Koeller, 2006). *Pandalus jordani* near Oregon and Northern California were reported to skip the male phase of their life-history and mature directly as females in response to reduced population-level reproductive capacity (Charnov et al., 1978); however, there is no evidence of Spot Prawns skipping their male phase (e.g., there is no bimodality in the length distributions of females from surveys to suggest the presence of an early-maturing cohort). It is not clear whether the estimated change in maturation timing for BC Spot Prawn affects reproductive capacity since all prawns in that cohort are assumed to be female once spawning begins (e.g., the timing of maturation in the summer has no effect on sex-ratio at spawning). The estimated trend toward later maturity for BC Spot Prawn does not appear to have affected size-at-spawning, as the weight or length of spawning females in SOG surveys has been relatively stable (Figures A12-A13).

We assumed that Spot Prawns had a 4-year lifespan in which Females spawned only in their first year as females and did not contribute to further spawning events; however, older life-history stages of Spot Prawn are highly uncertain. In BC, female Spot Prawns leave the rocky trapping area in the spring after eggs have hatched and are assumed to die soon after as they are not observed again (Butler, 1980). In contrast, a lifespan of at least 5 years was estimated from commercial landings of Spot Prawn in Southern California (Sunada 1986; Lowry 2007) and at least 6-7 years was estimated from tagging data for Spot Prawn in Prince William Sound Alaska (Kimker et al., 1996). Females were reported to have spawned multiple times in Prince William Sound (Lowry, 2007) and in captivity (Rensel and Prentice, 1977). The disappearance of

Females from fishing grounds in BC after first spawning may not be indicative of death, but rather an ontogenic migration to lower depths outside of fishing range (Lowry 2007), which have been reported for Pandalids (e.g., Shumway et al., 1985). It is also possible that multiparous females are landed in fisheries and surveys but are not sufficiently distinguishable from primiparous females due to reduced growth at older life-history stages. If multiparous females significantly contribute to overall fecundity, then our model will underestimate productivity.

The level of potential fishery exploitation depends on effort capacity. The WCVI and NS regions lack sensitivity to low Spawner Index targets because effort capacity is not high enough to deplete these stocks to their MMI within a single season unless the stock is at very low abundance, in which case effective fishing effort is higher because of hyperstability. Effort response across years was significant in SOG and WCVI, but not NVI and NS (Figure A.3). The lack of significance for the latter could be for different reasons. The NS region is remote and possibly simply lacks effort capacity as noted above. In contrast, the NVI region lies somewhat central to all other regions and often absorbs transient effort as harvesters move among the other areas.

### 4.1 Limitations

The specific grouping of PMFA areas into regions was to some degree arbitrary and could warrant further investigation. In particular, the NS region is an amalgamation of somewhat distinct areas, which contributes to a lack of clear depletion signals across some years. Parameter estimates for NS are also different than for other regions, as both catchability and the in-season transition rate are significantly lower in NS than in other regions, and maturation timing for age-3 prawns has been steady for the last two decades in contrast to later maturation in other regions. These estimates may accurately reflect differences arising from the NS ecosystem or may arise spuriously due to less informative data in that region or due to the area-composition of the NS region. Model fit and inference may be improved by partitioning the NS region into separate regions for Haida Gwaii (areas 1-2, 101-102, 142), the north coast (areas 3-6, 103-106), and the central coast (areas 7-10, 107-110) as fishery dynamics are relatively distinct among these areas.

There was little information in the fishery data about $M$, and estimating $M$ was intractable without an unreasonably tight prior, so we relied on model averaging to incorporate uncertainty about $M$ into our analysis, using estimates of $M$ derived from catch curves (Boutillier and Bond

2000) to weight the relative plausibility of each model. The principle of continuous model expansion suggests that inference should be made from the Full model, as this model contains the Base model as a special case. Given the relatively short time-series in our analysis, it is possible that the estimated effect of NPGO on recruitment by the Full model is overstated. We therefore resist making inferences solely from the Full model (as would be suggested by concept of continuous model expansion) or averaging the Base and Full models using formal metrics of model fit. It may instead be more appropriate to combine estimates from the Base and Full models in an unweighted ensemble (i.e., with equal weight given to both models).

Recreational fishery landings represent a data gap in our analysis; data prior to 2012 was unavailable and there was no age/stage-composition data. We therefore needed to make assumptions about the magnitude and variance of recreational catches prior to 2012 and about recreational fishery selectivity. Better fits to the observed recreational landings could be achieved by allowing more variance in recreational fishing rates, however this could lead to the recreational fishing mortality absorbing noise from other processes, rendering estimates unreliable.

The Government of Canada is currently undertaking an extensive initiative to create spatial conservation areas (e.g., MPA Network 2022). We did not incorporate conservation areas into our modelling framework. Coupling spatial closures with the current fishery management system would be required to ensure stock and fishery sustainability.

## 4.2 Lessons for harvest strategy simulations

*Transition period from historical estimates to future projections*

Transitioning from historical stock status (i.e., stock assessment) to future projections (i.e., harvest strategy simulation) requires careful treatment of stochastic processes connecting the two. For instance, recruitment estimates in the last few years of the historical period are typically highly uncertain, yet are also highly influential on the first few years or even decades of harvest strategy projections and expected performance. This is important because, in theory, harvest strategy choices should be made based on long-term expected performance, but in practice, the first few years are often most critical to harvesters and fishery managers who ultimately must adopt formal harvest strategies.

We estimated a negative effect of NPGO on recruitment from the Full model, which, in response to record low NPGO values in the terminal years of the analysis, produced record high recruitment in the opening projection years, leading to decades of elevated biomass. This effect is possibly overstated or entirely spurious, therefore using this model to inform management decisions may create overharvesting risks. The propensity of correlations between population processes and environmental factors to break down as new data is added suggests that harvest strategies relying on these models need to be updated more often than otherwise.

*Implications for stock status and future strategy*

Spot Prawn biomass in each region has been in the Cautious zone in recent years, with WCVI biomass near the LRP. Simulations show that the *de facto* rule should allow biomass to recover, particularly if environmental factors exert little influence over recruitment. The *de facto* rule should also allow biomass to recover if SST and NPGO affect recruitment as estimated in the Full model; however, the probability of biomass dropping below the LRP was higher when environmental factors were considered. Validating the estimated environmental effects with additional data will be important for understanding conservation risks in each region.

Overcompensation in recruitment suggests that higher productivity and catches could be achieved in some regions with slight effects on conservation risks. For instance, if the March MMI target in NS was shifted from the *de facto* 3.8 spawners per trap to 3.5 spawners per trap, average annual projected landings in the Base model were 76% higher while probability of dropping below the LRP increased by 0.04. Similarly, shifting the WCVI March MMI target from 4.9 spawners per trap to 4.5 spawners per trap increased average annual projected landings by 18% while the probability of dropping below the LRP decreased by 0.03.

## Acknowledgements

The Pacific Prawn Fishermen's Association and the BC Salmon Restoration and Innovation Fund contributed funding to this work. We thank Emma Atkinson for providing valuable feedback on earlier versions of this manuscript.

**Tables**

*Table 1. List of regions in which BC prawns are commercially landed, along with the Pacific Fishery Management Areas that comprise each region.*

| $r$ | Code | Region Name | PFMAs |
|---|---|---|---|
| 1 | SOG | Strait of Georgia | 13-19, 28-29 |
| 2 | NVI | North Vancouver Island | 11-12 |
| 3 | WCVI | West Coast Vancouver Island | 20-27, 121, 123-127 |
| 4 | NS | Northern Shelf | 1-10, 101-111, 130, 142 |

*Table 2. Operating model notation.*

| Symbol | Value | Description |
| --- | --- | --- |
| *Indices* | | |
| $Y$ | 20 | Number of years |
| $W$ | 12 | Number of weeks in summer fishing season |
| $r$ | 1, 2, 3, 4 | Region (1=SOG, 2=NVI, 3=WCVI, 4=NS) |
| $y$ | 1, 2, …, $Y$ | Year (1=2000, 2=2001, …, $Y$=2019) |
| $w$ | 1, 2, …, $W$+1 | Week |
| $s$ | 1, 2, 3, 4 | Stage (1=Juvenile, 2=Male, 3=Transitional, 4=Female) |
| *Inputs* | | |
| $I_{r,y,w}$ | | Observed region $r$ spawner index in week $w$ of year $y$ |
| $C^{(F)}_{r,y,w}$ | | Observed commercial catch in region $r$ in week $w$ of year $y$ ('000 lbs) |
| $C^{(G)}_{r,y}$ | | Observed landings in region $r$ in year $y$ (millions) |
| $u_{s,r,y,w}$ | | Observed stage-composition by region, year, and week |
| $\kappa^{(G)}_s$ | {0,1,1,1} | Recreational fishery selectivity-at-stage |
| $X_{i,r,y}$ | | Recruitment covariates by region and year |
| $\tau^{(G)}_y$ | 0.5 for $y$<2002, 0.2 for $y$≥2002 | Recreational catch observation error variance |
| $b_t$ | range: 0.75-0.79 | Proportion of postseason elapsed at spawning (March 1) |
| *Parameters* | | |
| $M$ | 0.9 | Natural mortality rate (year$^{-1}$) |
| $h_r$ | Est. | Stock-recruitment steepness |
| $R^{(0)}_r$ | Est. | Unfished recruitment (millions) |
| $x_{i,r}$ | Est. | Recruitment effect of covariate $i$ in region $r$ |
| $\psi_r$ | Est. | Hyperstability parameter |

| $g_{r,y}$ | Est. | Recreational fishing mortality by region and year (year$^{-1}$) |
| $p_{r,y}$ | Est. | Proportion of age-3 prawns that are female at commercial fishery opening by region and year |
| $t_r$ | Est. | In-season transition rate by region (week$^{-1}$) |
| $\varepsilon_{r,y}^{(R)}$ | Est. | Recruitment process errors by region and year |
| $\varepsilon_{r,y}^{(f)}$ | Est. | Fishing efficiency process errors by region and year |
| $\varepsilon_{s,r,y}^{(init)}$ | Est. | Initial abundance process errors |
| $\kappa_{s,r}^{(F)}$ | Est. | Commercial fishery selectivity-at-stage in region $r$ |
| $\tau_r^{(I)}$ | Est. | Spawner indices observation error variance |
| $\tau^{(F)}$ | Est. | Commercial catch observation error variance |
| $\tau^{(u,F)}$ | Est. | Commercial stage-composition variance |
| $\sigma_{p,r}$ | Est. | Proportion female at fishery opening process error standard deviation |
| $\sigma_{R,r}$ | Est. | Recruitment process error standard deviation |
| $\sigma_{f,r}$ | Est. | Fishing efficiency process error standard deviation |
| $\sigma_g$ | 0.025 | Recreational fishing mortality process error standard deviation |

*Latent variables*

| $N_{s,r,y,w}$ | | Abundance at the beginning of week $w$ by stage, region and year (millions) |
| $F_{s,r,y,w}$ | | Instantaneous commercial fishing mortality rate by stage, region, year, and week (year$^{-1}$) |
| $G_{s,r,y}$ | | Instantaneous recreational fishing mortality rate by stage, region, and year (week$^{-1}$) |
| $D_{s,r,y,w}$ | | In-season deaths by stage, region, year, and week (millions) |
| $Z_{s,r,y}^{(post)}$ | | Instantaneous postseason mortality rate by stage, region and year |
| $f_{r,y}$ | | Commercial trap efficiency by region and year (trap$^{-1}$) |

| | |
|---|---|
| $B_{r,y}$ | Spawning biomass by year and region ('000 lbs) |
| $\omega_s$ | Weight-at-stage ('000 lbs) |
| $\omega^{(B)}$ | Spawner weight at time of spawning ('000 lbs) |
| $\phi$ | Unfished spawner biomass per recruit ('000 lbs) |
| $\alpha_r$ | Maximum recruits per spawner in region $r$ |
| $\beta_r$ | Ricker density-dependence in region $r$ |
| $\hat{I}_{r,y,w}$ | Predicted spawner index by region, year and week (females/trap) |
| $C^{(F)}_{s,r,y,w}$ | Predicted commercial catch by stage, region, year and week (millions) |
| $\hat{C}^{(F)}_{r,y,w}$ | Predicted commercial catch in region $r$ in week $w$ of year $y$ ('000 lbs) |
| $\hat{C}^{(G)}_{r,y}$ | Predicted recreational catch in region $r$ in year $y$ (millions) |
| $\hat{u}_{s,r,y,w}$ | Predicted stage-composition by region, year, and week |
| $\varepsilon^{(p)}_{r,y}$ | Transition probability process errors by region and year |
| $\varepsilon^{(g)}_{r,y}$ | Recreational fishing mortality process errors by region and year |

*Table 3. Operating model equations.*

| Model quantity | Formula |
|---|---|
| *Life-history / demographics* | |
| (OM.1) Unfished spawner biomass per recruit | $\phi = \exp(-3M)\,\omega^{(B)}$ |
| (OM.2) Maximum recruits per spawner | $\alpha_r = 5h_r^{5/4}/\phi$ |
| (OM.3) Ricker density-dependence | $\beta_r = \dfrac{\log\left(5h_r^{5/4}\right)}{R_r^{(0)}\phi}$ |
| *Initialization of commercial fishing season* | |
| (OM.4) Juvenile/male abundance at start of week 1, $y=1$, $s<3$ | $N_{s,r,y=1,w=1} = R_r^{(0)} \exp\left(-(s-1)M + \varepsilon_{s,r,y=1}^{(\text{init})}\right)$ |
| (OM.5) Transitional abundance at start of week 1, $y=1$ | $N_{s=3,r,y=1,w=1} = R_r^{(0)} \exp\left(-2M + \varepsilon_{s=3,r,y=1}^{(\text{init})}\right)(1 - p_{r,y=1})$ |
| (OM.6) Female abundance at start of week 1, $y=1$ | $N_{s=4,r,y=1,w=1} = R_r^{(0)} \exp\left(-2M + \varepsilon_{s=4,r,y=1}^{(\text{init})}\right) p_{r,y=1}$ |
| (OM.7) Juvenile abundance at start of week 1, $y=2$ | $N_{s=1,r,y=2,w=1} = R_r^{(0)} \exp\left(\varepsilon_{s=1,r,y=2}^{(\text{init})}\right)$ |
| (OM.8) Juvenile abundance at start of week 1, $y>2$ | $N_{s=1,r,y,w=1} = \alpha_r B_{r,y-1} \exp\left(-\beta_r B_{r,y-2} + \sum_i x_{i,r} X_{i,r,y-1} + \varepsilon_{r,y}^{(R)}\right)$ |
| (OM.9) Male abundance at start of week 1, $y>1$ | $N_{s=2,r,y,w=1} = N_{1,r,y-1,n_w+1} \exp\left(-Z_{s=1,r,y=y-1}^{(\text{post})}\right)$ |
| (OM.10) Transitional abundance at start of week 1, $y>1$ | $N_{s=3,r,y,w=1} = N_{2,r,y-1,n_w+1} \exp\left(-Z_{s=2,r,y=y-1}^{(\text{post})}\right)(1 - p_{r,y})$ |
| (OM.11) Female abundance at start of week 1, $y>1$ | $N_{s=4,r,y,w=1} = N_{2,r,y-1,n_w+1} \exp\left(-Z_{s=2,r,y=y-1}^{(\text{post})}\right) p_{r,y}$ |

*In-season dynamics (May-Aug)*

| (OM.12) Fishing efficiency | $f_{r,y} = \dfrac{\exp\left(\varepsilon_{r,y}^{(f)}\right)}{\psi_r \omega^{(B)} \sum_{s=3}^{4} N_{s,r,y,w}}$ |
| --- | --- |
| (OM.13) Commercial fishing mortality rate | $F_{s,r,y,w} = \kappa_{s,r}^{(F)} f_{r,y} E_{r,y,w}$ |
| (OM.14) Recreational fishing mortality rate | $G_{s,r,y} = \kappa_s^{(G)} g_{r,y}/52$ |
| (OM.15) Total weekly mortality | $Z_{s,r,y,w} = F_{s,r,y,w} + G_{s,r,y} + M/52$ |
| (OM.16) Total weekly deaths | $D_{s,r,y,w} = N_{s,r,y,w}\left(1 - \exp(-Z_{s,r,y,w})\right)$ |
| (OM.17) Juvenile/male abundance at start of week $w$, $w > 1$, $s < 3$ | $N_{s,r,y,w} = N_{s,r,y,w-1} \exp(-Z_{s,r,y,w-1})$ |
| (OM.18) Transitional abundance at start of week $w$, $w > 1$ | $N_{s=3,r,y,w} = N_{3,r,y,w-1} \exp(-Z_{3,r,y,w-1})(1 - t_r)$ |
| (OM.19) Female abundance at start of week $w$, $w > 1$, $s < 3$ | $N_{s=4,r,y,w} = N_{4,r,y,w-1} \exp(-Z_{4,r,y,w-1})$ $+ N_{3,r,y,w-1} \exp(-Z_{3,r,y,w-1}) t_r$ |

*Post-season dynamics (Sep-Mar)*

| (OM.20) Total postseason mortality | $Z_{s,r,y}^{(\text{post})} = \left(\kappa_s^{(G)} g_{r,y} + M\right)\left(52 - n^{(w)}\right)/52$ |
| --- | --- |
| (OM.21) Spawning biomass | $B_{r,y} = \sum_{s=3}^{4} N_{s,r,y,w=n^{(w)}} \exp\left(-b_t Z_{s,r,y}^{(\text{post})}\right) \omega^{(B)}$ |

*Predicted observations*

| (OM.22) Spawners caught per trap | $\hat{I}_{r,y,w} = q_r \sum_{s=3}^{4} C_{s,r,y,w}^{(F)}/E_{r,y,w}$ |
| --- | --- |
| (OM.23) Weight of weekly commercial catch | $\hat{C}_{r,y,w}^{(F)} = \sum_s D_{s,r,y,w} \omega_s F_{s,r,y,w}/Z_{s,r,y,w}$ |
| (OM.24) Predicted fishery stage-composition | $\hat{u}_{s,r,y,w} = \dfrac{C_{s,r,y,w}^{(F)}}{\sum_j C_{j,r,y,w}^{(F)}}$ |

(OM.25) Number of prawns caught by recreational fishery

$$\hat{C}^{(G)}_{r,y} = \sum_s \sum_w C^{(G)}_{s,r,y,w}$$

*Process errors*

(OM.26) Proportion female at beginning of year process errors

$$\varepsilon^{(p)}_{r,y} = \text{logit}\, p_{r,y} - \text{logit}\, p_{r,y-1}$$

(OM.27) Recreational fishing mortality process errors

$$\varepsilon^{(g)}_{r,y} = \log g_{r,y} - \log g_{r,y-1}$$

*Table 4. Likelihood and prior distributions for BC spot prawn operating model. Note that standard deviations are provided instead of variances in some Normal distribution specifications where conflicting super-scripts would occur or where we use numerical constants.*

| Model component | Function |
| --- | --- |
| | *Data likelihood* |
| (L.1) Spawner indices | $\log I_{r,y,w} \sim N\left(\log \hat{I}_{r,y,w}, \tau_r^{(I)}\right)$ |
| (L.2) Commercial landings | $\log C_{r,y,w}^{(F)} \sim N\left(\log \hat{C}_{r,y,w}^{(F)}, \tau_r^{(F)}\right)$ |
| (L.3) Recreational landings | $\log C_{r,y}^{(G)} \sim N\left(\log \hat{C}_{r,y}^{(G)}, \tau_y^{(G)}\right)$ |
| (L.4) Commercial stage-composition | $\hat{\boldsymbol{u}}_{r,y,w} \sim P\left(N\left(\boldsymbol{u}_{r,y,w}, \tau_{r,y}^{(u)}\right)\right)$ |
| | *Prior distribution* |
| (P.1) Recruitment deviations | $\varepsilon_{r,y}^{(R)} \sim N(0, \sigma_{R,r}^2)$ |
| (P.2) Commercial trap efficiency deviations | $\varepsilon_{r,y}^{(f)} \sim N(0, \sigma_{f,r}^2)$ |
| (P.3) Prop. female deviations | $\varepsilon_{r,y}^{(p)} \sim N(0, \sigma_{p,r}^2)$ |
| (P.4) Recreational fishing mortality deviations | $\varepsilon_{r,y}^{(g)} \sim N(0, \sigma_g^2)$ |
| (P.5) Steepness | $h_r \sim N(0.8, 0.2)$ |
| (P.6) Unfished recruitment | $R_r^{(0)} \sim N(300, 300)$ |
| (P.7) Initial abundance deviations | $\varepsilon_{a,r,y}^{(\text{init})} \sim N(0,1)$ |
| (P.8) Fishing efficiency density-dependence | $\psi_r \sim N(0.00075, 0.1)$ |
| (P.9) Recruitment deviation SD | $\sigma_{R,r} \sim \Gamma(2, 0.5)$ |
| (P.10) Prop. female deviation SD | $\sigma_{p,r} \sim \Gamma(2, 0.5)$ |
| (P.11) Trap efficiency deviation SD | $\sigma_{f,r} \sim \Gamma(2, 0.5)$ |

(P.12) Environmental effects $\quad x_{i,r} \sim N(0,1)$

Table 5. Management procedure performance metrics computed from operating model projections over 1000 random parameter draws from the joint posterior distribution. The indicator function $1(X)$ takes value 1 when X is true and 0 otherwise. The variable j indexes posterior draw and r indexes region. Projections begin in 2020 (y=21) and end in 2049 (y=50).

| Metric | Definition |
|---|---|
| $pLRP_r$ | $P(\boldsymbol{B}_r < LRP) = \sum_{y=21}^{50} \sum_j \frac{1(B_{j,r,y} < 0.4\, B_{MSY,r})}{30 \cdot 2000}$ |
| $pHealthy_r$ | $P(\boldsymbol{B}_r > USR) = \sum_{y=21}^{50} \sum_j \frac{1(B_{j,r,y} > 0.8\, B_{MSY,r})}{30 \cdot 2000}$ |
| $avgC10_r$ | $\overline{C10}_r = \text{median} \left\{ \frac{\sum_{y=21}^{30} \sum_w \widehat{C}^{(F)}_{j,r,y,w}}{10} \right\}_{j=1}^{j=1000}$ |
| $avgC30_r$ | $\overline{C30}_r = \text{median} \left\{ \frac{\sum_{y=21}^{50} \sum_w \widehat{C}^{(F)}_{j,r,y,w}}{30} \right\}_{j=1}^{j=1000}$ |
| $AAV_r$ | $AAV_r = \text{median} \left\{ \frac{\sum_{y=22}^{50} \left| \sum_w \widehat{C}^{(F)}_{j,r,y,w} - \sum_w \widehat{C}^{(F)}_{j,r,y-1,w} \right|}{\sum_{y=21}^{50} \sum_w \widehat{C}^{(F)}_{j,r,y,w}} \right\}_{j=1}^{j=1000}$ |
| $nOpen_r$ | $nOpen_r = \text{median} \left\{ \frac{\sum_{y=21}^{50} \{\max w : I_{j,r,y,w} > I^*_w\}}{30} \right\}_{j=1}^{j=1000}$ |
| $eff_r$ | $eff_r = \text{median} \left\{ \frac{\sum_{y=21}^{50} \sum_w E_{j,r,y,w}}{30} \right\}_{j=1}^{j=1000}$ |

$$\overline{C10}_r = \text{median}\left\{\frac{\sum_{y=21}^{30} \sum_w \widehat{C}^{(F)}_{j,r,y,w}}{10}\right\}_{j=1}^{j=1000}$$

$$\overline{C10}_r = \text{median}_j \frac{\sum_{y=21}^{30} \sum_w \widehat{C}^{(F)}_{j,r,y,w}}{10}$$

*Table 6. Estimated standard deviations for process and observation errors. Estimates of $\tau^{(u,F)}$ represent means across years.*

| | Process errors | | | Observation errors | | |
|---|---|---|---|---|---|---|
| | Recruitment | Proportion female | Fishing efficiency | Weekly spawner indices | Weekly fishery landings | Stage proportions |
| | $\sigma^2_{R,r}$ | $\sigma^2_{p,r}$ | $\sigma^2_{f,r}$ | $\tau^{(I)}_r$ | $\tau^{(F)}_r$ | $\tau^{(u)}_r$ |
| **Base OM** | | | | | | |
| SOG | 0.45 | 0.46 | 0.24 | 0.15 | 0.09 | 0.52 |
| NVI | 0.29 | 0.96 | 0.33 | 0.32 | 0.09 | 0.63 |
| WCVI | 0.71 | 0.76 | 0.24 | 0.23 | 1.71 | 0.77 |
| NS | 0.50 | 0.33 | 0.23 | 0.24 | 0.13 | 0.52 |
| | | | | | | |
| **Full OM** | | | | | | |
| SOG | 0.42 | 0.48 | 0.24 | 0.15 | 0.09 | 0.49 |
| NVI | 0.33 | 0.95 | 0.33 | 0.32 | 0.09 | 0.59 |
| WCVI | 0.67 | 0.77 | 0.25 | 0.23 | 1.71 | 0.63 |
| NS | 0.34 | 0.34 | 0.22 | 0.23 | 0.13 | 0.49 |

Table 7. Leading parameter estimates (posterior mode with associated central 95% credibility interval) for the Base OM and Full OM.

| Parameter | SOG | NVI | WCVI | NS |
|---|---|---|---|---|
| **Base Model** | | | | |
| $h$ | 1.04 (0.79, 1.32) | 0.98 (0.75, 1.25) | 0.94 (0.73, 1.30) | 1.00 (0.78, 1.29) |
| $R^{(0)}$ | 313 (268, 783) | 95 (82, 146) | 163 (121, 306) | 356 (284, 484) |
| $\psi$ (x10e-4) | 6.3 (5.5, 7.0) | 5.9 (5.0, 6.8) | 6.6 (5.3, 8.8) | 7.6 (6.7, 8.4) |
| $p_{y=1}$ | 0.52 (0.39, 0.65) | 0.87 (0.81, 0.90) | 0.96 (0.93, 0.97) | 0.40 (0.32, 0.49) |
| $p_{y=Y}$ | 0.29 (0.15, 0.58) | 0.39 (0.08, 0.88) | 0.40 (0.11, 0.82) | 0.41 (0.25, 0.58) |
| $t$ | 0.15 (0.13, 0.16) | 0.14 (0.12, 0.16) | 0.12 (0.09, 0.15) | 0.06 (0.04, 0.07) |
| $q$ | 0.87 (0.84, 0.90) | 0.75 (0.71, 0.81) | 0.90 (0.69, 1.17) | 0.57 (0.54, 0.60) |
| **Full model** | | | | |
| $h$ | 0.67 (0.51, 0.95) | 0.69 (0.53, 0.95) | 0.83 (0.57, 1.11) | 0.49 (0.35, 0.72) |
| $R^{(0)}$ | 422 (308, 751) | 133 (103, 279) | 136 (95, 216) | 354 (286, 496) |
| $\psi$ (x10e-4) | 6.3 (5.6, 7.1) | 6.0 (5.2, 7.3) | 8.2 (6.0, 10.8) | 7.8 (6.9, 8.8) |
| $p_{y=1}$ | 0.55 (0.42, 0.66) | 0.86 (0.81, 0.90) | 0.96 (0.93, 0.97) | 0.40 (0.32, 0.49) |
| $p_{y=Y}$ | 0.32 (0.14, 0.59) | 0.38 (0.08, 0.86) | 0.34 (0.12, 0.79) | 0.41 (0.25, 0.59) |
| $t$ | 0.14 (0.13, 0.16) | 0.14 (0.12, 0.16) | 0.13 (0.09, 0.15) | 0.06 (0.04, 0.07) |
| $q$ | 0.87 (0.84, 0.90) | 0.75 (0.71, 0.81) | 0.87 (0.68, 1.15) | 0.57 (0.54, 0.60) |
| $x_{SST}$ | 0.04 (-0.22, 0.26) | -0.18 (-0.43, 0.04) | -0.40 (-0.80, -0.06) | -0.41 (-0.65, -0.21) |
| $x_{NPGO}$ | -0.24 (-0.41, -0.11) | -0.59 (-0.88, -0.36) | -0.48 (-0.77, 0.22) | -0.51 (-0.68, -0.32) |

Table 8. Estimated biological reference points for the Base and Full operating models by region and coastwide (CW). March MMI is the target March MMI implied by the rule that produced MSY. Estimates were averaged over three models with varying M values. Units for MSY and BMSY are thousands of pounds.

| Model | SOG | NVI | WCVI | NS | CW |
|---|---|---|---|---|---|
| **MSY** | | | | | |
| Base | 1986 | 770 | 451 | 1107 | 4032 |
| Full | 2043 | 717 | 373 | 905 | 3722 |
| | | | | | |
| **BMSY** | | | | | |
| Base | 496 | 170 | 316 | 804 | 1946 |
| Full | 634 | 182 | 295 | 692 | 2388 |
| | | | | | |
| **March MMI** | | | | | |
| Base | 3.2 | 3.2 | 1.2 | 1.0 | 2.9 |
| Full | 3.6 | 3.1 | 1.4 | 1.2 | 3.2 |

Table 9. Simulated performance over 1000 stochastic simulation trials of the Status Quo spawner index rule for operating model scenarios involving environmental effects on stock-recruitment (S-R) and sea surface temperature trend (SST). "Const" indicates scenarios in which the mean SST was constant while "Inc" indicates scenarios increase mean SST linearly by one degree Celsius over the 30-year projection. All results represent ensemble averages over natural morality scenarios.

| S-R | SST | pLRP | pHealthy | avgC10 | avgC30 | AAV | nOpen | eff |
|---|---|---|---|---|---|---|---|---|
| **SOG** | | | | | | | | |
| Base | - | 0.007 | 0.661 | 1715 | 1739 | 0.362 | 6.9 | 1.32 |
| Full | Const | 0.019 | 0.564 | 1809 | 1930 | 0.346 | 7.5 | 1.43 |
| Full | Inc | 0.019 | 0.601 | 1843 | 1944 | 0.347 | 7.6 | 1.4 |
| **NVI** | | | | | | | | |
| Base | - | 0.005 | 0.817 | 609 | 616 | 0.276 | 5.2 | 0.42 |
| Full | Const | 0.104 | 0.566 | 644 | 722 | 0.425 | 5.9 | 0.47 |
| Full | Inc | 0.126 | 0.496 | 558 | 680 | 0.431 | 5.7 | 0.45 |
| **WCVI** | | | | | | | | |
| Base | - | 0.110 | 0.546 | 293 | 299 | 0.494 | 7.3 | 0.22 |
| Full | Const | 0.298 | 0.394 | 213 | 248 | 0.684 | 6.2 | 0.18 |
| Full | Inc | 0.345 | 0.312 | 167 | 221 | 0.710 | 5.7 | 0.17 |
| **NS** | | | | | | | | |
| Base | - | 0.037 | 0.807 | 407 | 424 | 0.377 | 3.9 | 0.30 |
| Full | Const | 0.085 | 0.656 | 358 | 433 | 0.538 | 4.0 | 0.31 |
| Full | Inc | 0.180 | 0.475 | 243 | 361 | 0.567 | 3.5 | 0.27 |
| **CW** | | | | | | | | |
| Base | - | 0.001 | 0.880 | 3067 | 3113 | 0.228 | 6.9 | 2.26 |
| Full | Const | 0.023 | 0.580 | 3073 | 3381 | 0.289 | 7.5 | 2.41 |
| Full | Inc | 0.057 | 0.457 | 2864 | 3247 | 0.293 | 7.6 | 2.33 |

**Figures**

*Figure 1. Pacific Fishery Management Areas (numbers) in British Columbia in which actual Spot Prawn fisheries are managed along with shaded areas indicating the "regions" defining the operating model.*

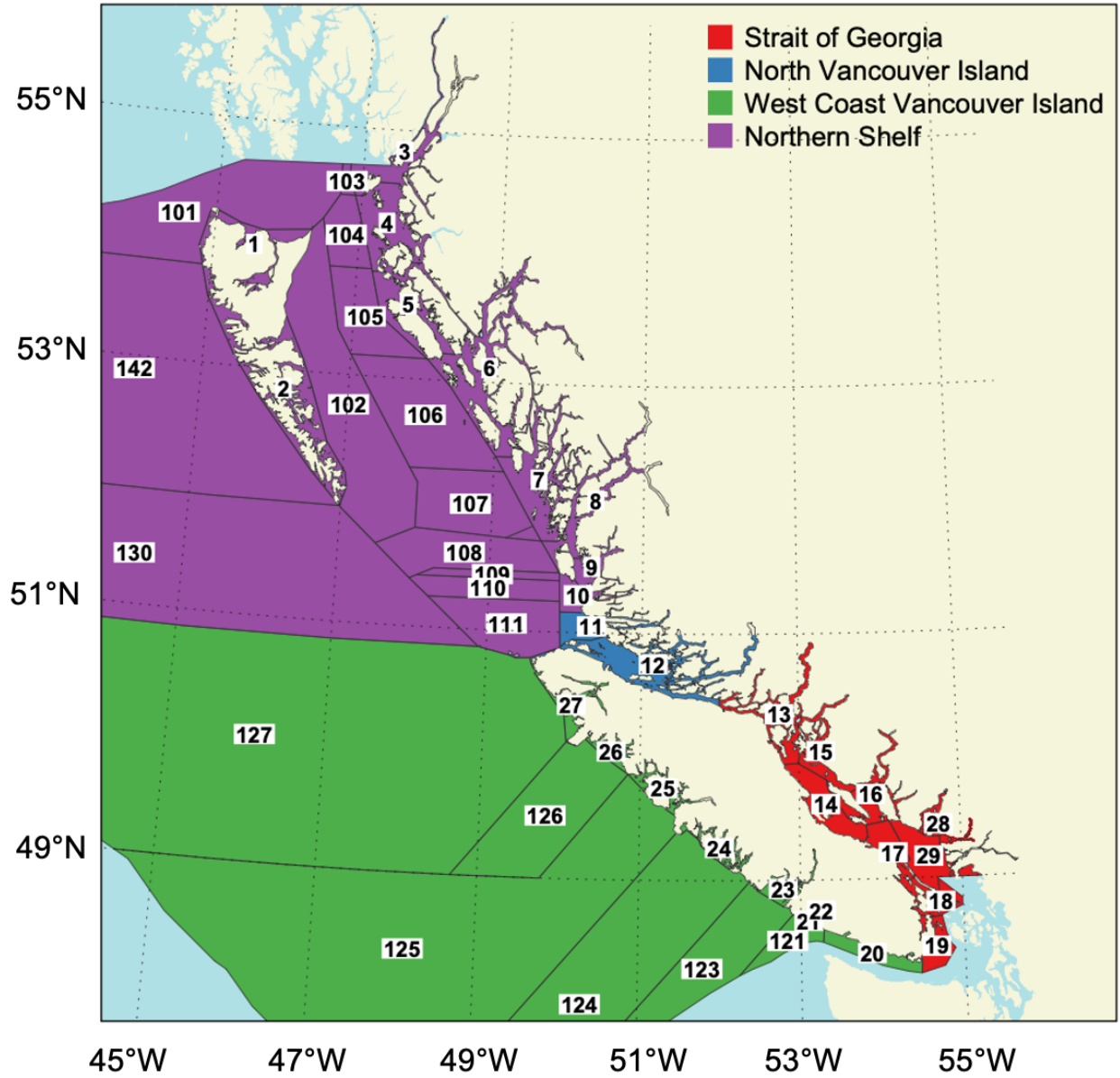

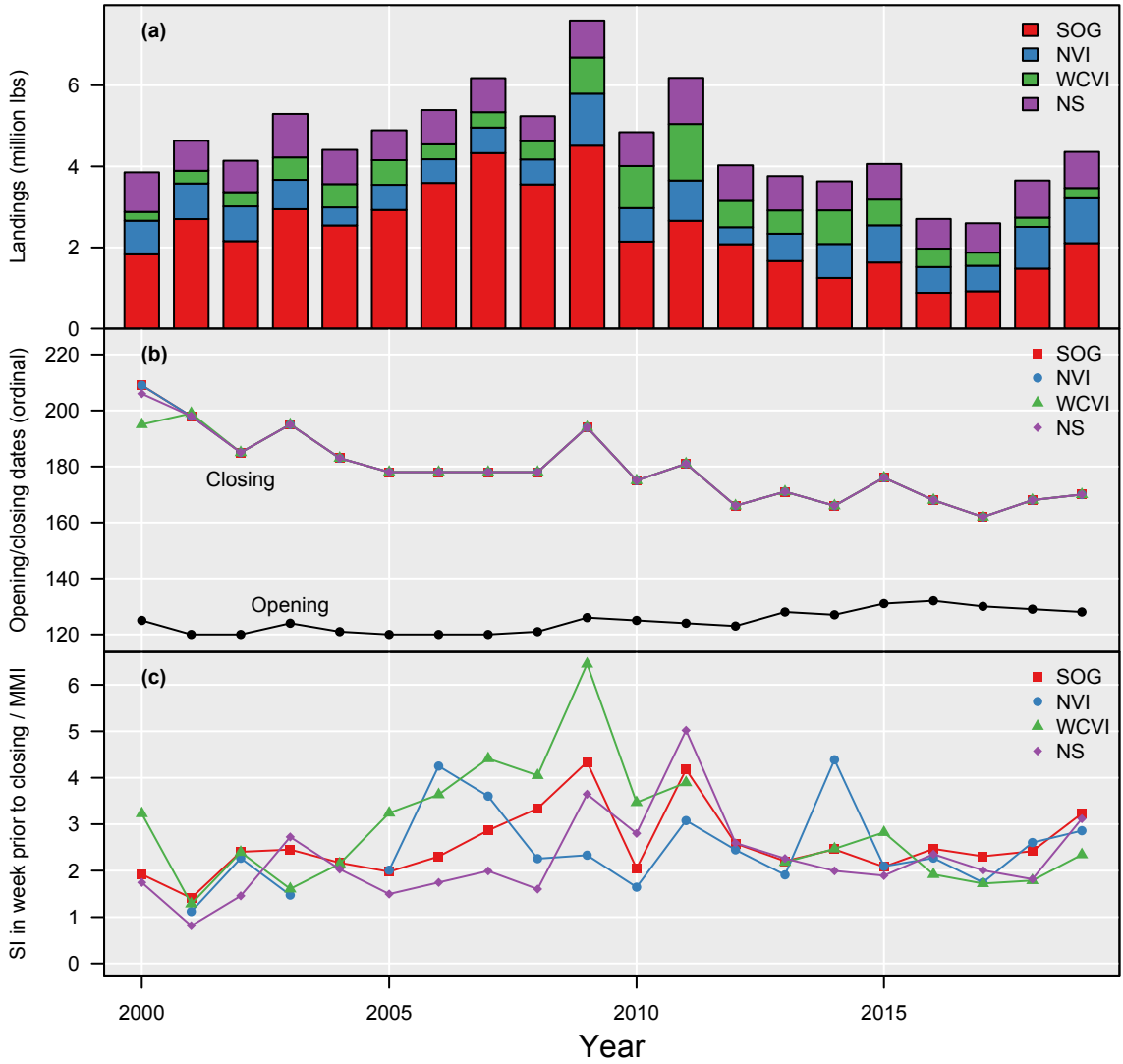

*Figure 2. Annual Spot Prawn commercial fishery region-level summary for 2000-2019, including: (a) Landings. (b) Opening and closing dates. The broken horizontal line indicates July 1 on a non-leap year. (c) Spawner indices relative to the MMI in the week prior to fishery closure in the region.*

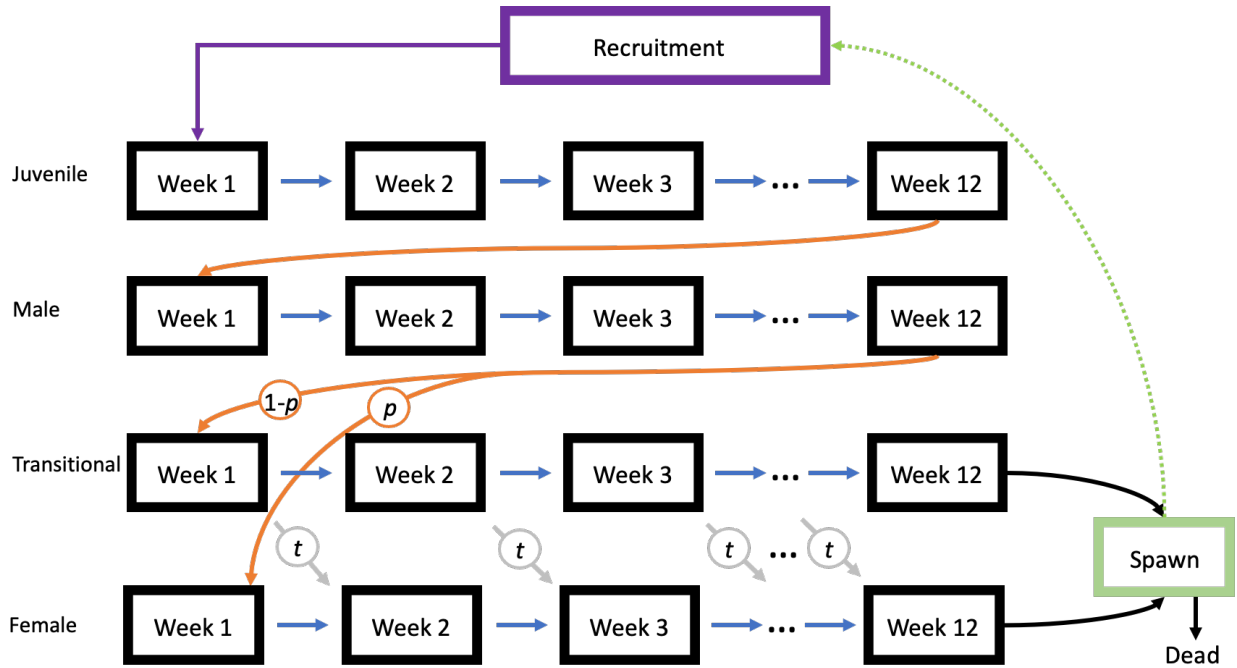

*Figure 3. Schematic representation of the stage-structured model for BC Spot Prawn. The weekly boxes represent the in-season model phase generally over May-June, while lines connecting Spawn-Recruitment, Recruitment-Juvenile, and Week 12-Transitional/Female represent continuous processes over the July-April period. Parameter p is the annual proportion of males that transition to female over the continuous phase and 1-p is the proportion that transition during the in-season phase with parameter t representing the weekly transition rate.*

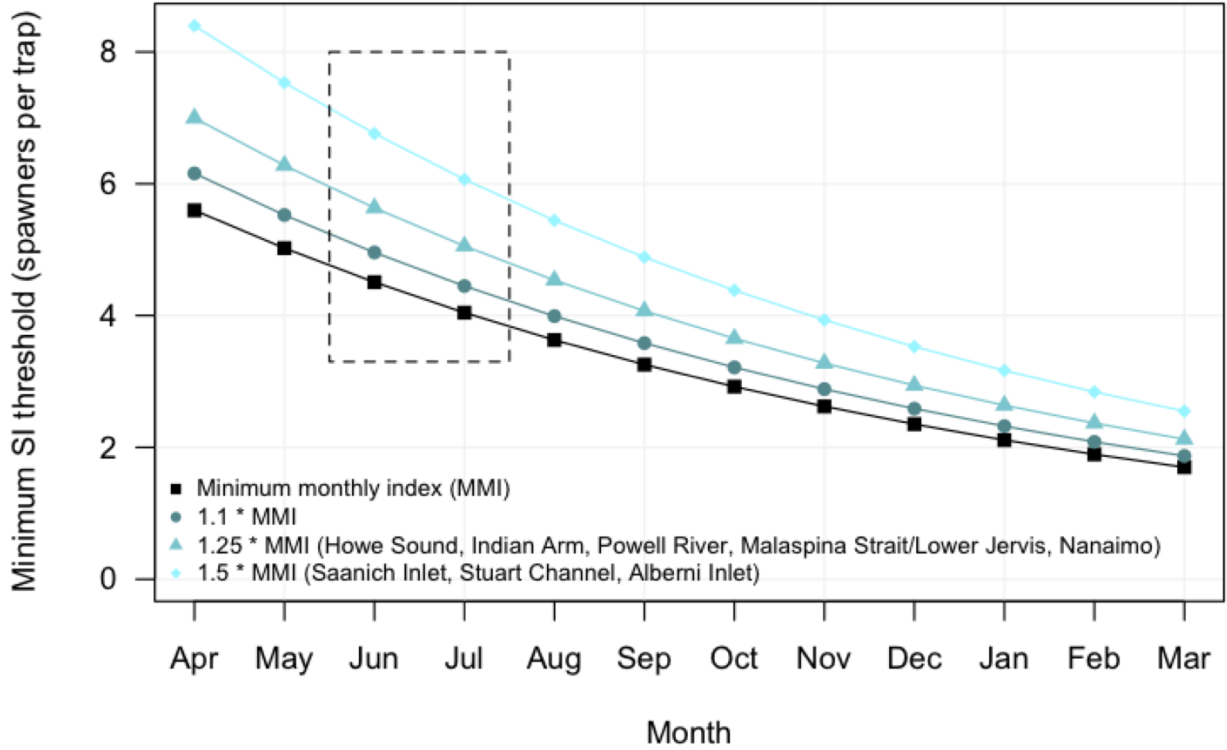

*Figure 4. Minimum monthly index (MMI) threshold for implementing the spawner index rule in the Spot Prawn fishery. The baseline MMI (black line and squares) aims to maintain a March 31st spawner index greater than 1.7 spawners per trap. MMI values from April to the following March are back-calculated from the March 31st threshold assuming $M=1.3$ yr$^{-1}$. The box highlights June and July when commercial fisheries typically close. Three alternative buffer sizes used in selected sub-areas are also shown for reference.*

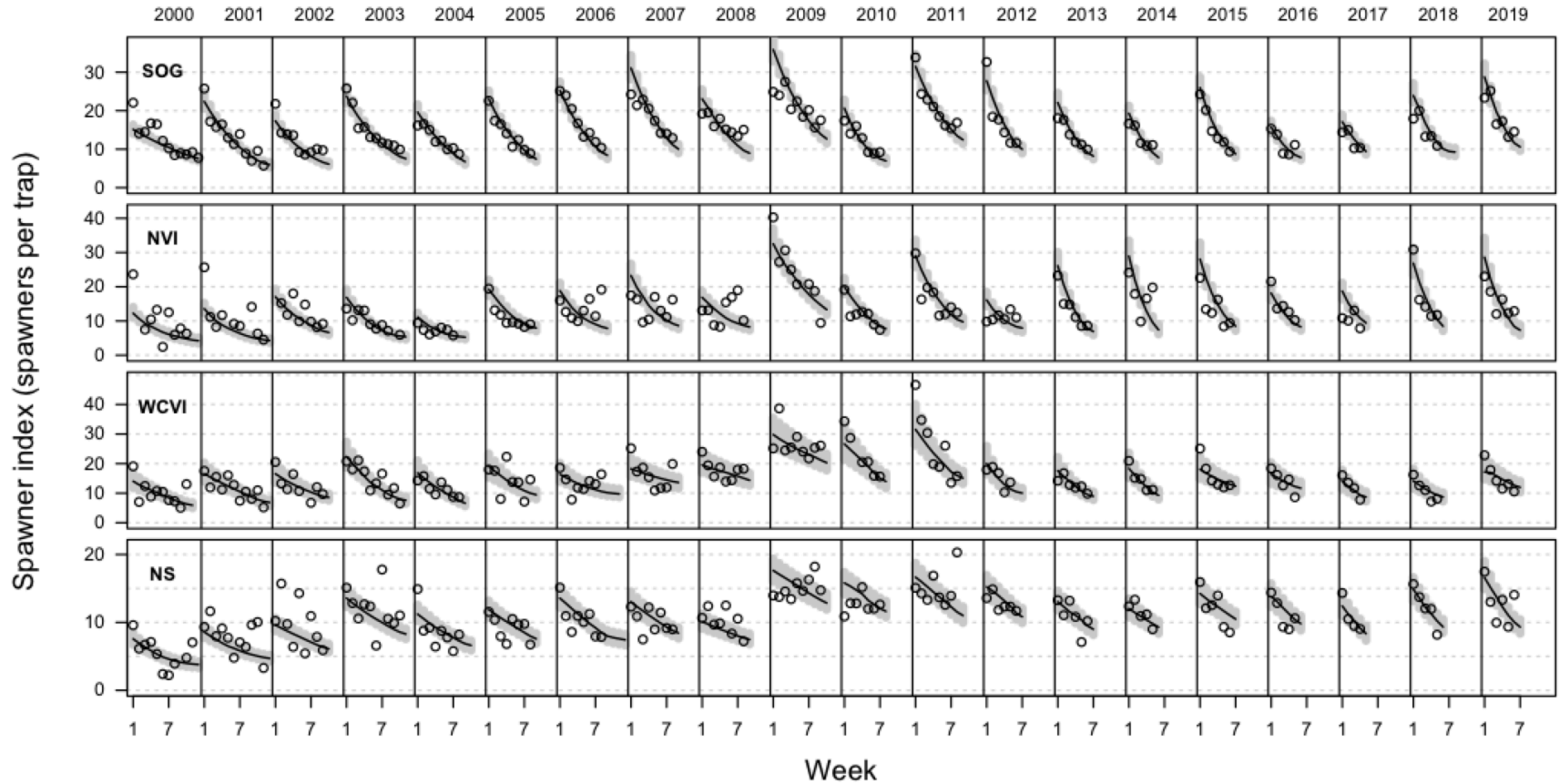

*Figure 5. Observed spawner index (circles) and model-predicted spawner index (black lines) posterior modes. Shaded region represents central 95% uncertainty interval from the Base operating model by region (rows) and year (column). Each year is represented by a 12-week subset of the commercial in-season period.*

*Figure 6. Estimated relationship between vulnerable biomass and fishing efficiency by region. Circles represent posterior modes. The underlying relationship (i.e., without process error) is represented by the red line (posterior mode) and shaded region (central 95% uncertainty interval).*

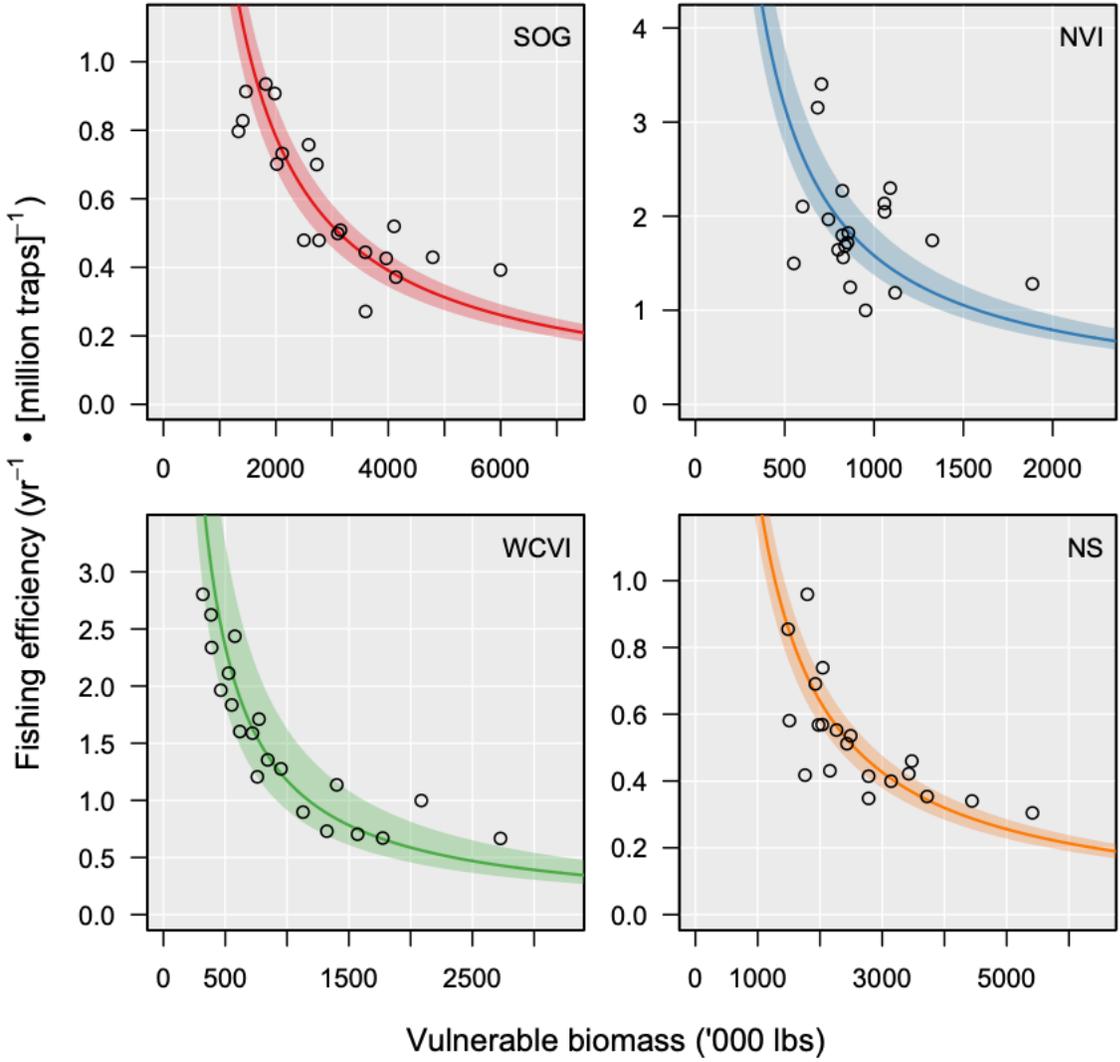

*Figure 7. Estimated posterior mode (lines) and 95% uncertainty interval (shaded area) for the proportion of age-3 prawns that are female on June 1 (top row), spawning biomass (middle row), and harvest rate for age-3 prawns (bottom row). The upper and lower horizontal lines in SSB plots represent the USR and LRP, respectively.*

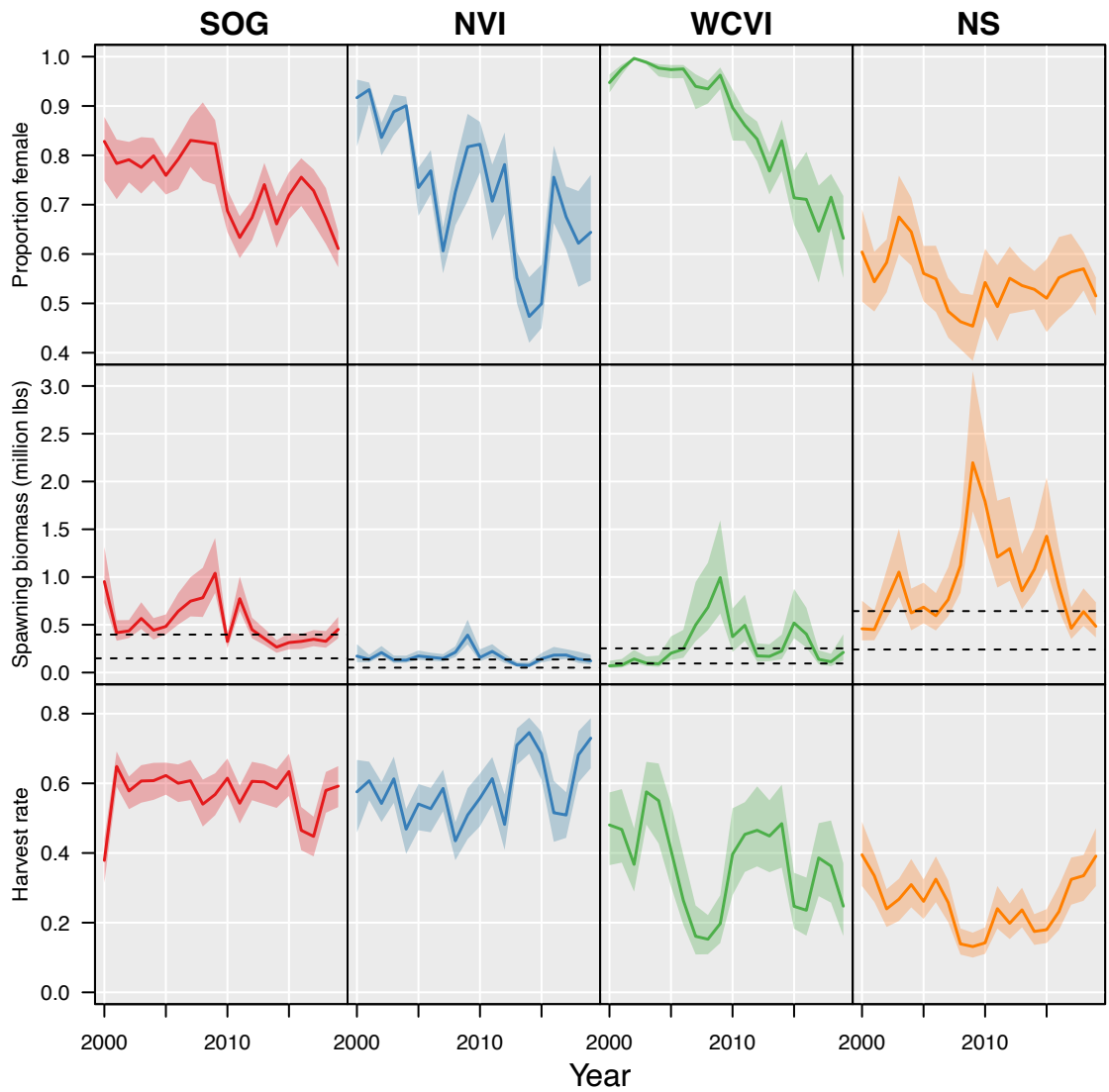

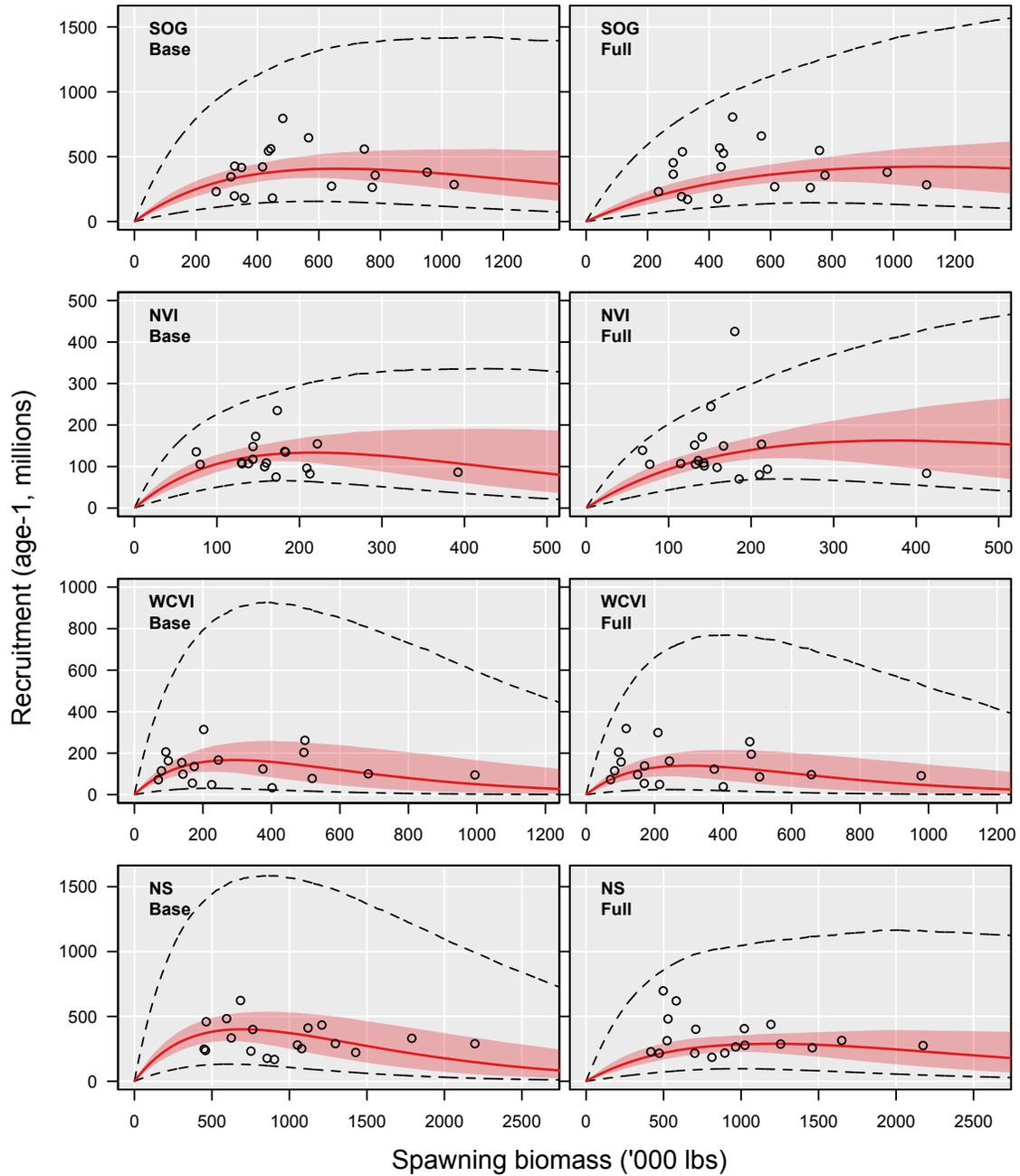

*Figure 8. Estimated relationship between prawn spawning biomass in year t and the resulting recruitment of age-1 juveniles in year t+2 assuming either no environmental effects (left column) or assuming SST and NPGO as covariates (right column). Circles represent posterior modes. The underlying relationship (i.e., without process error) is represented by the red line (posterior mode) and shaded region (central 95% uncertainty interval). Horizontal broken lines indicate the central 95% uncertainty interval for the full recruitment distribution.*

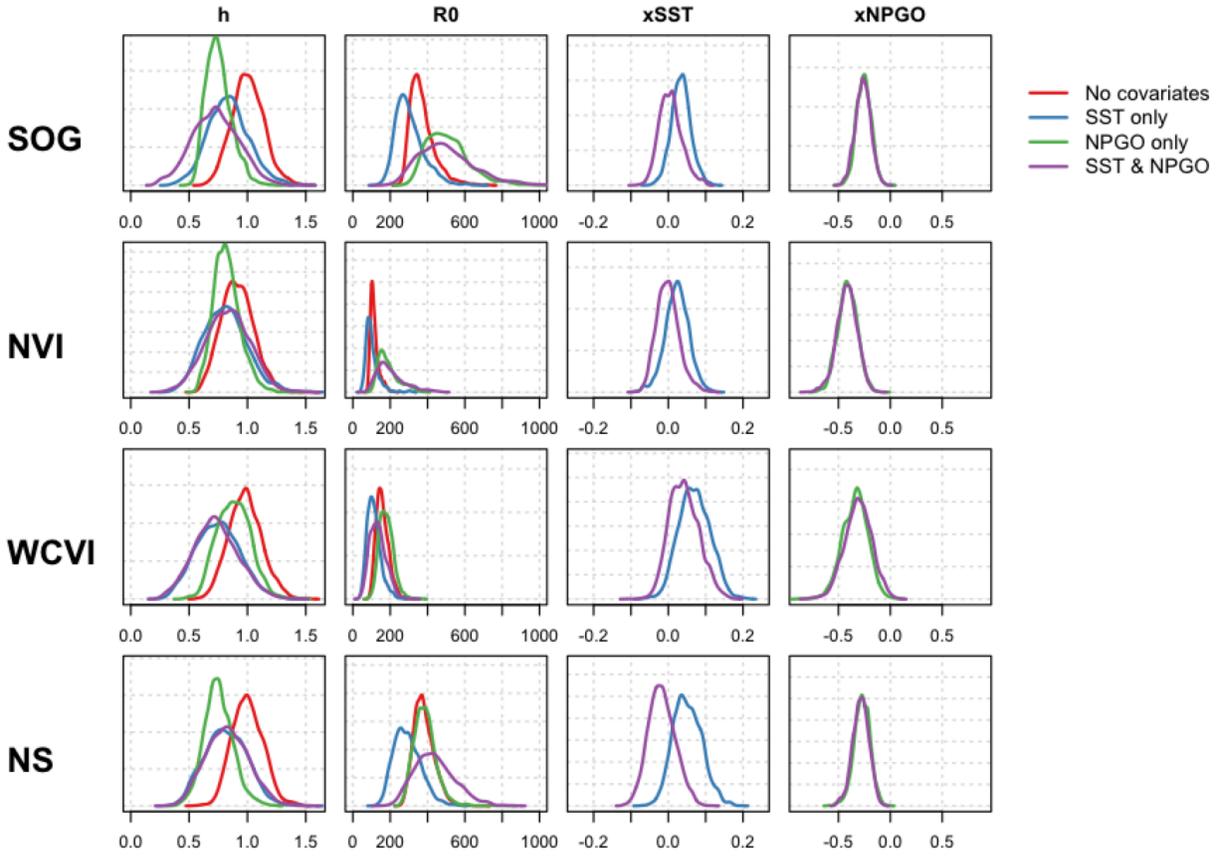

*Figure 9. Marginal posterior distributions for stock-recruitment model parameters: (red) – the Base operating model with no environmental covariates; (blue) – an "SST only" model that included only sea surface temperature as an environmental covariate, (green) – an "NPGO only" model that included only the North Pacific Gyre Oscillation as an environmental covariate; and (purple) – the Full operating model with "SST & NPGO" both included as environmental covariates. The four parameters are steepness (h), unfished recruitment (R0), the effect size multiplier for SST (xSST), and the effect size multiplier for the NPGO (xNPGO).*

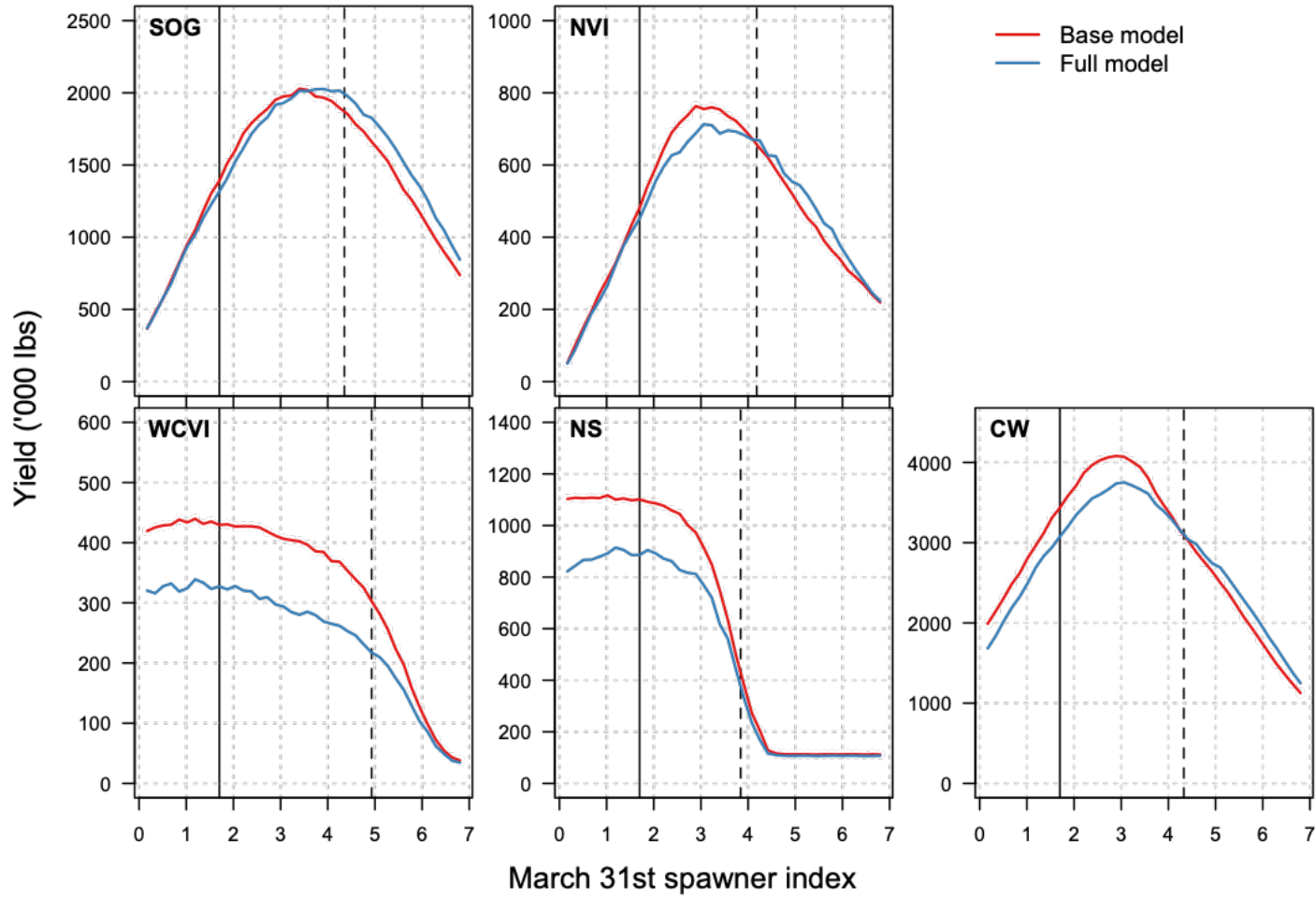

*Figure 10. Yield curves for Base and Full operating models averaged over natural mortality scenarios. Yield curves are computed by simulating 1000 trajectories over a grid of March 31$^{st}$ spawner index targets and computing the average yield and spawner abundance over the last 20 years. The solid vertical line indicates the de jure March MMI target of 1.7 female spawners per trap while the broken vertical line indicates the de facto target used between 2000-2019.*

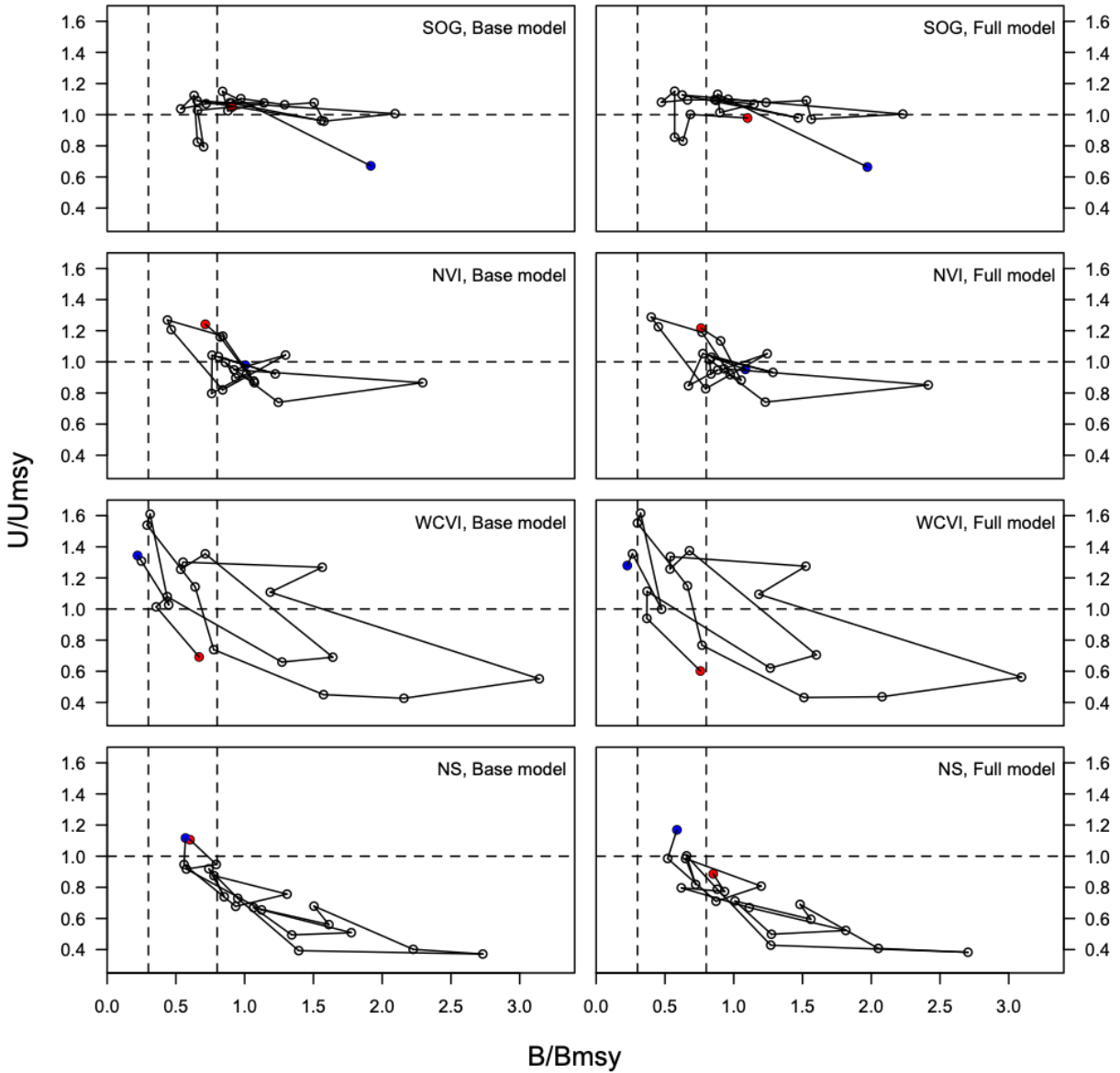

*Figure 11. Posterior mode stock status trajectories from the Base (left column) and Full operating models (right column) with respect to Canadian default biological references points (commonly known as a Kobe plot) for each region. Blue and red points indicate estimates from 2000 and 2019, respectively. Vertical lines indicate LRP (0.3BMSY) and USR (0.8BMSY) values defining the Critical, Cautious, and Healthy zones for March 31$^{st}$ spawning biomass.*

*Figure 12. Projected biomass by region for the Base OM (left column), Full OM with constant mean SST (center column), and Full OM with mean SST increasing linearly by one degree Celsius over the 30-year projection (right column) with M=0.9 yr$^{-1}$. The upper and lower horizontal lines represent the USR and LRP, respectively, under each model and corresponding $B_{MSY}$ is indicated by the open circle. Black lines represent posterior modes for the historical period (2000-2019), while coloured lines indicate modes of the projections (2020-2049). Grey lines indicate individual randomly sampled replicate trajectories. Shaded regions cover the central 95% uncertainty interval over 1000 posterior parameter draws.*

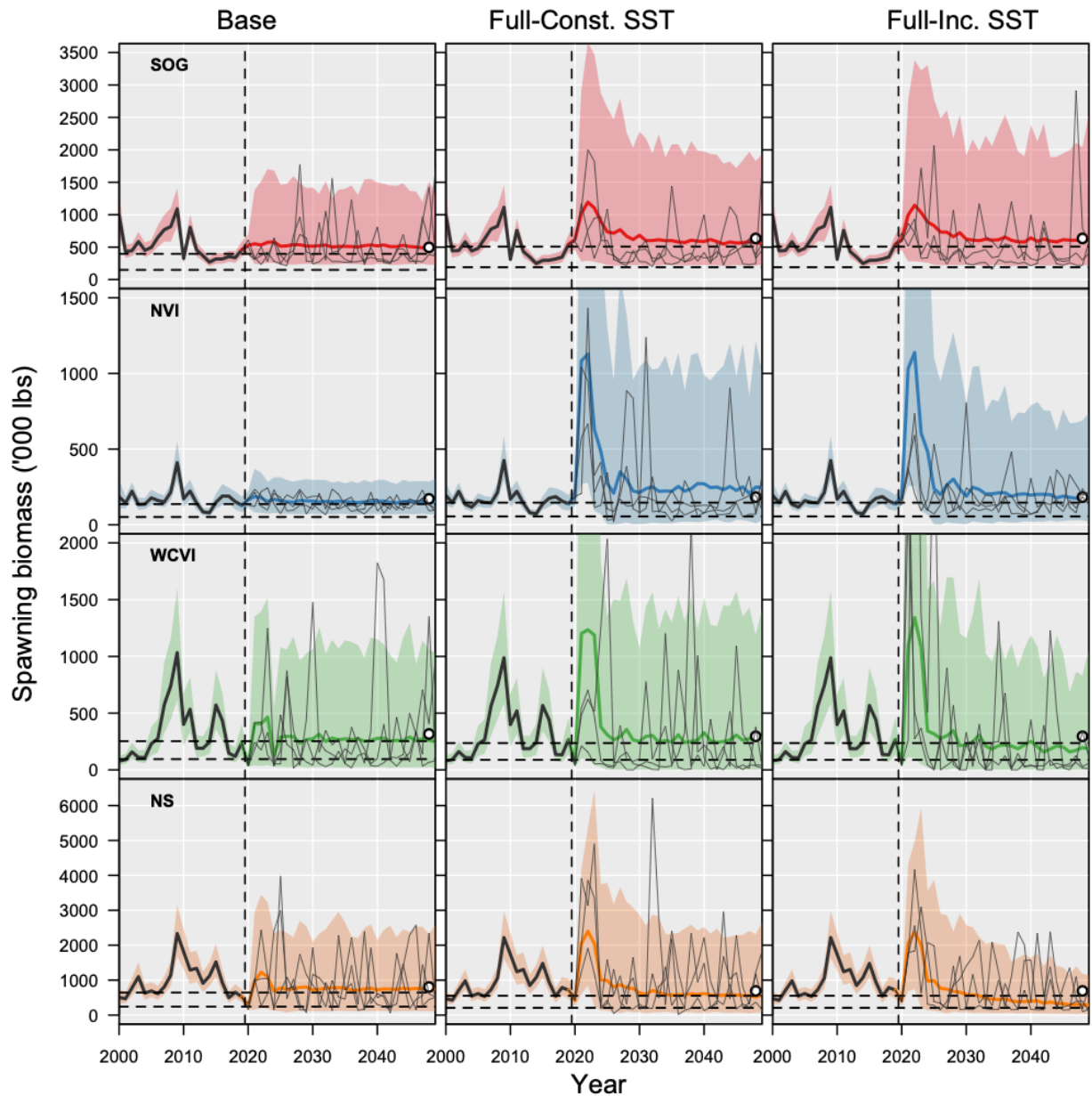

**Appendix A**

**Supplementary Figures for evaluating the sustainability of a *de facto* harvest strategy for British Columbia's Spot Prawn (*Pandalus platyceros*) fishery in the presence of environmental drivers of recruitment and hyperstable catch rates**

Steven P. Rossi, Sean P. Cox, Samuel D.N. Johnson, Ashleen J. Benson

*Figure A.1. Indices for the Pacific Decadal Oscillation (PDO), North Pacific Gyre Oscillation (NPGO) and sea surface temperature (SST), 2000-2019. Each time-series was standardize to have mean 0 and variance 1.*

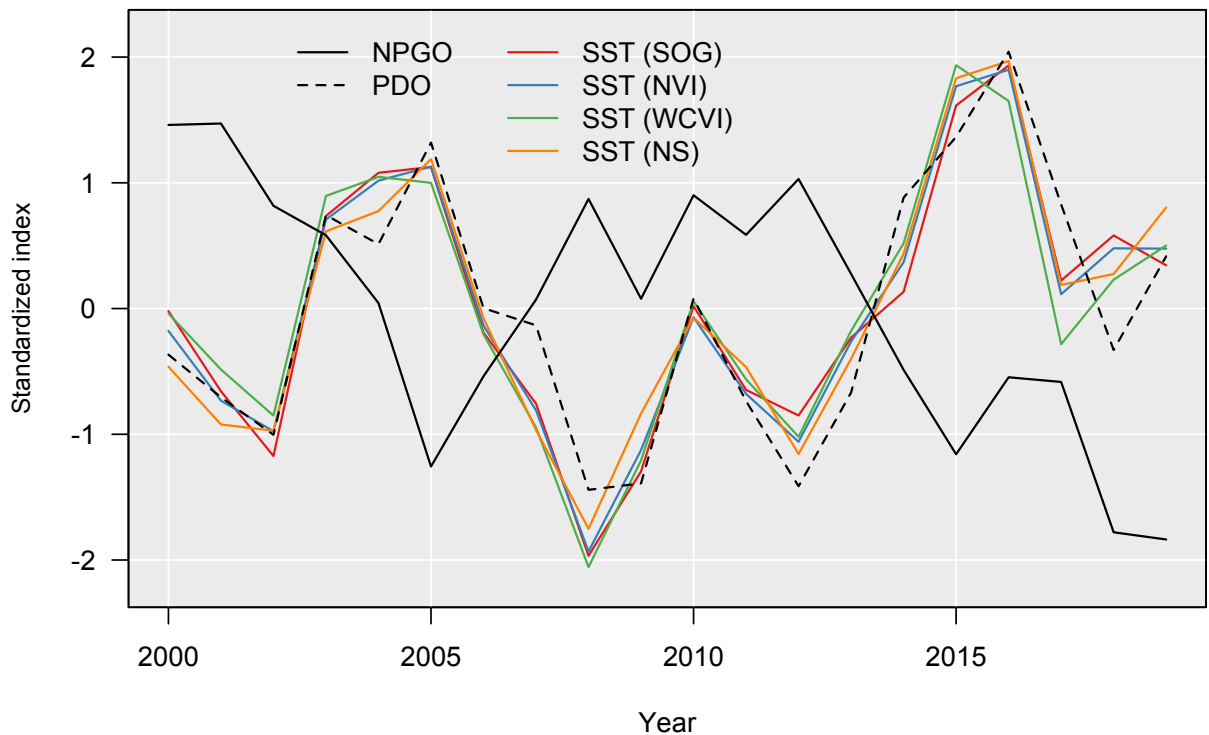

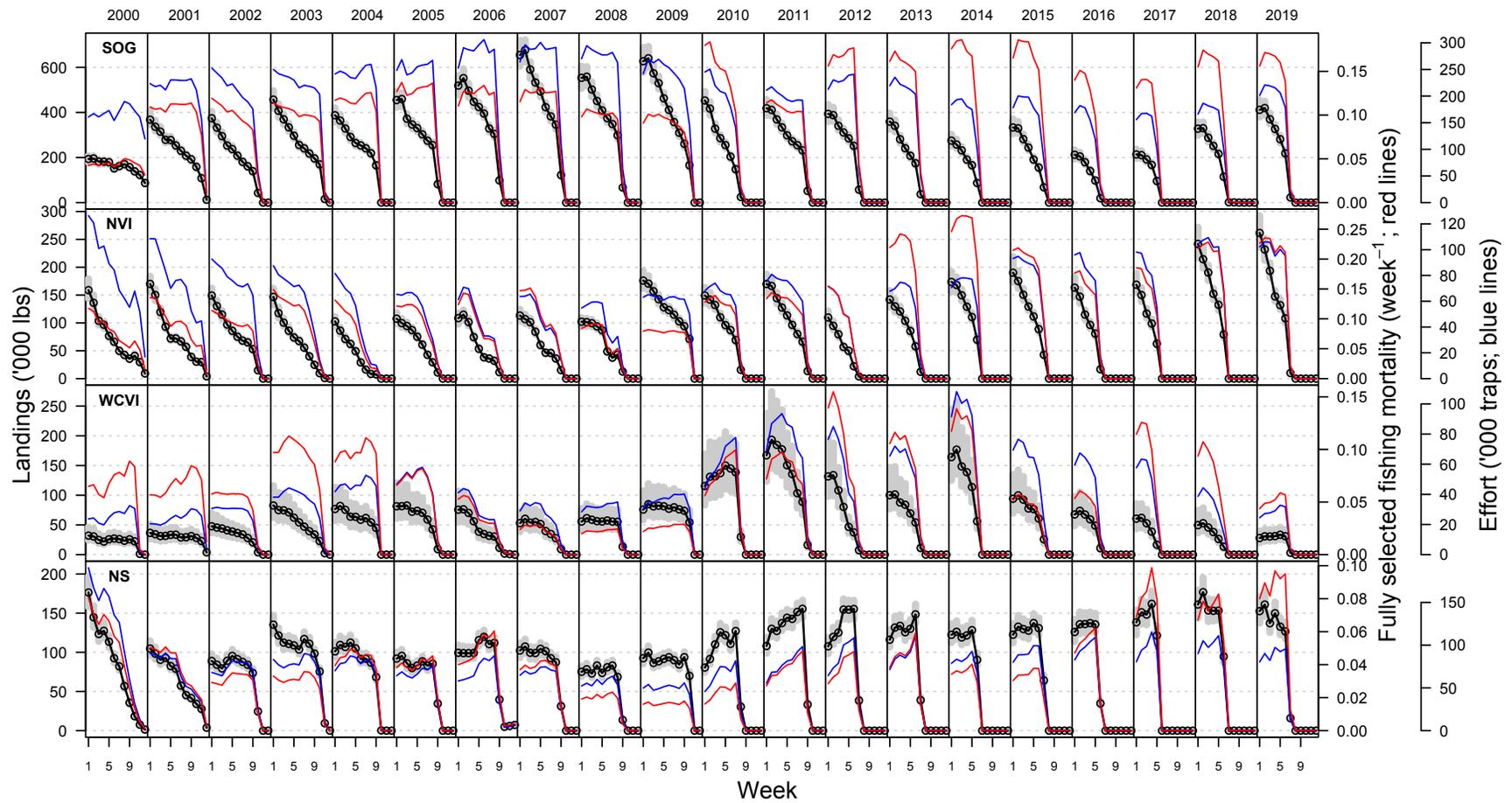

*Figure A.2. Observed (circles) and model-predicted (black lines) posterior modes and central 95% uncertainty intervals for weekly commercial Spot Prawn landings, estimated fully selected fishing mortality (red lines), and weekly fishing effort (blues lines) by region (rows) and year (columns).*

*Figure A.3. Relationships between total commercial fishery landings in a given year and mean weekly effort over the first four weeks of the fishery in the following year. The plotted numbers represent the years (e.g., "01" is 2001). The line represents the relationship between landings and effort that was used to generate expected weekly effort in the projections. For SOG and WCVI, this line represents an ordinary linear regression fitted to the historical relationship between landings and effort. For NVI and NS, the slope parameter in the fitted ordinary linear regression was not significant, so projected effort was equal to mean historical effort.*

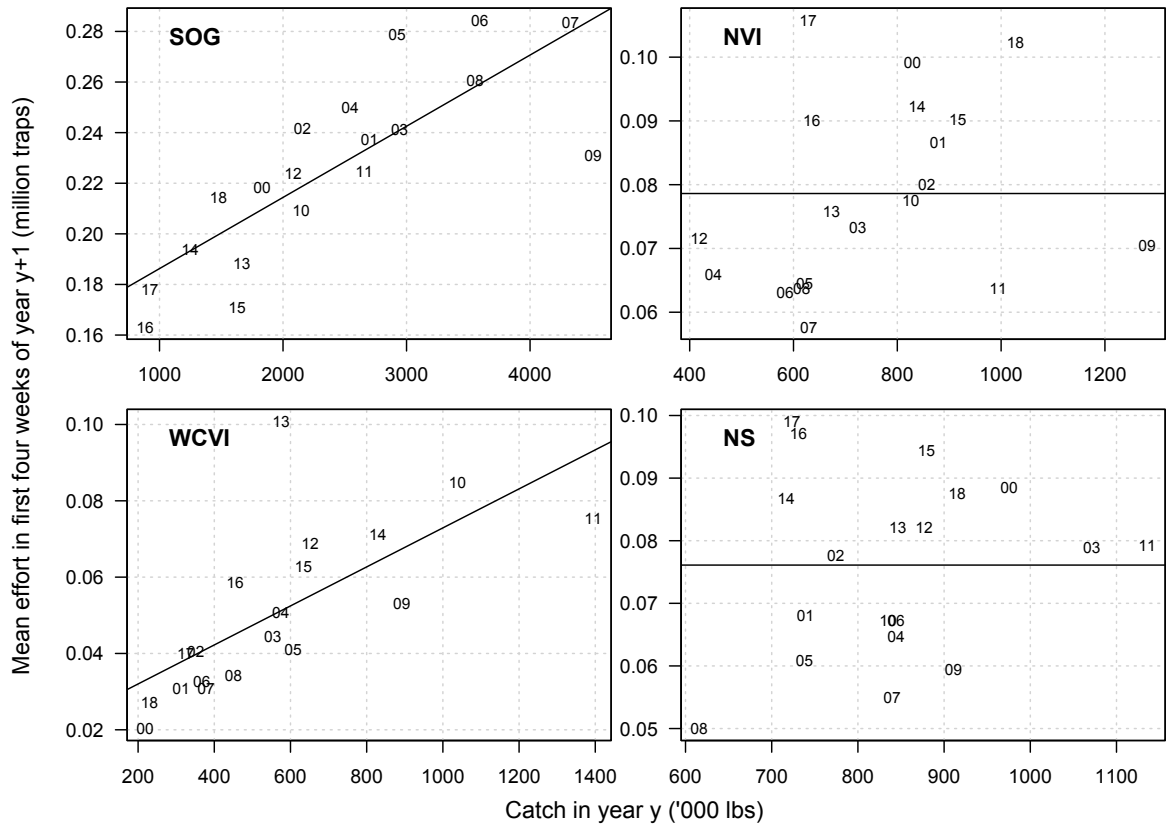

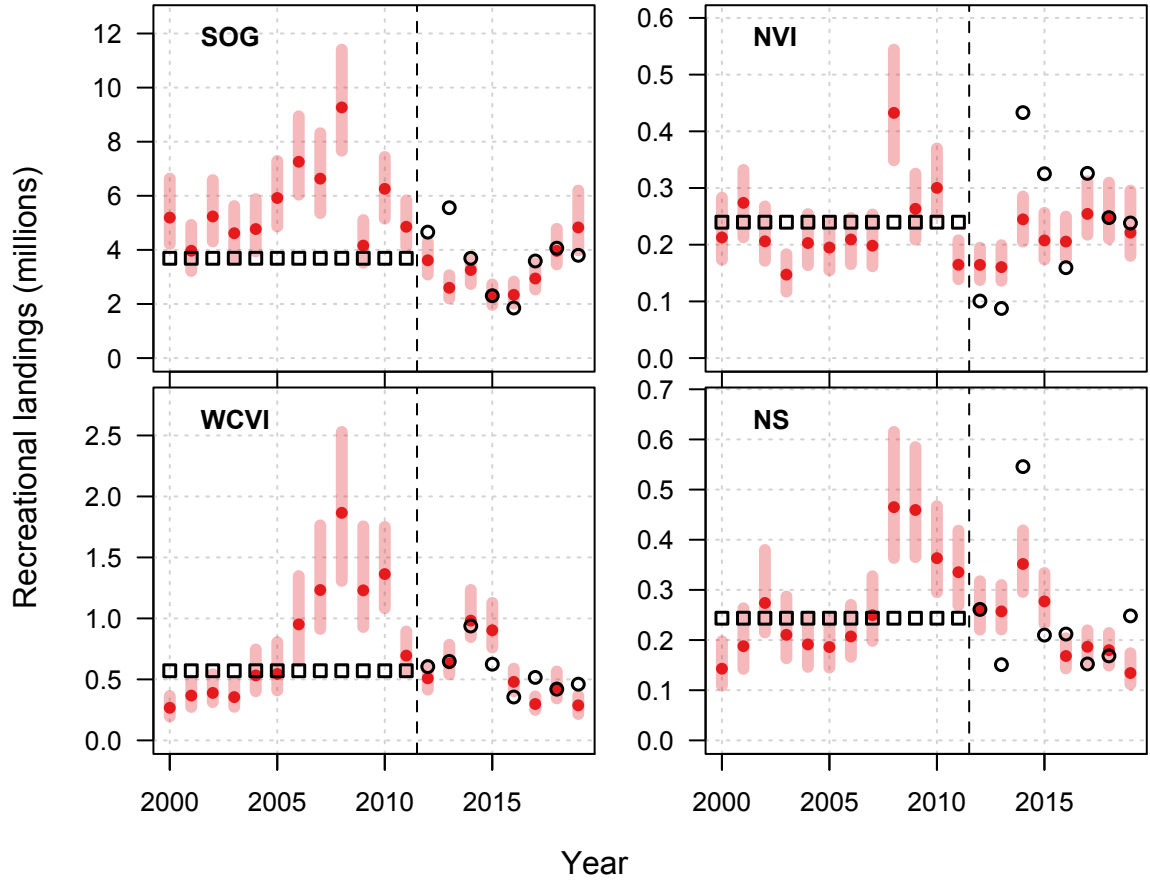

*Figure A.4. Observed (open circles) and model-fitted (red circles representing posterior modes with red lines representing the associated central 95% uncertainty interval) recreational fishery landings. Recreational landings data were unavailable prior to 2012, so we used mean observed landings from 2012-2019 as "observations" in 2000-2011 (open boxes).*

*Figure A.5. Observed (bars) and predicted (circles) stage-composition for juveniles (J), males (M), transitionals (T), and females (F) in commercial fishery biological samples. For plotting purposes, stage-compositions were combined across all weeks and years and weighted by sample size. Few egged/spent females were observed in the commercial stage-composition so we combined the female and egged/spent female stages for these samples.*

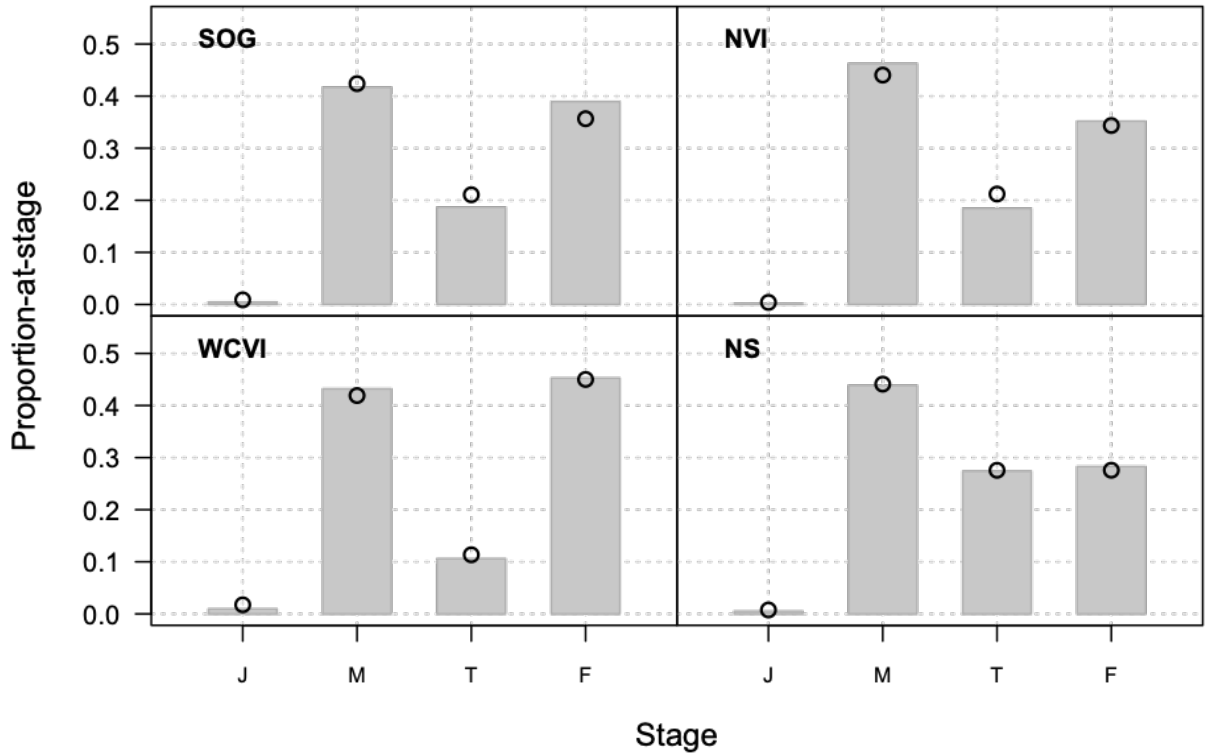

*Figure A.6. Observed (letters) and predicted (lines and shading) stage-composition for juveniles (J), males (M), transitionals (T), and females (F) in commercial fishery biological samples in the SOG region. Lines represent posterior modes, while shading represents the central 95% uncertainty interval.*

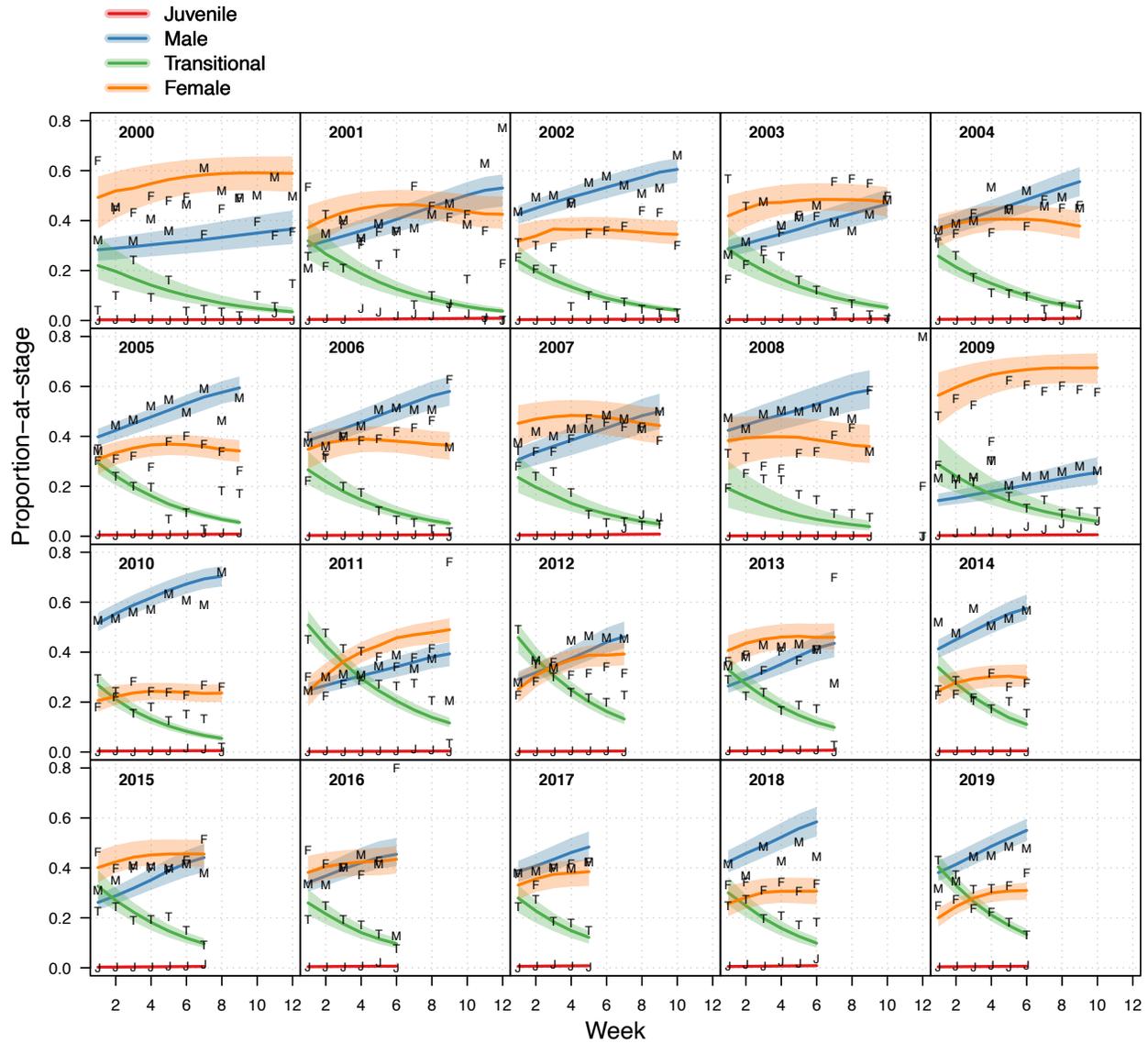

*Figure A.7. Observed (letters) and predicted (lines and shading) stage-composition for juveniles (J), males (M), transitionals (T), and females (F) in commercial fishery biological samples in the NVI region. Lines represent posterior modes, while shading represents the central 95% uncertainty interval.*

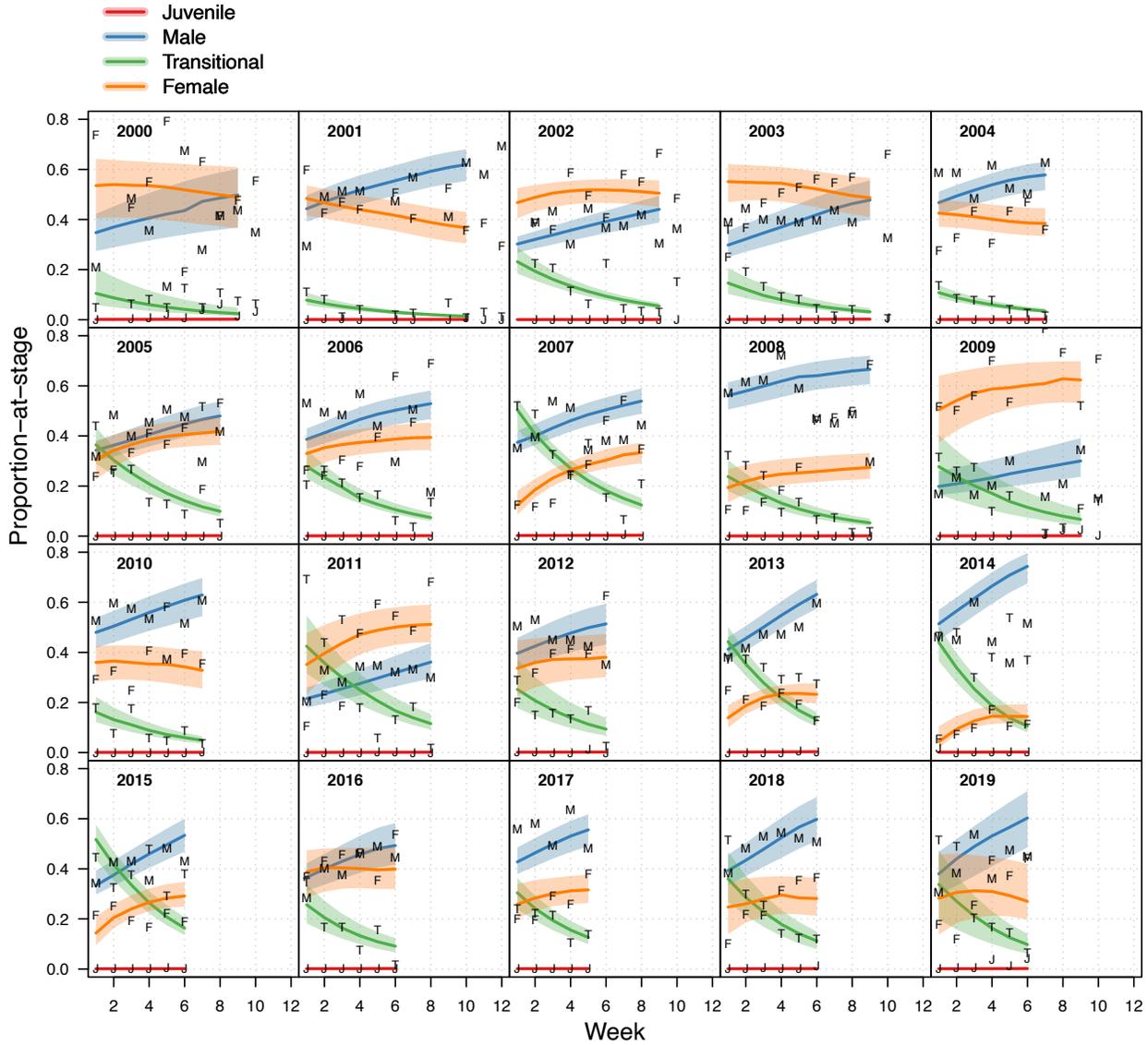

*Figure A.8. Observed (letters) and predicted (lines and shading) stage-composition for juveniles (J), males (M), transitionals (T), and females (F) in commercial fishery biological samples in the WCVI region. Lines represent posterior modes, while shading represents the central 95% uncertainty interval.*

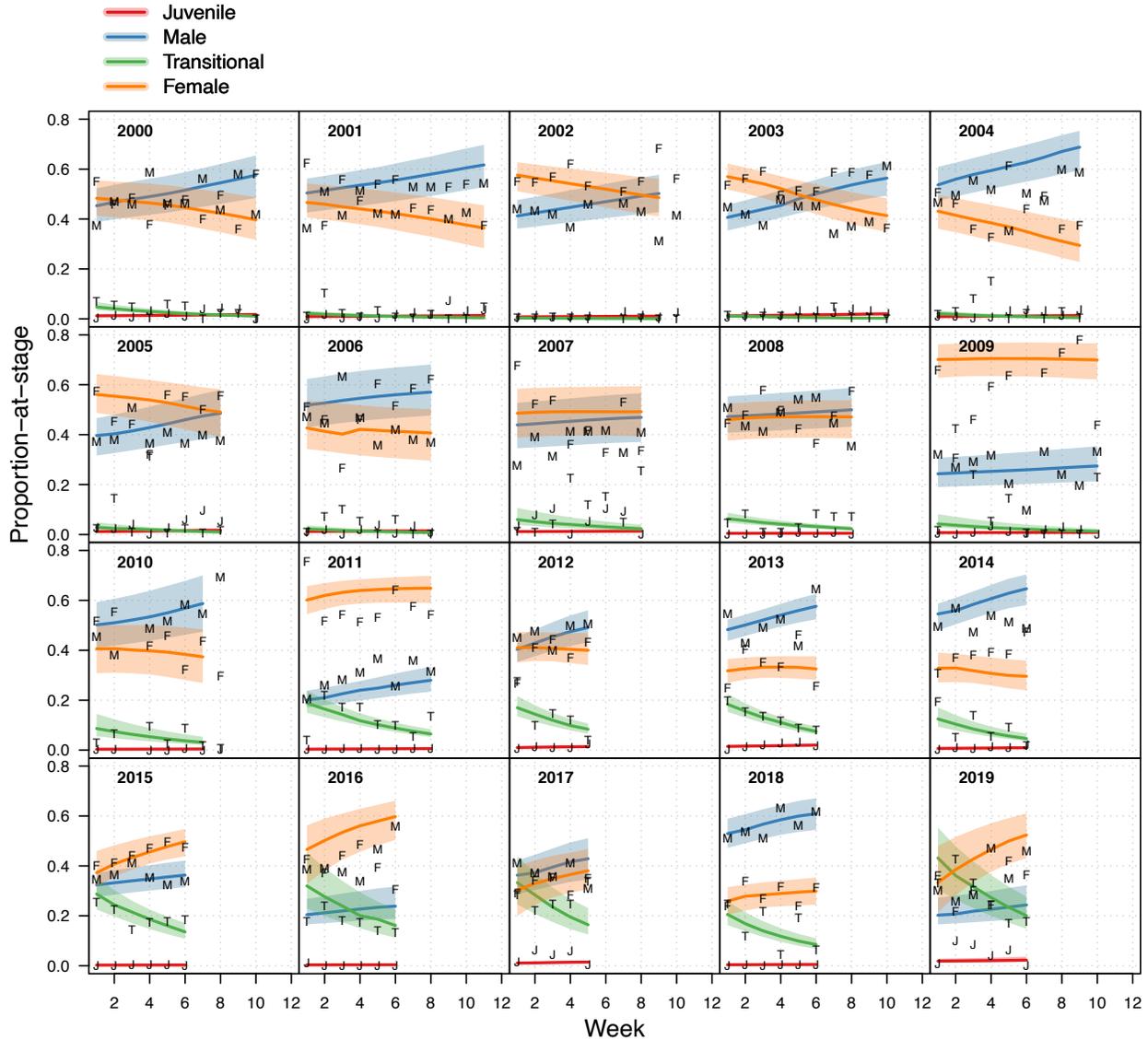

*Figure A.9. Observed (letters) and predicted (lines and shading) stage-composition for juveniles (J), males (M), transitionals (T), and females (F) in commercial fishery biological samples in the NS region. Lines represent posterior modes, while shading represents the central 95% uncertainty interval.*

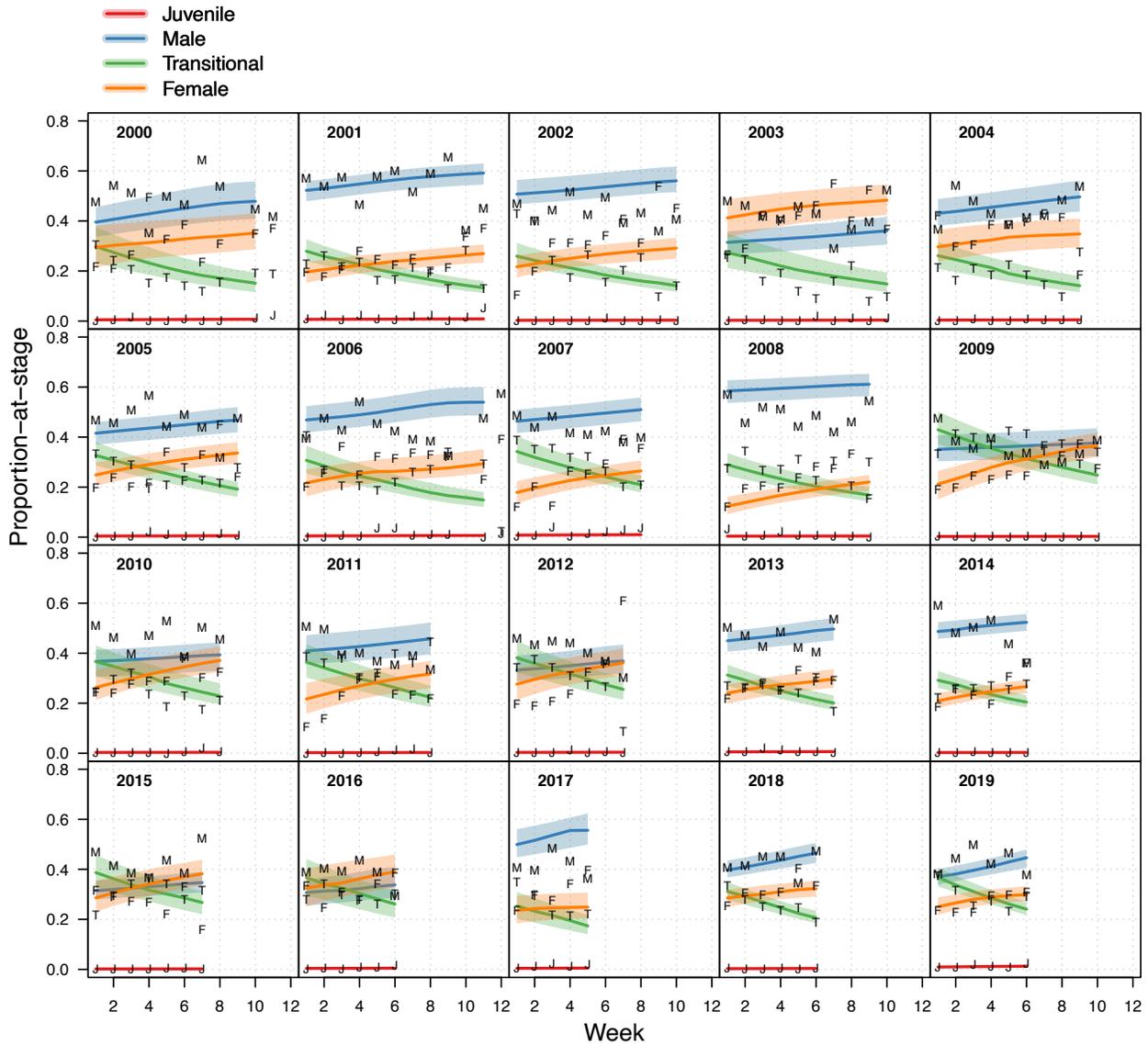

*Figure A.10. Estimated recruitment (left column) and recruits-per-spawner (right column) by region (rows), 2000-2019. Coloured lines and shaded regions indicate the posterior mode and central 95% uncertainty interval for the Full model, respectively. The solid and broken black lines represent the posterior mode and the limits of central 95% uncertainty interval for the Base model, respectively*

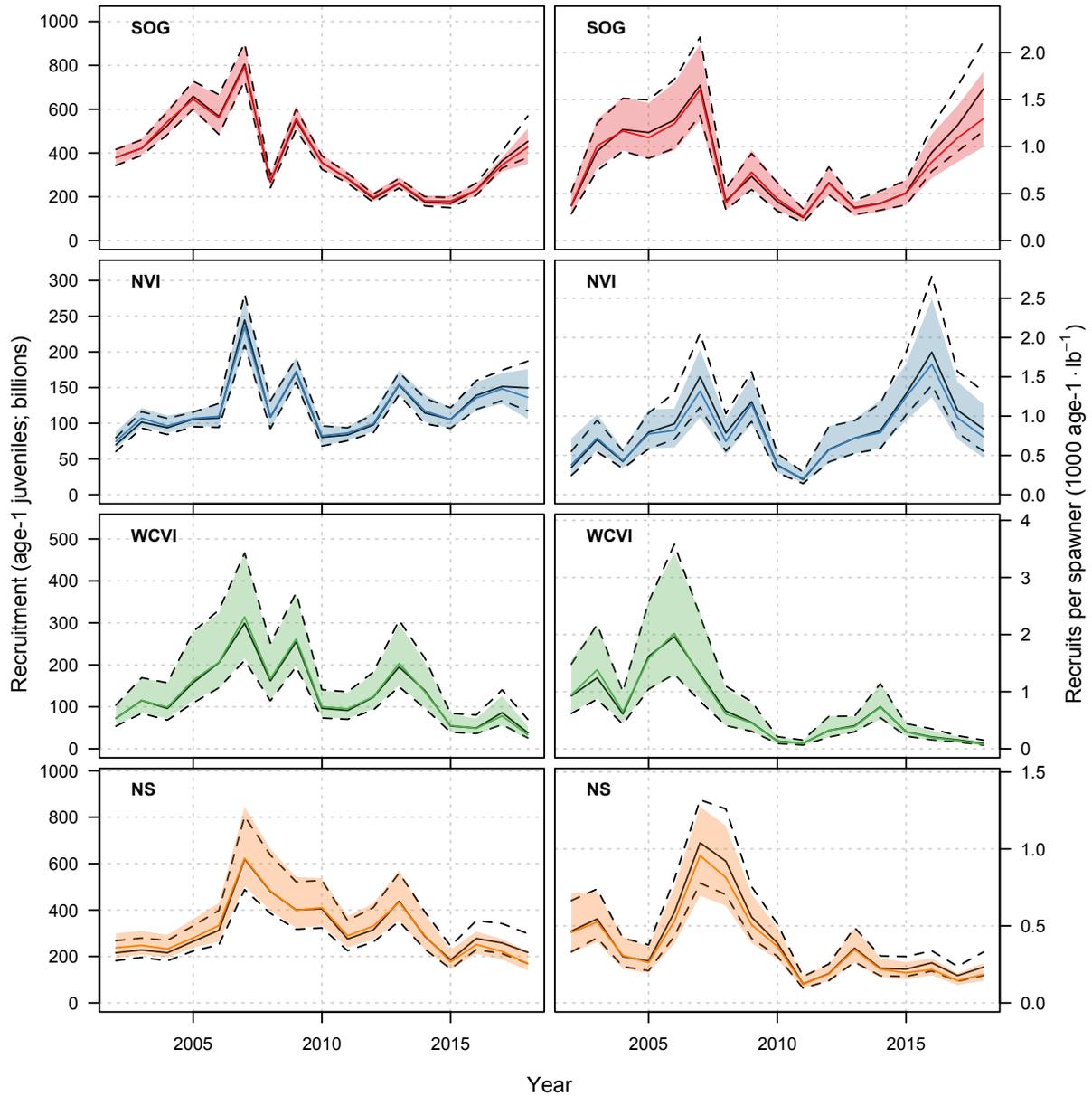

*Figure A.11. Relationship between sea surface temperature and estimated fishing efficiency deviations from the Base model with M=0.9. The line represents a fitted linear model while p and R^2 denote p-value and coefficient of determination, respectively, for the fitted model.*

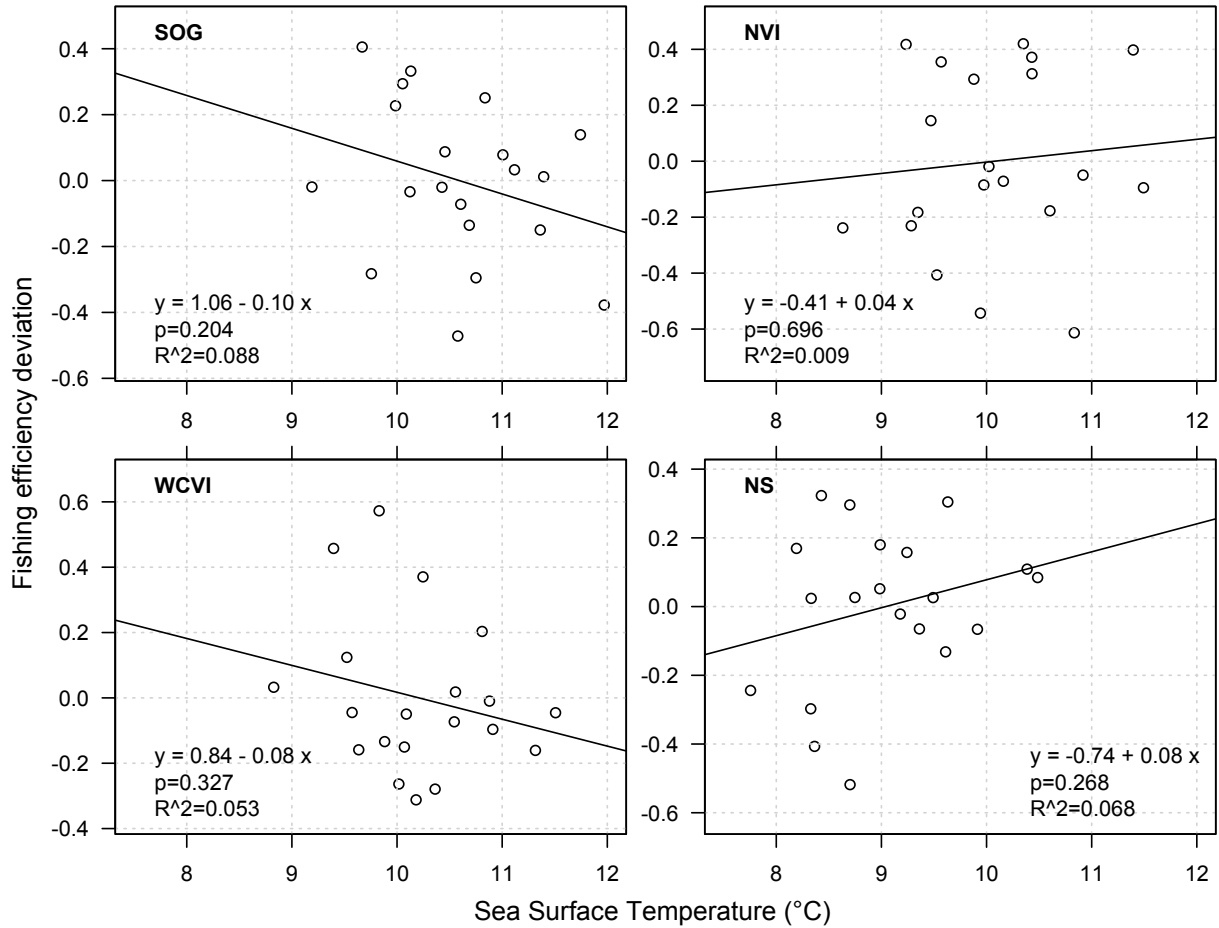

*Figure A.12. Mean weight for six Spot Prawn stages observed in annual spring (black circles) and fall (red circles) survey in the Strait of Georgia, 2001-2019. Years with less than 25 samples are omitted. Horizontal broken lines indicate the mean stage weight across all samples from both surveys.*

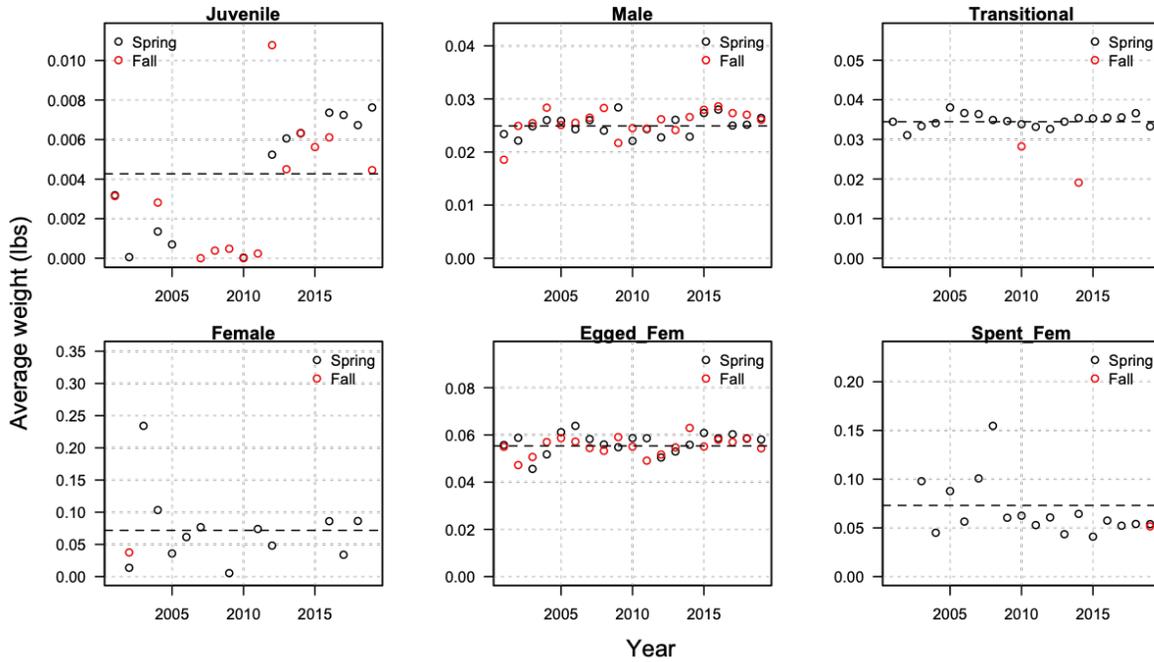

*Figure A.13. Length of egged females observed in annual spring/preseason (black) and fall/postseason (green) survey in the Strait of Georgia, 2001-2019. Light and dark colored lines indicate the central 50% and 95% intervals, respectively, while points represent modes.*

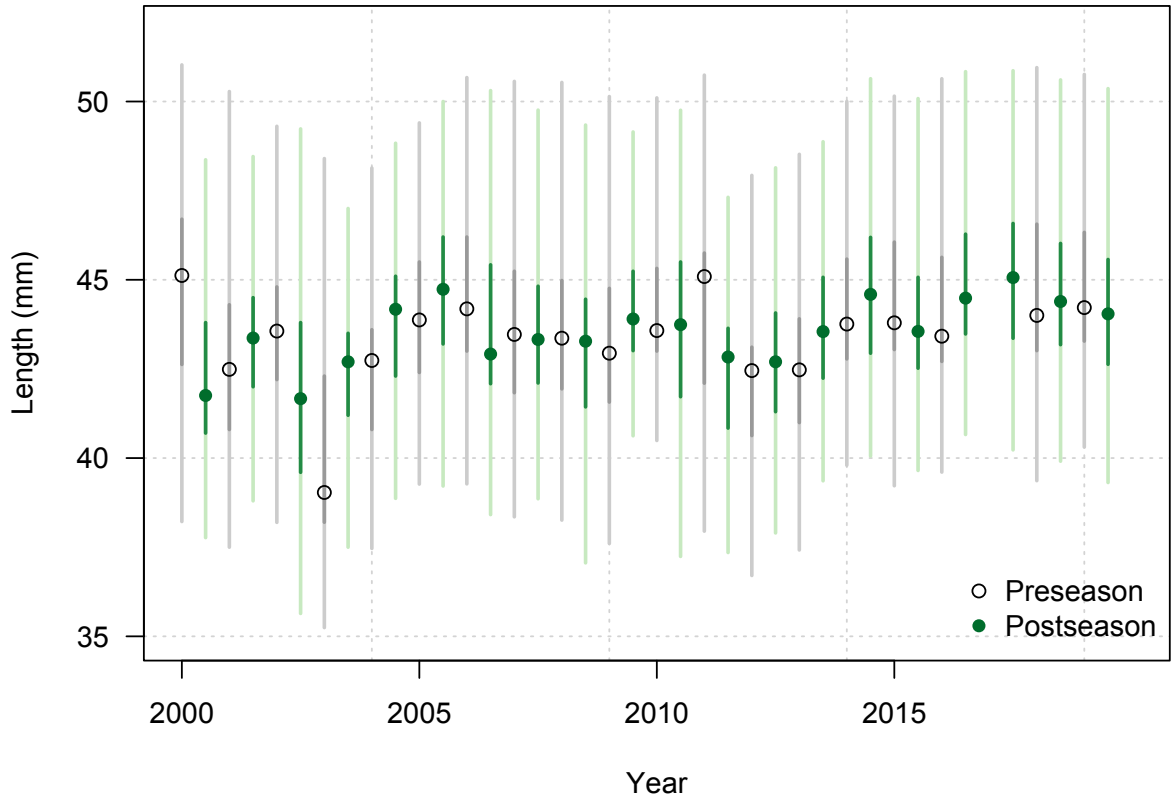

**Appendix B**

**Daily commercial landings by region with approximated "closing" dates**

Steven P. Rossi, Sean P. Cox, Samuel D.N. Johnson, Ashleen J. Benson

The annual commercial fishery closing date in each region is an important variable in our analysis that enables us to infer the de facto region-level rules governing the fishery. There is ambiguity as to when the fishery could be considered closed at the region level, since closures actually occur at the area or subarea level, and because small, inconsequential landings may occur in some areas after all other effort in that region has ceased. We considered a fishery in a given year and region "closed" on the first day of the year in which daily landings declined by at least 80%. Figures B.1-B.20 show daily landings by regions for 2000-2019 with our inferred closing dates.

74  *Figure B.1. Daily landings in 2000 by region. The dashed line indicates the first day on which*
75  *daily landings declined by at least 80%.*

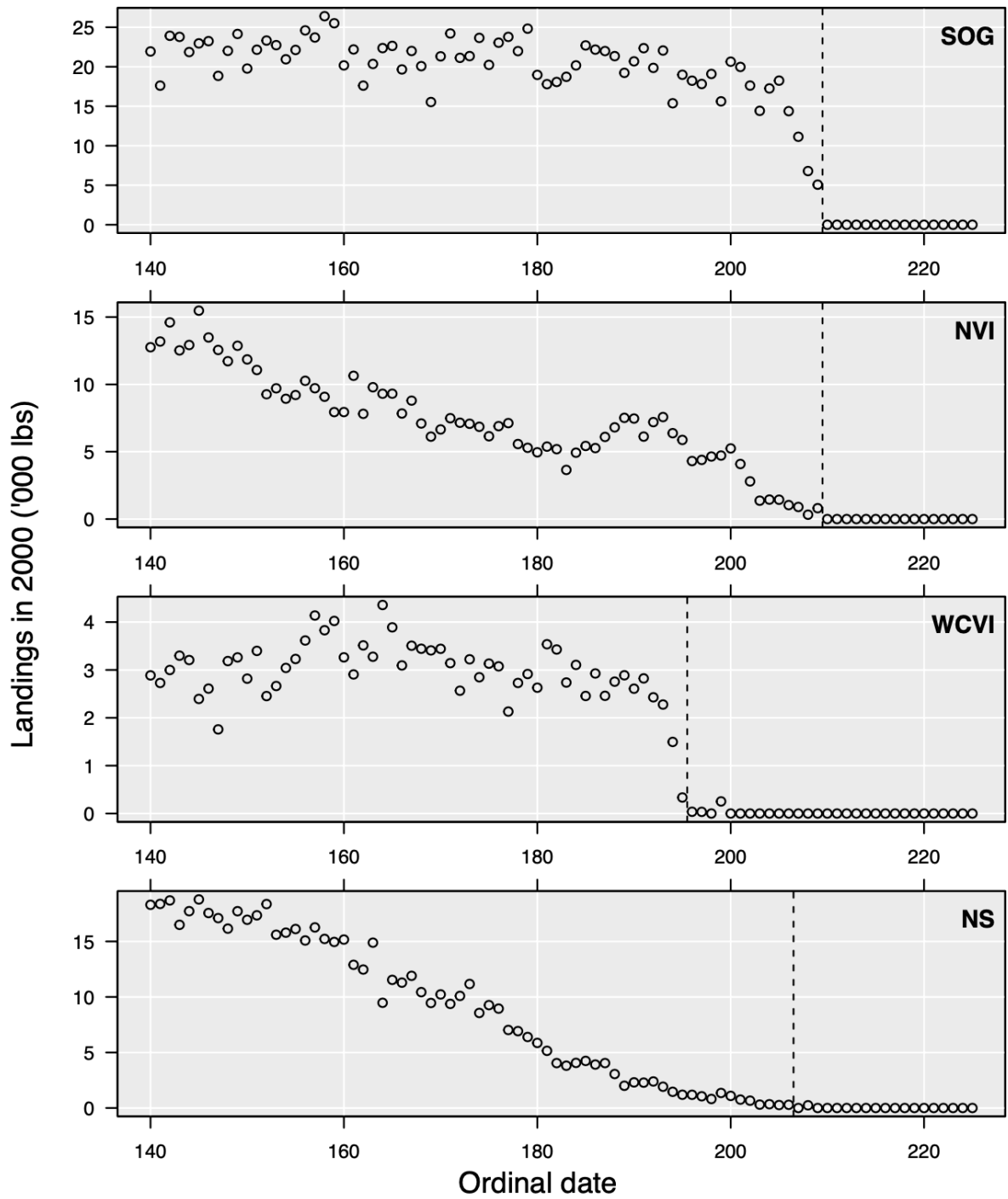

76
77

78  *Figure B.2. Daily landings in 2001 by region. The dashed line indicates the first day on which*
79  *daily landings declined by at least 80%.*

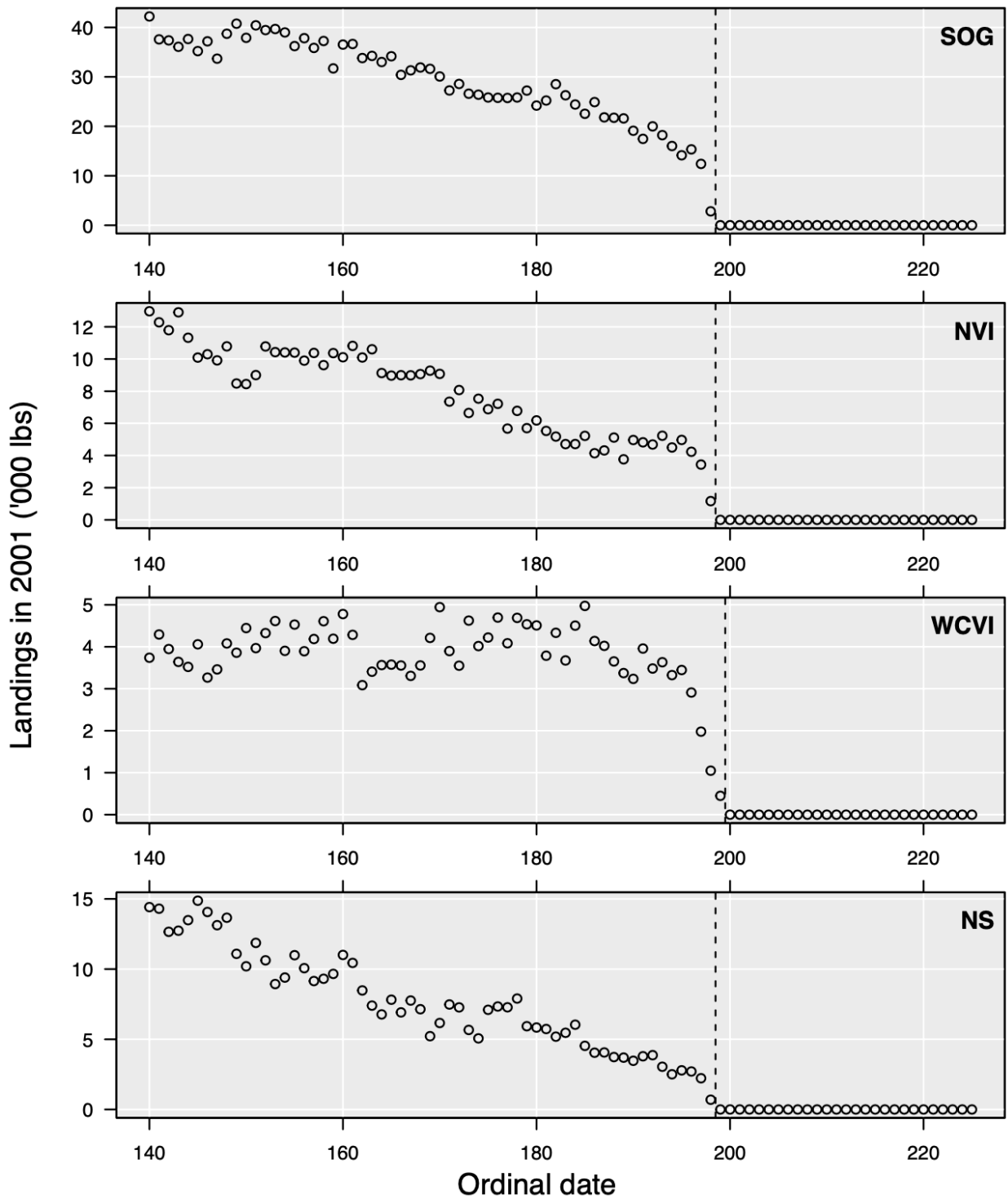

81  *Figure B.3. Daily landings in 2002 by region. The dashed line indicates the first day on which*
82  *daily landings declined by at least 80%.*

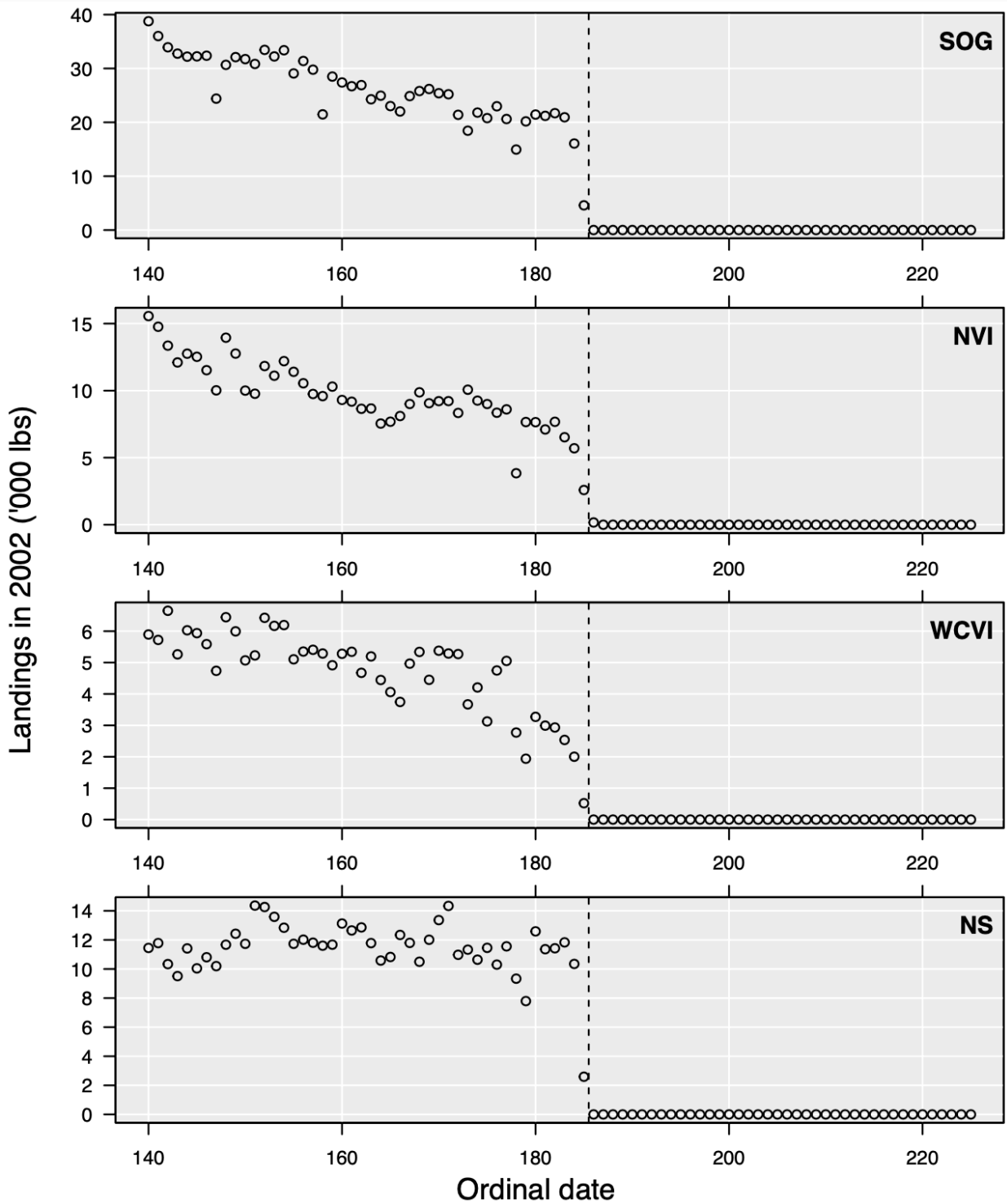

84  *Figure B.4. Daily landings in 2003 by region. The dashed line indicates the first day on which*
85  *daily landings declined by at least 80%.*

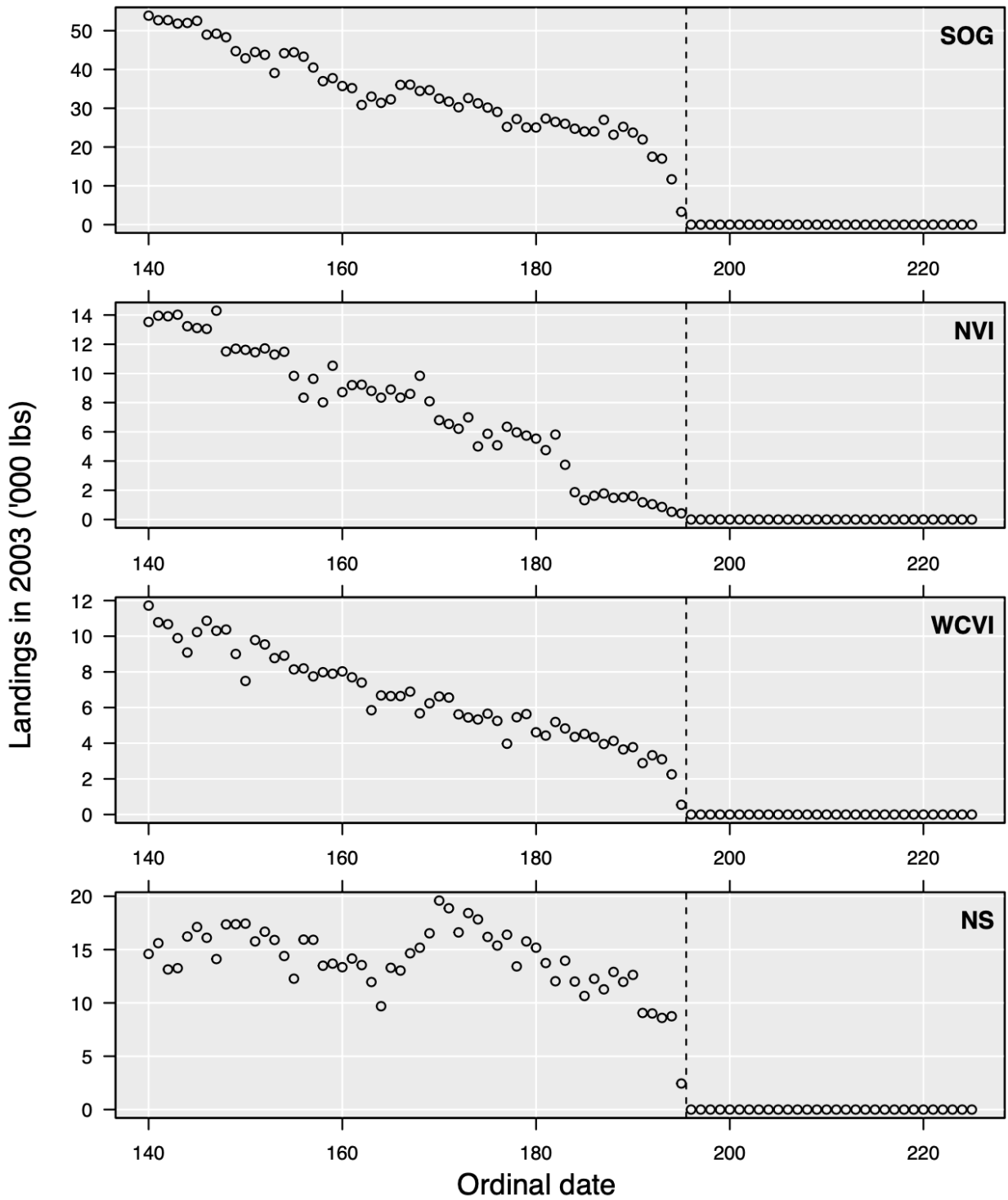

87  *Figure B.5. Daily landings in 2004 by region. The dashed line indicates the first day on which*
88  *daily landings declined by at least 80%.*

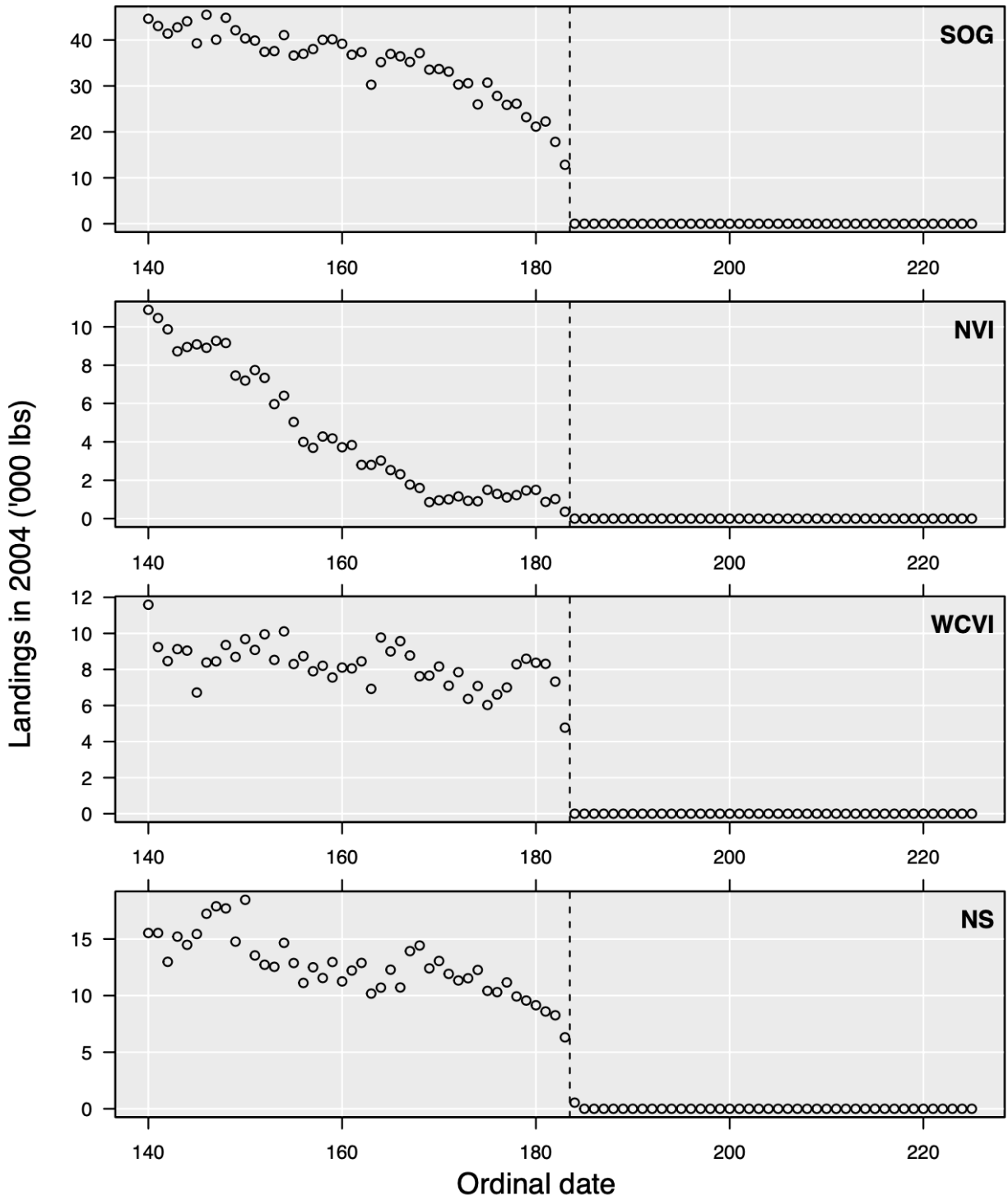

89

90   *Figure B.6. Daily landings in 2005 by region. The dashed line indicates the first day on which*
91   *daily landings declined by at least 80%.*

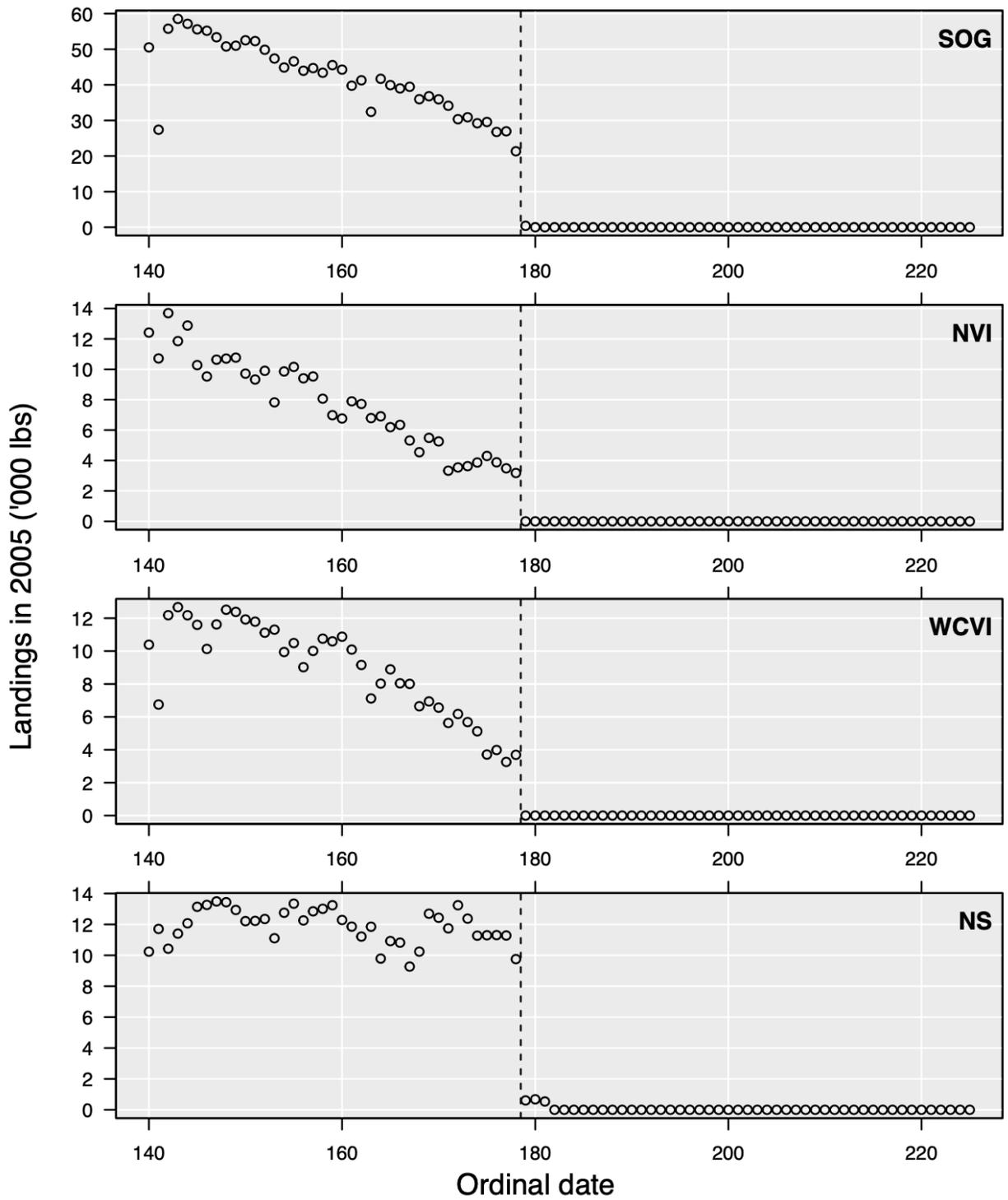

93   *Figure B.7. Daily landings in 2006 by region. The dashed line indicates the first day on which*
94   *daily landings declined by at least 80%.*

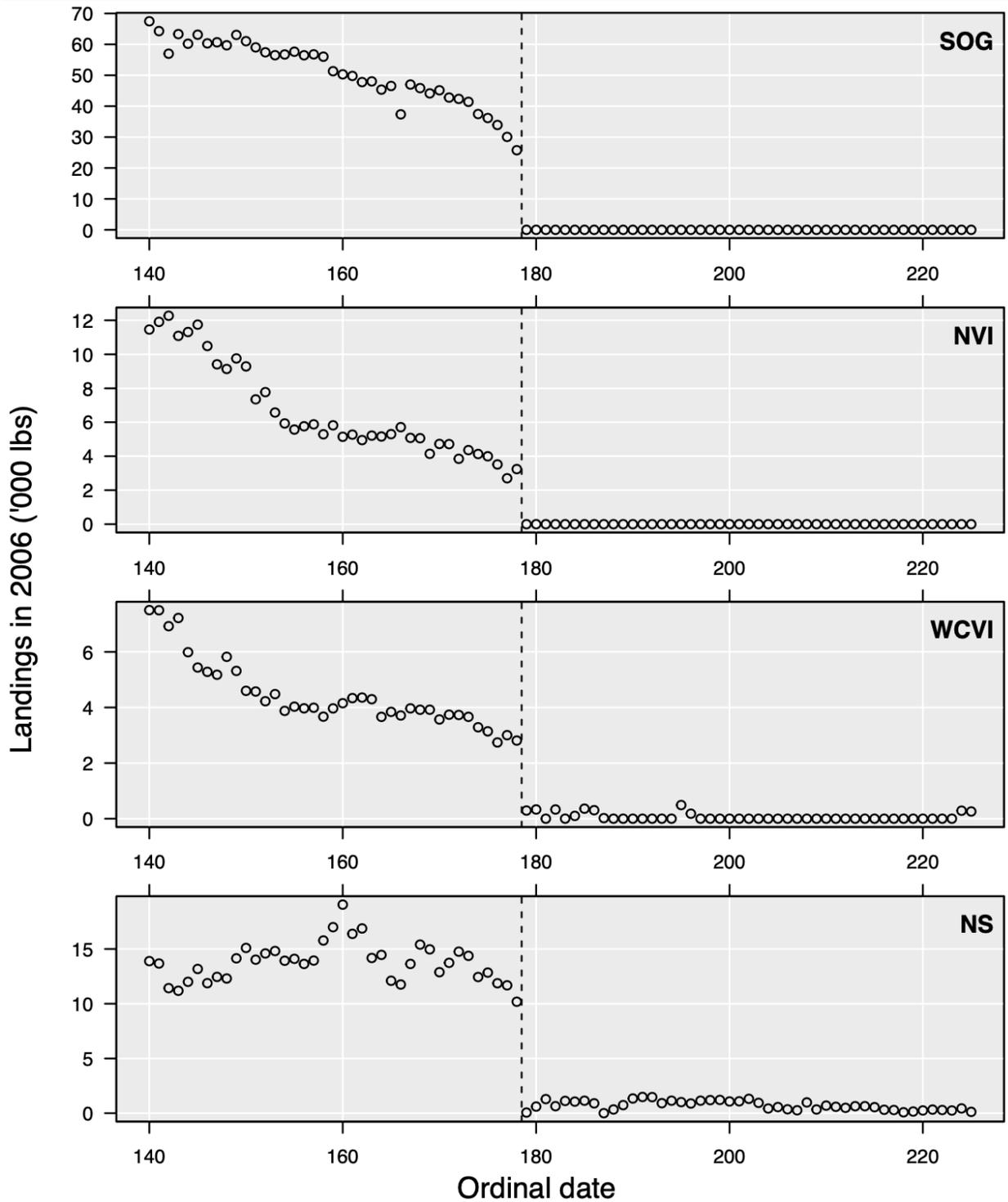

96  *Figure B.8. Daily landings in 2007 by region. The dashed line indicates the first day on which*
97  *daily landings declined by at least 80%.*

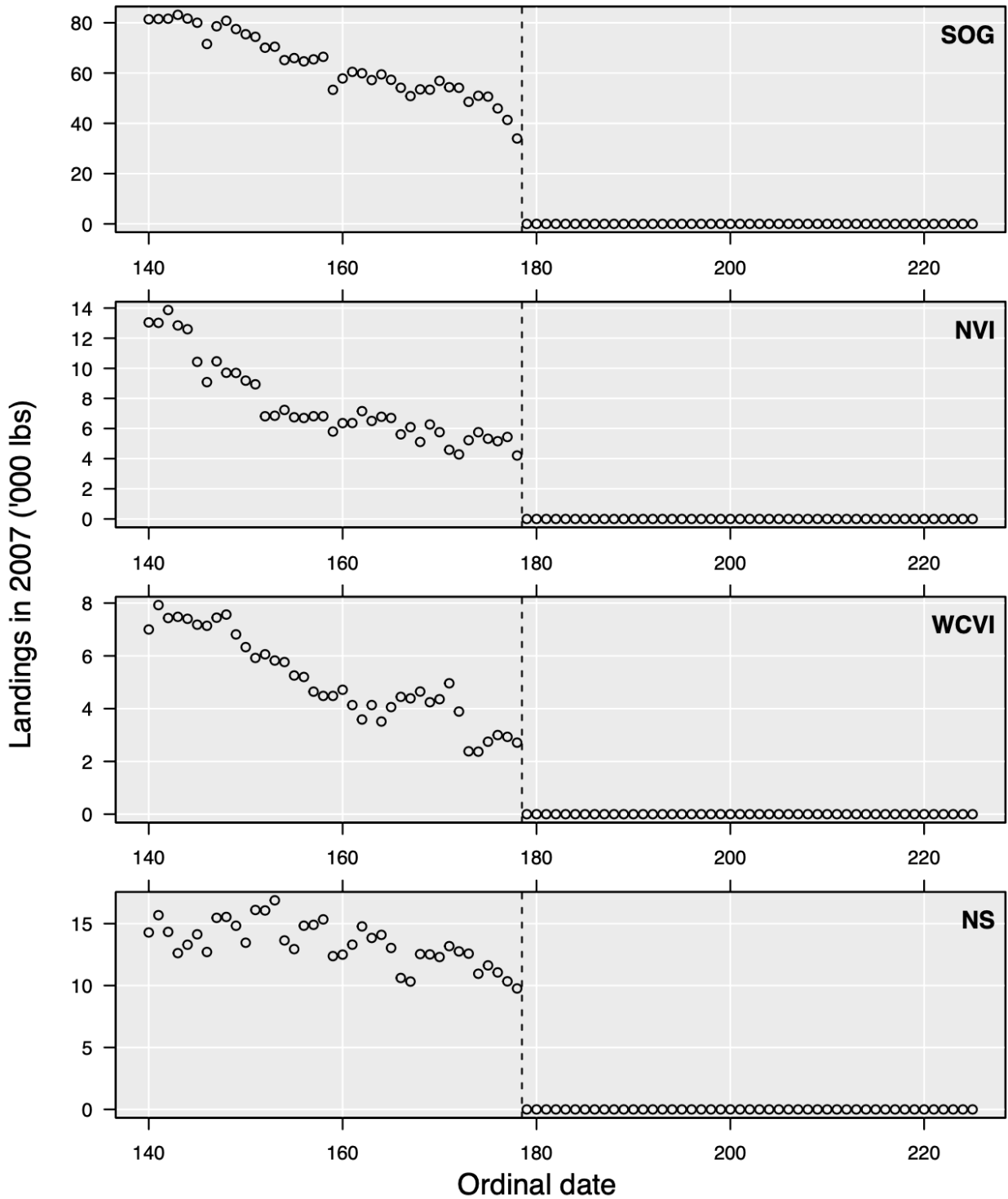

98

99  *Figure B.9. Daily landings in 2008 by region. The dashed line indicates the first day on which*
100 *daily landings declined by at least 80%.*

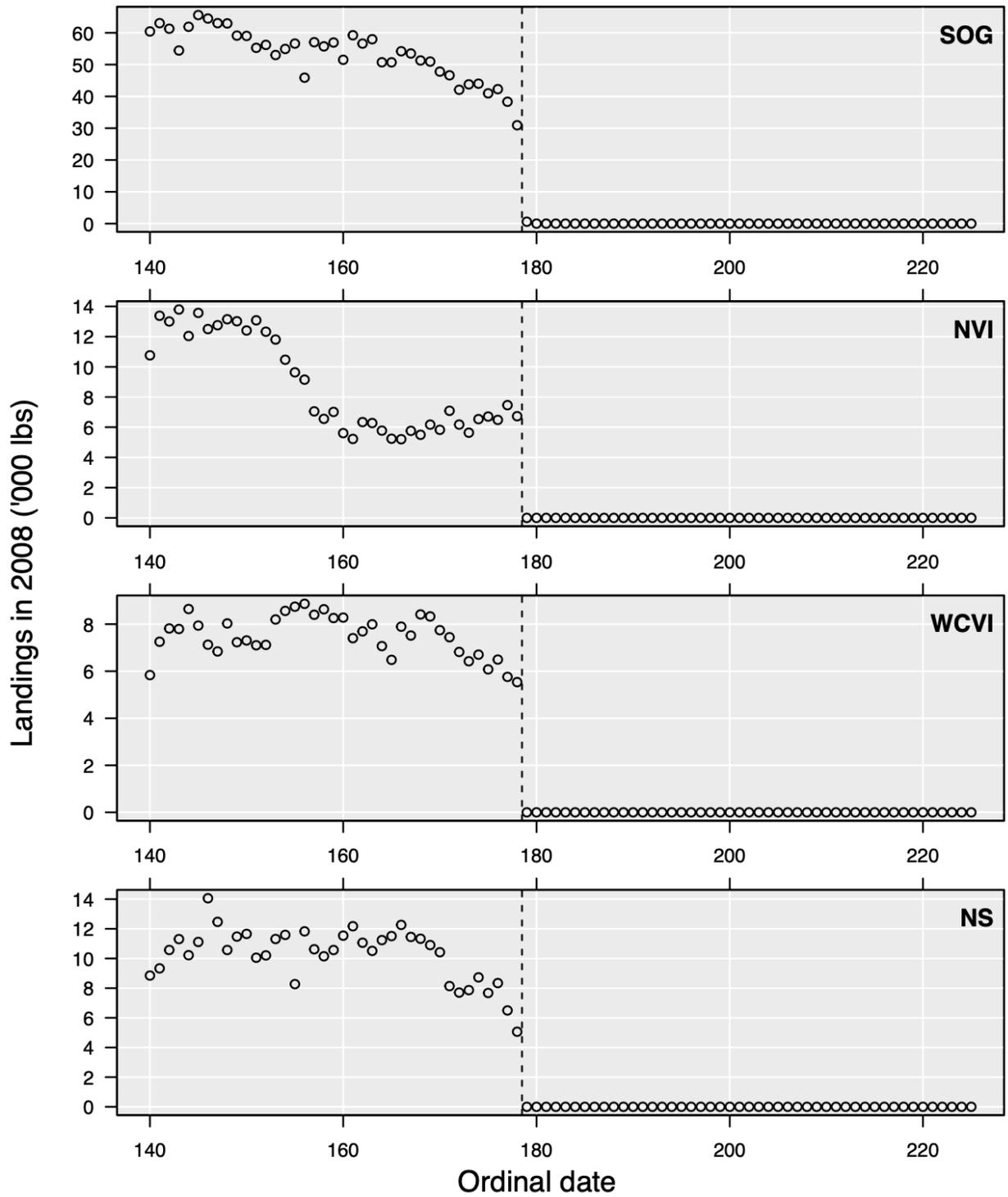

101

102  *Figure B.10. Daily landings in 2000 by region. The dashed line indicates the first day on which*
103  *daily landings declined by at least 80%.*

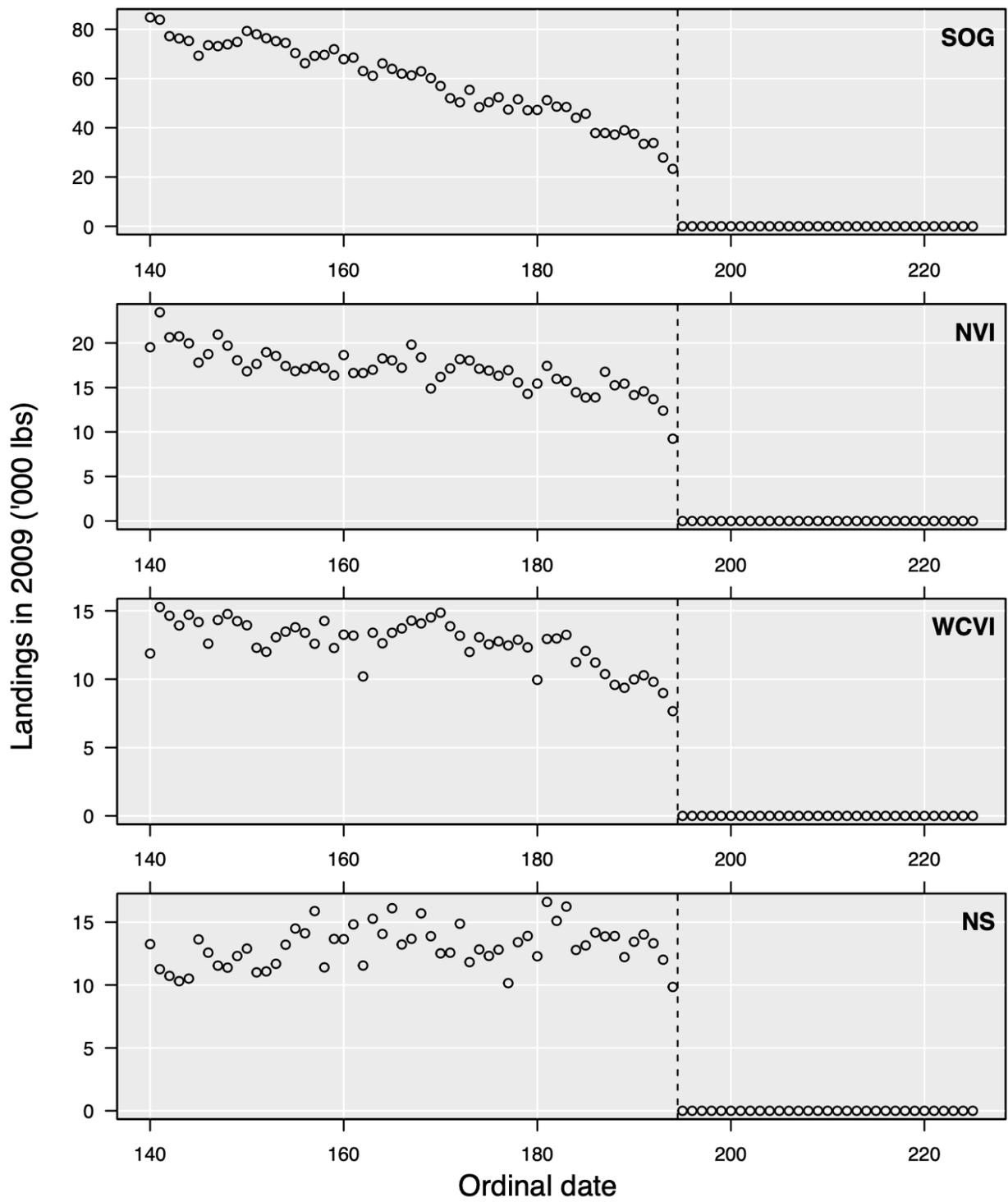

104

105  *Figure B.11. Daily landings in 2010 by region. The dashed line indicates the first day on which*
106  *daily landings declined by at least 80%.*

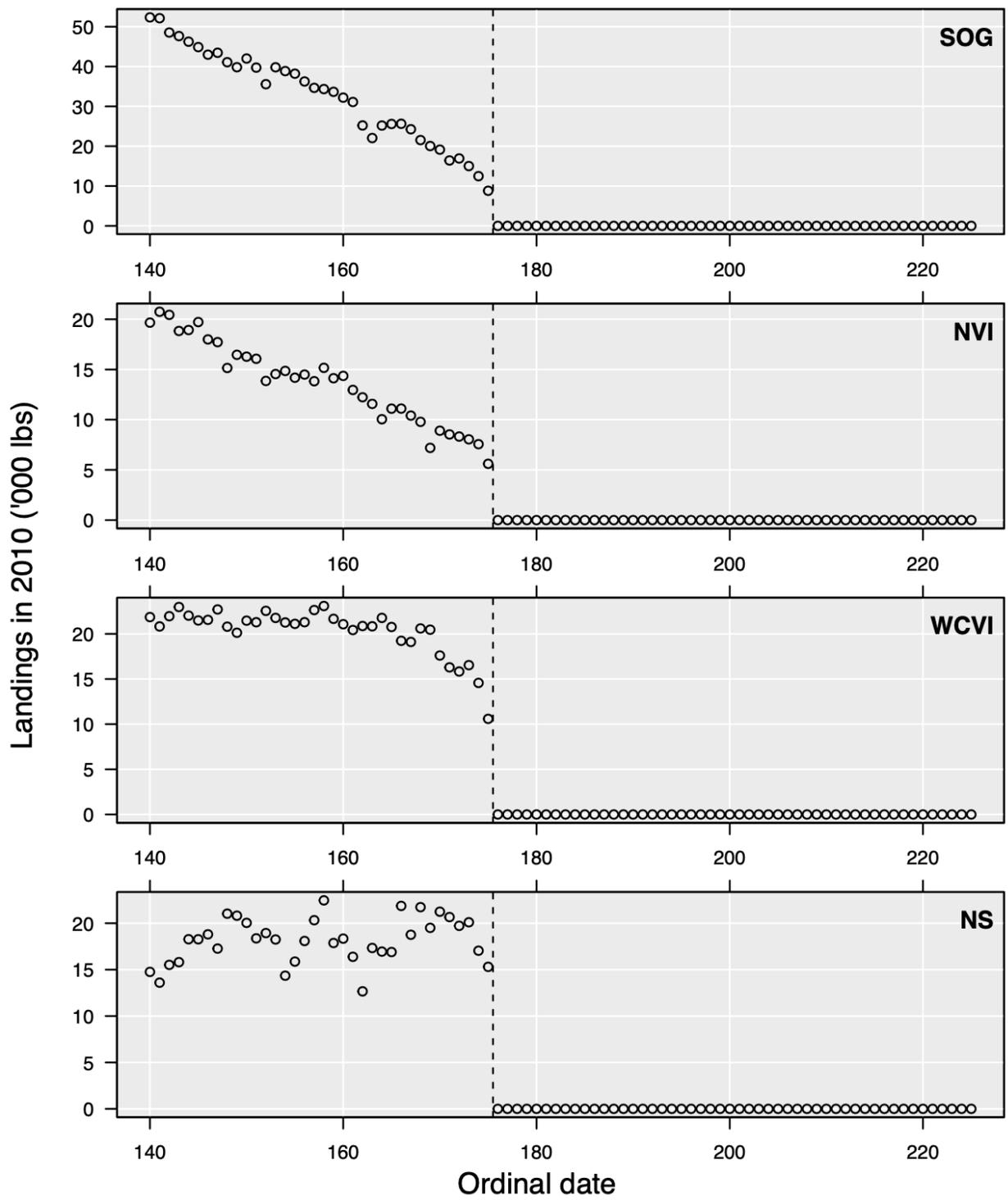

107

108  *Figure B.12. Daily landings in 2011 by region. The dashed line indicates the first day on which*
109  *daily landings declined by at least 80%.*

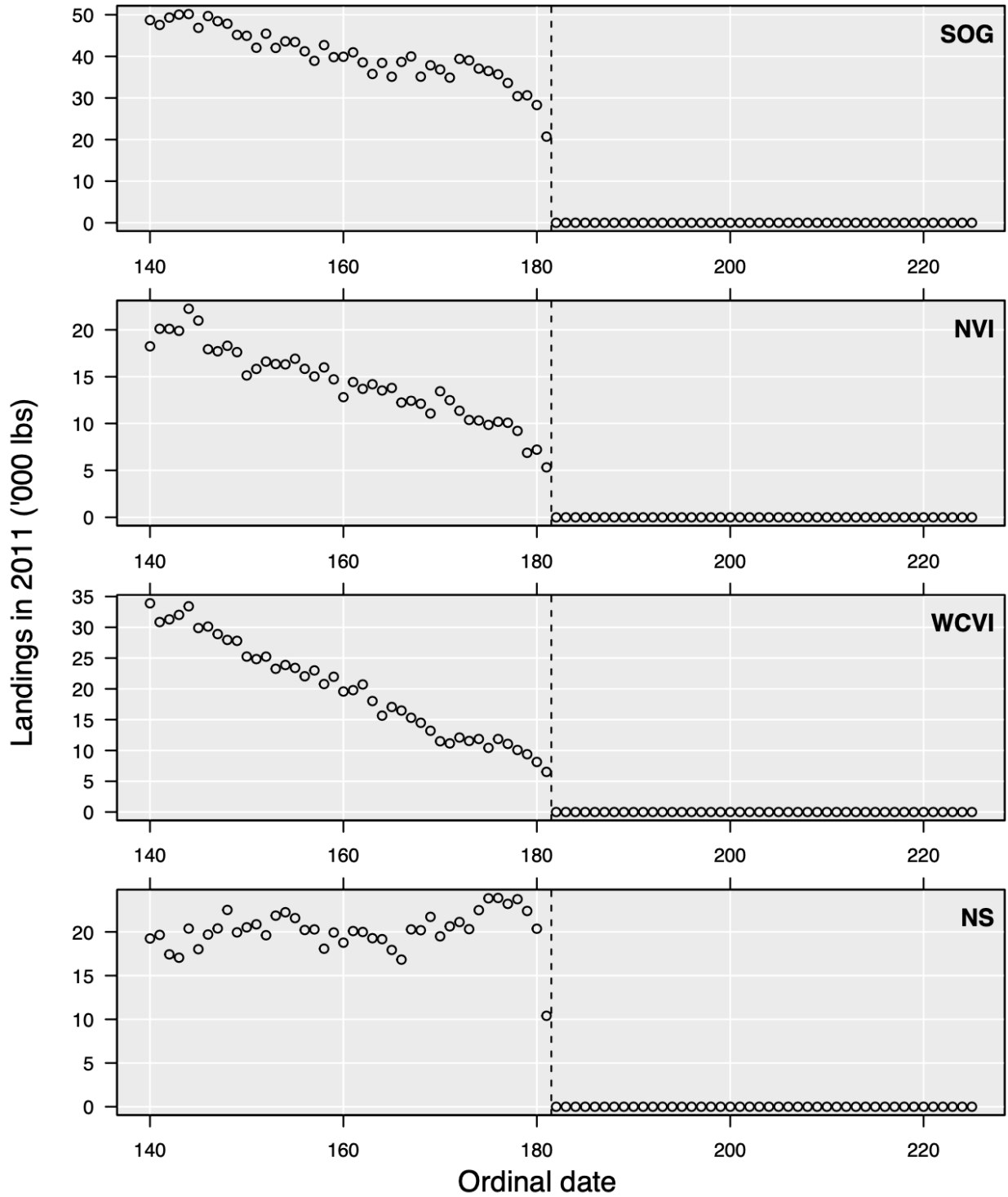

110

111 *Figure B.13. Daily landings in 2012 by region. The dashed line indicates the first day on which*
112 *daily landings declined by at least 80%.*

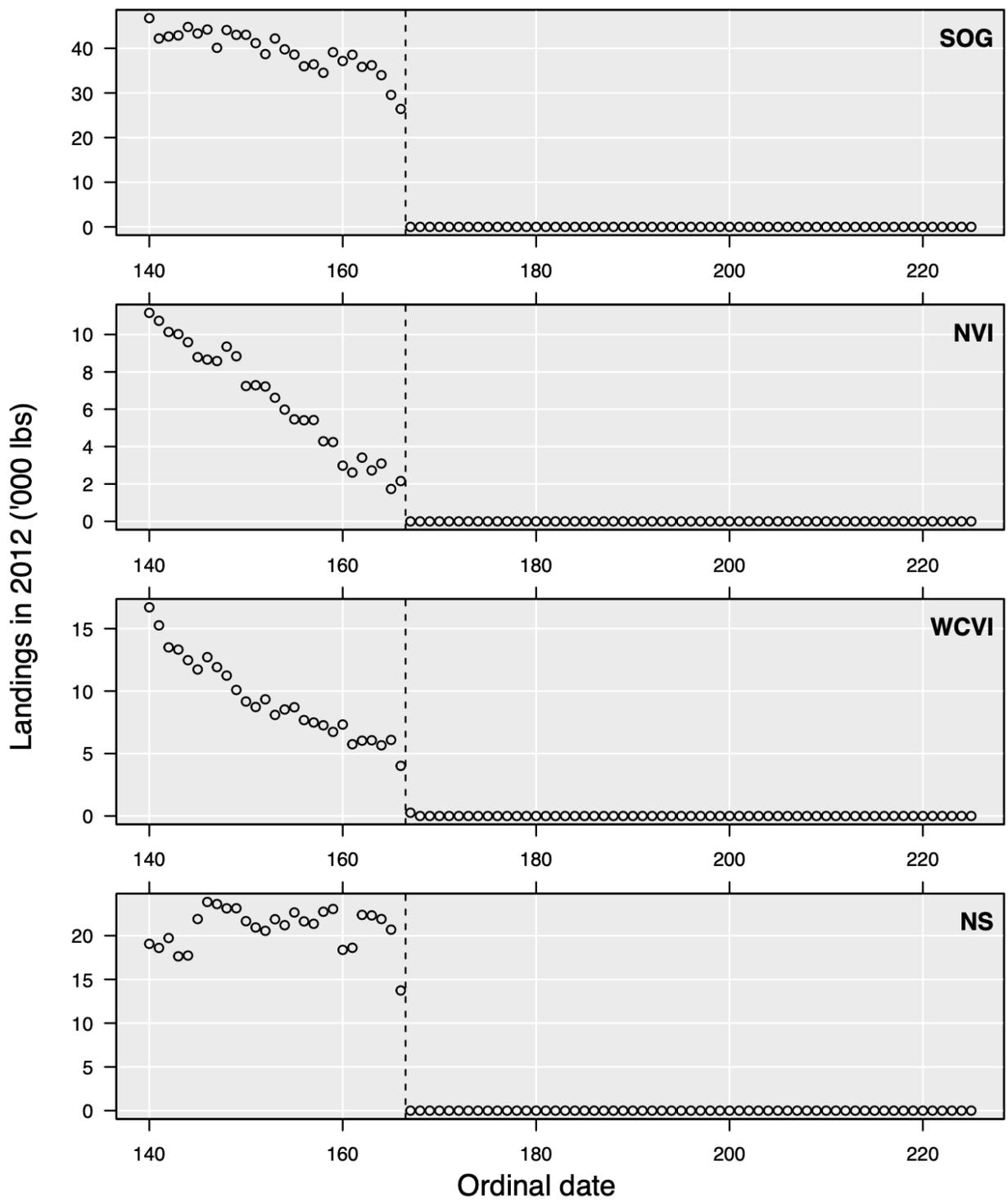

113

114  *Figure B.14. Daily landings in 2013 by region. The dashed line indicates the first day on which*
115  *daily landings declined by at least 80%.*

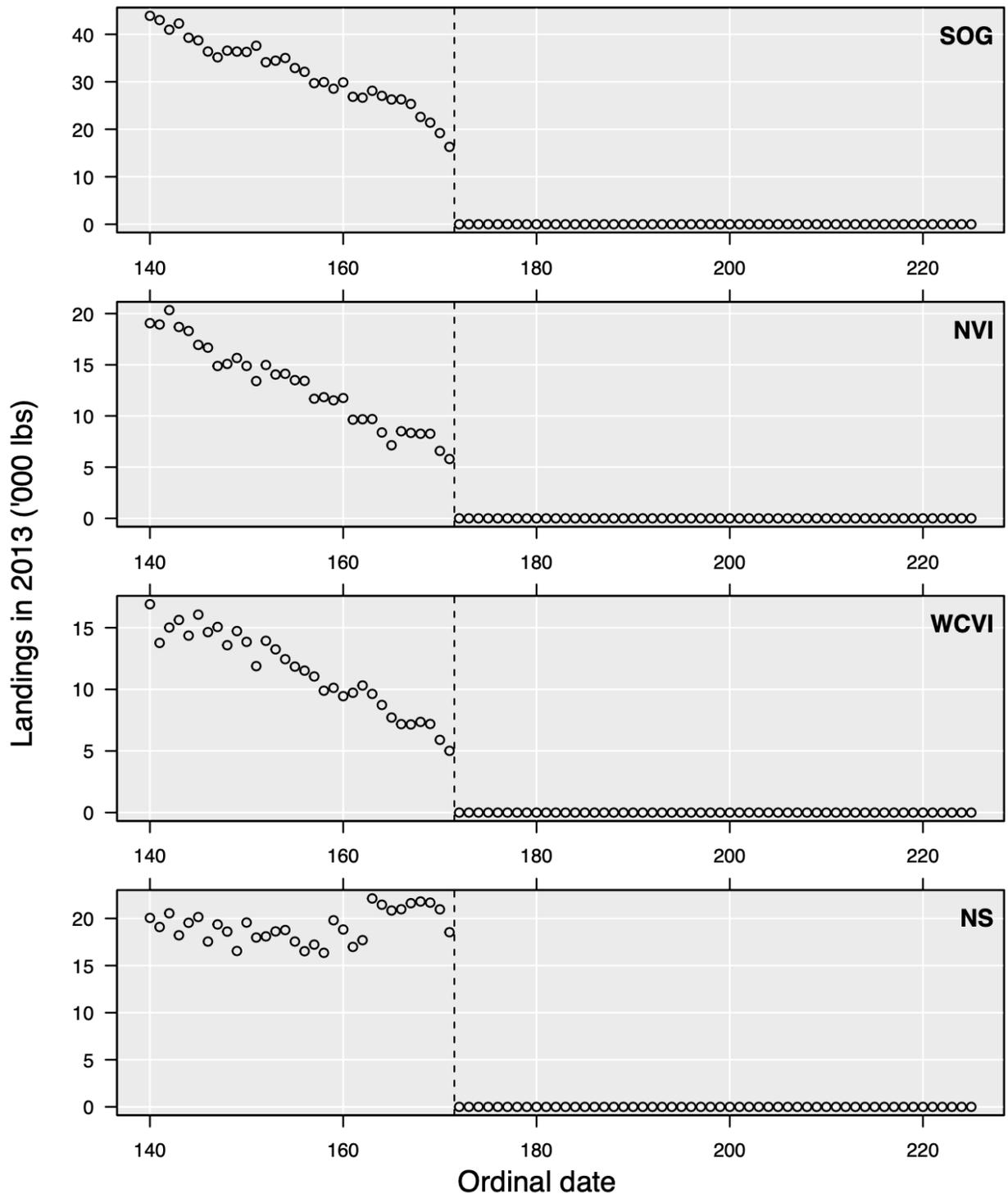

116

117  *Figure B.15. Daily landings in 2014 by region. The dashed line indicates the first day on which*
118  *daily landings declined by at least 80%.*

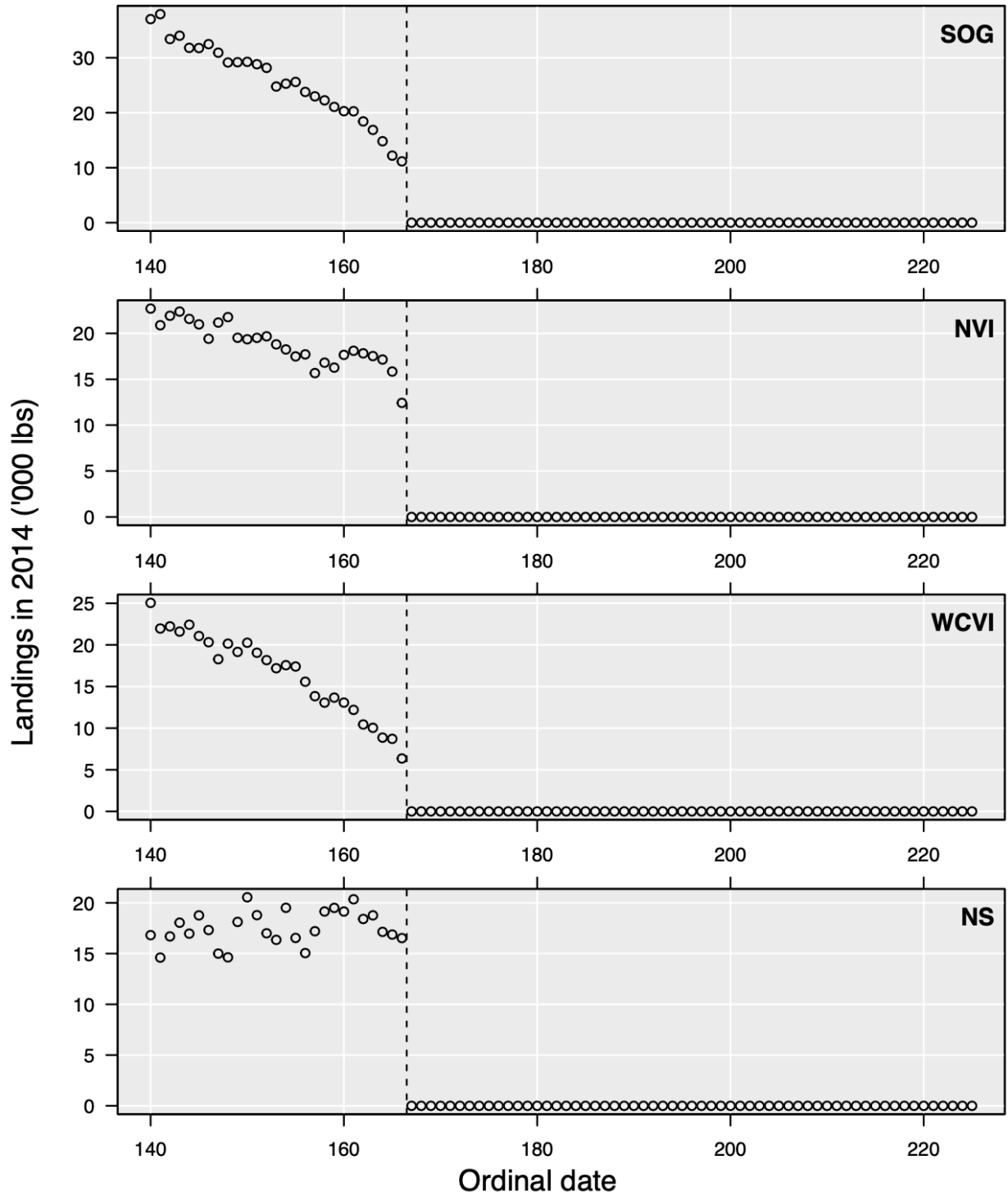

119

120  *Figure B.16. Daily landings in 2015 by region. The dashed line indicates the first day on which*
121  *daily landings declined by at least 80%.*

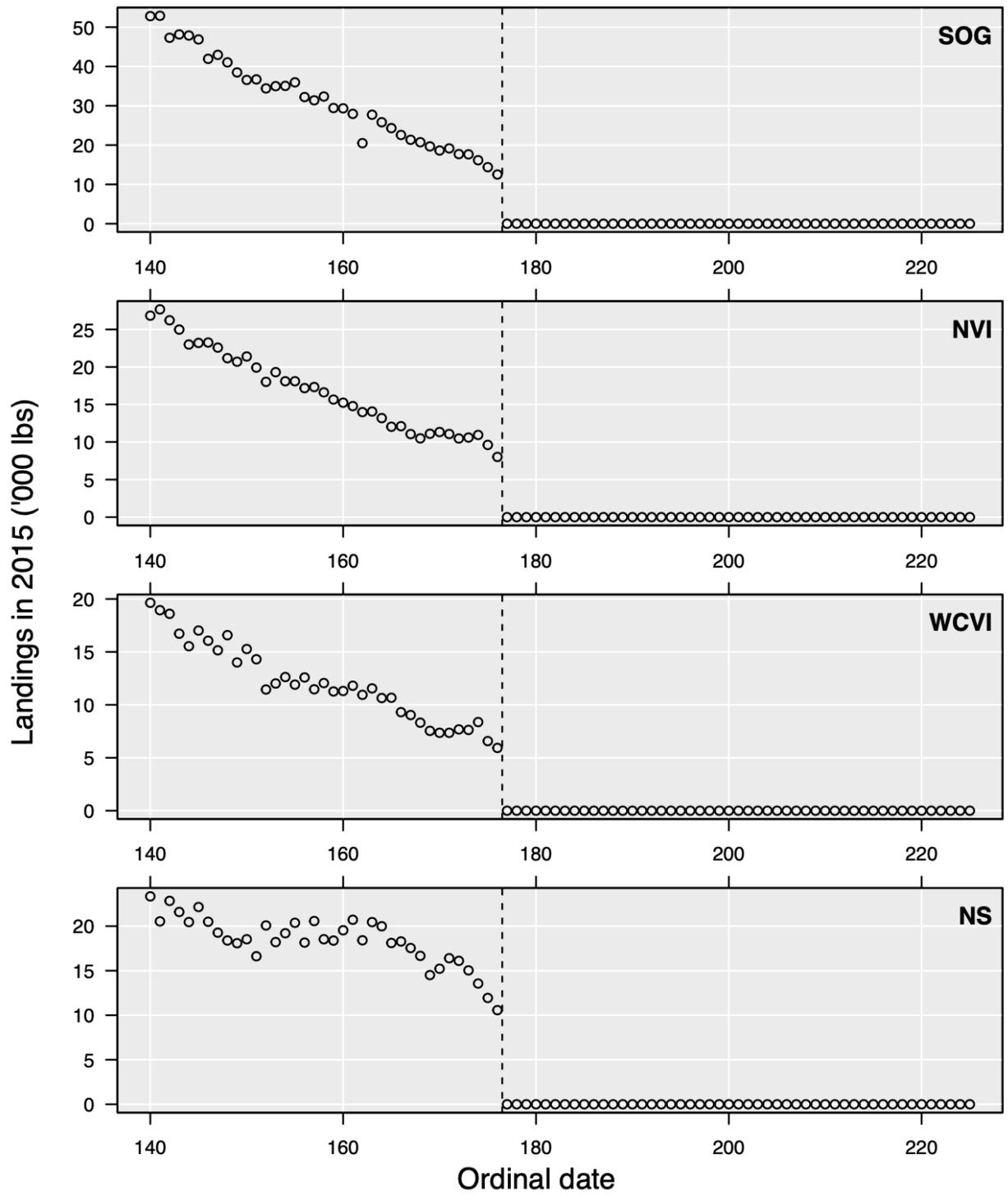

122

123  *Figure B.17. Daily landings in 2016 by region. The dashed line indicates the first day on which*
124  *daily landings declined by at least 80%.*

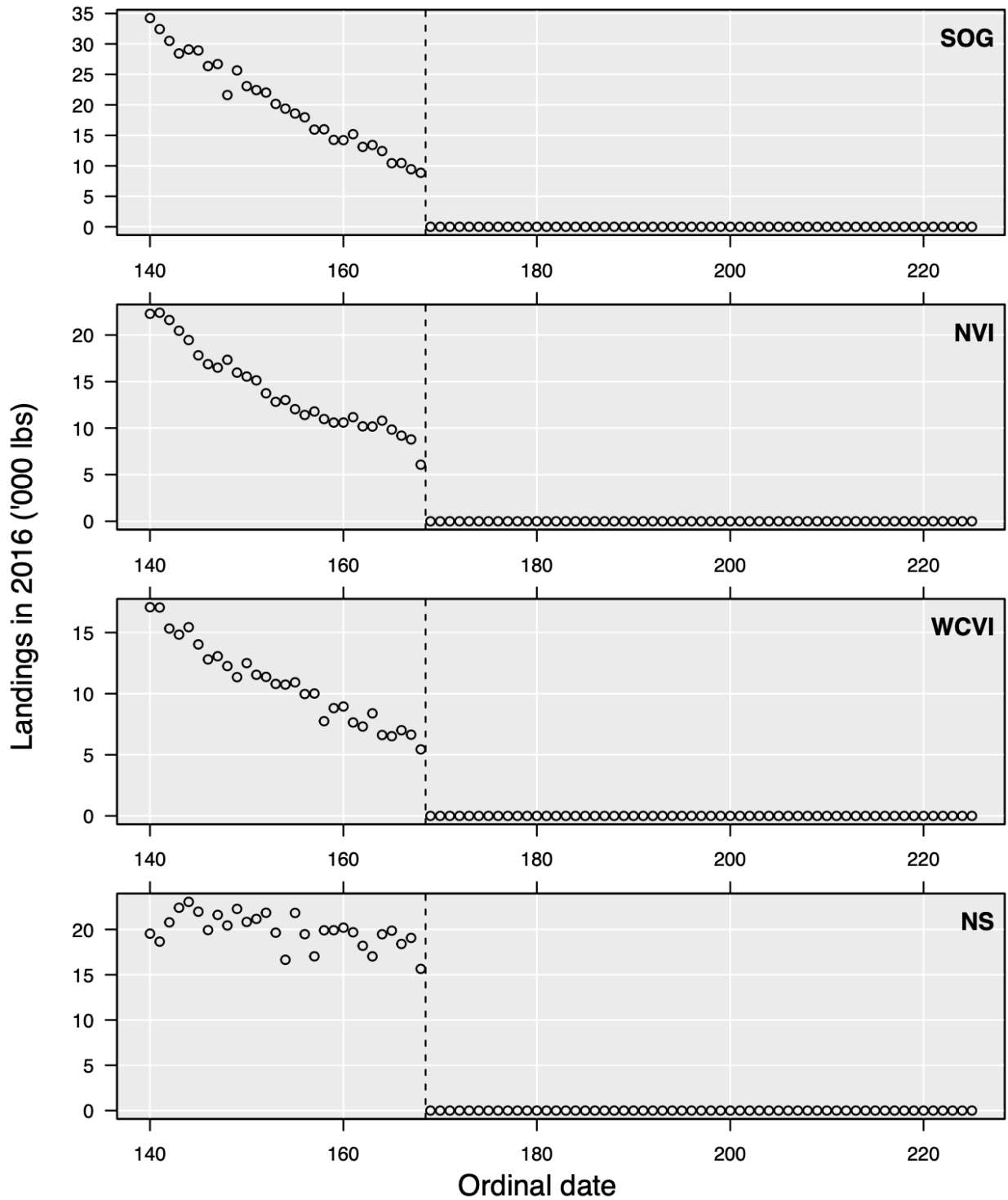

125

126  *Figure B.18. Daily landings in 2017 by region. The dashed line indicates the first day on which*
127  *daily landings declined by at least 80%.*

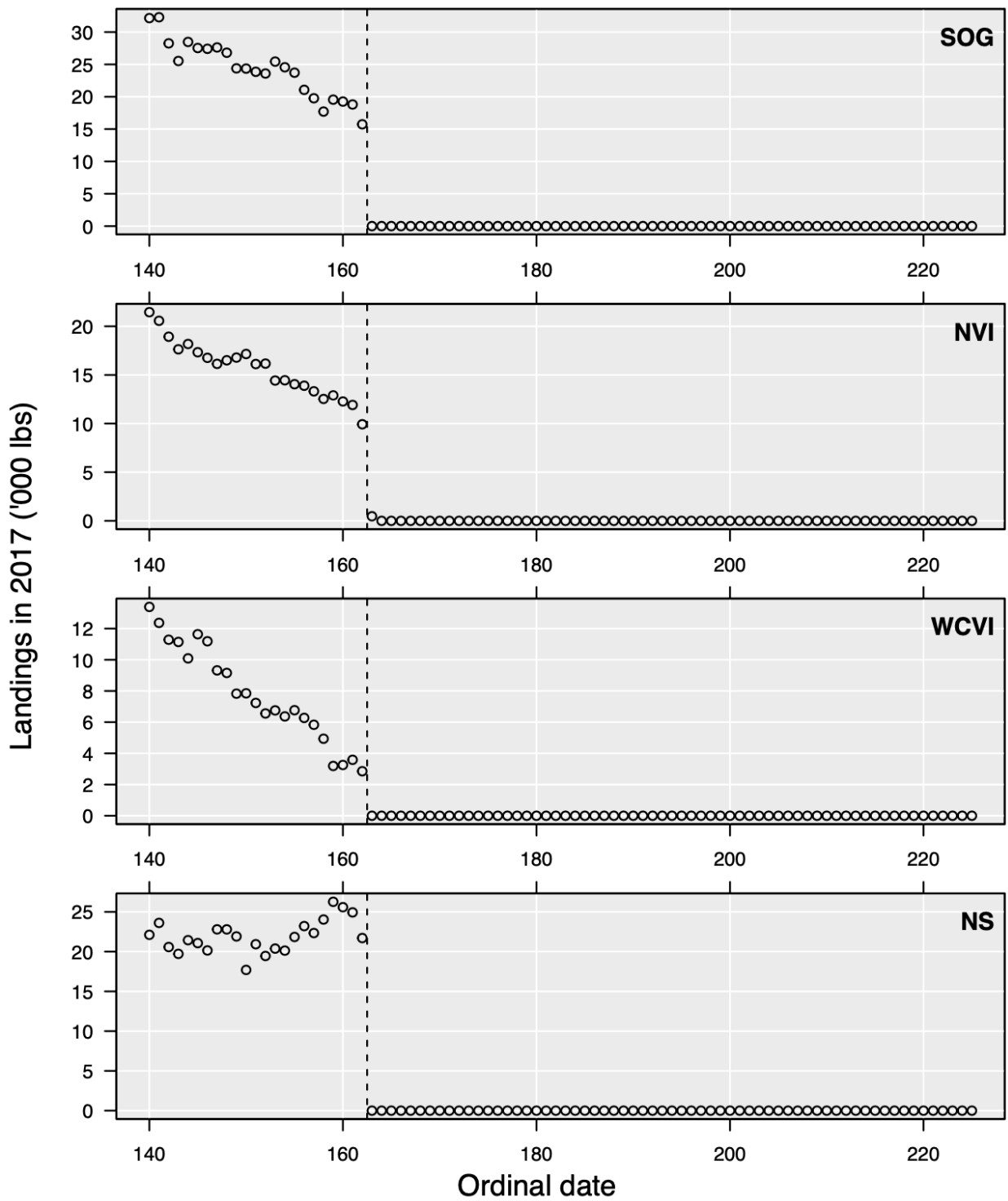

128

129  *Figure B.19. Daily landings in 2018 by region. The dashed line indicates the first day on which*
130  *daily landings declined by at least 80%.*

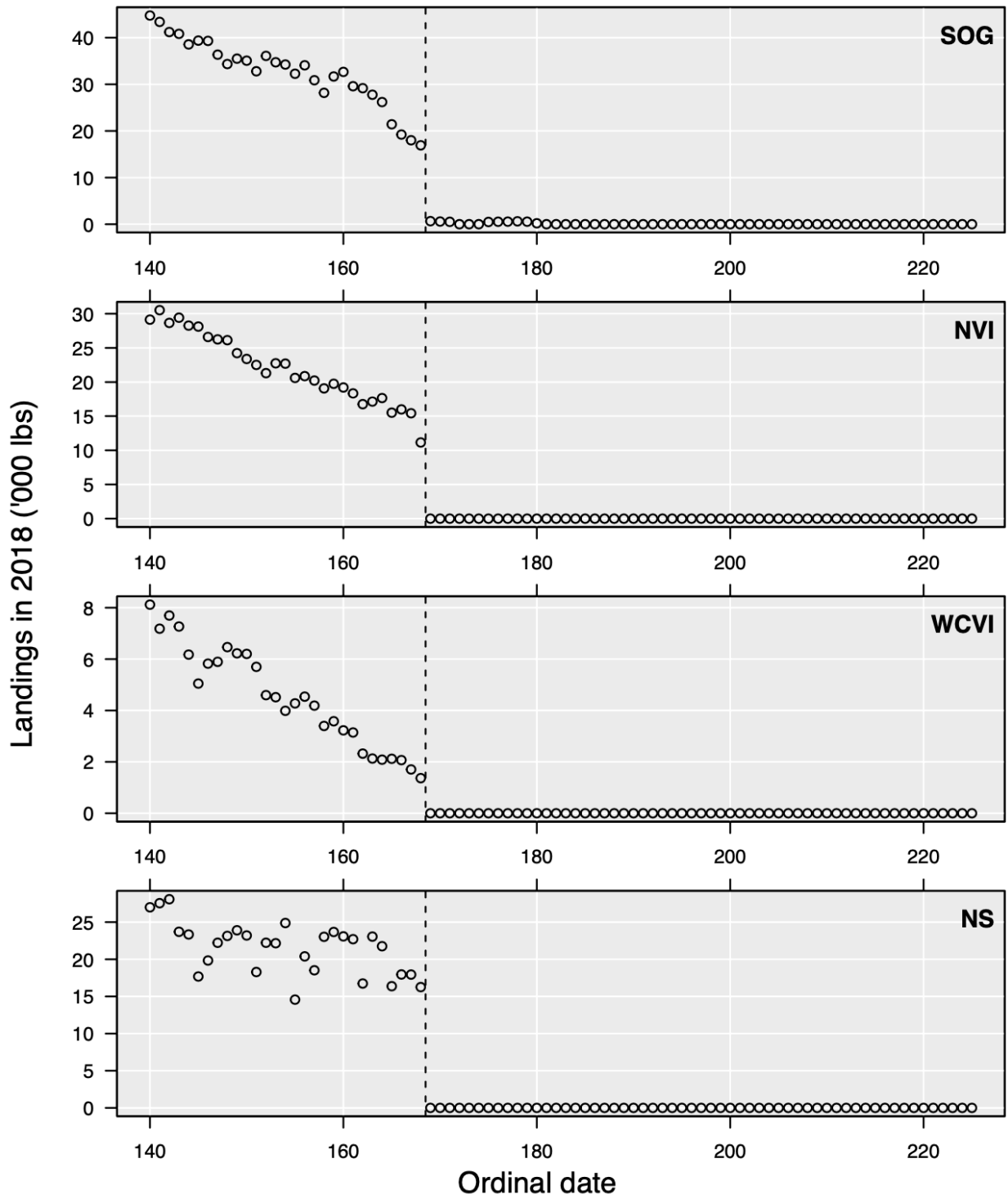

131

132  *Figure B.20. Daily landings in 2019 by region. The dashed line indicates the first day on which*
133  *daily landings declined by at least 80%.*

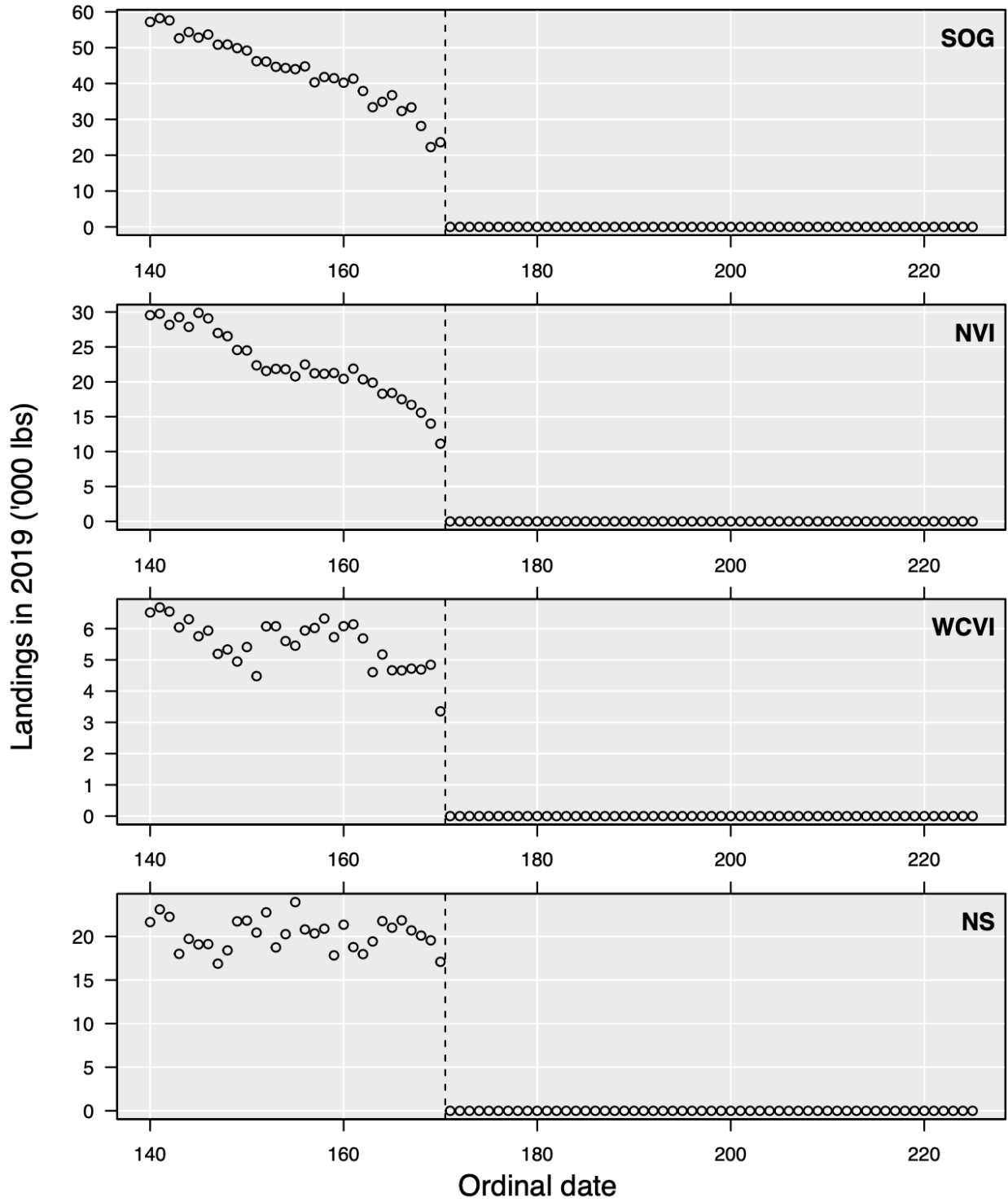

134
135